\documentclass[twocolumn]{aastex631}

\usepackage{xcolor}
\usepackage[normalem]{ulem}
\usepackage{comment}
\usepackage{amsmath}
\usepackage{subfigure}
\usepackage{graphicx}
\usepackage{multirow}
\usepackage{amssymb}
\usepackage{longtable}
\usepackage[figuresright]{rotating}

\newcommand{\arc}{$^{\prime\prime}$}
\newcommand{\feii}{[Fe\,{\scriptsize II}]}
\newcommand{\sii}{[S\,{\scriptsize II}]}
\newcommand{\ngi}{[N\,{\scriptsize I}]}
\newcommand{\ci}{C\,{\scriptsize I}}
\newcommand{\hei}{He\,{\scriptsize I}}
\newcommand{\hi}{H\,{\scriptsize I}}
\newcommand{\pii}{[P\,{\scriptsize II}]}
\newcommand{\neii}{[Ne\,{\scriptsize II}]}
\newcommand{\neiii}{[Ne\,{\scriptsize III}]}
\newcommand{\niii}{[Ni\,{\scriptsize II}]}
\newcommand{\arii}{[Ar\,{\scriptsize II}]}
\newcommand{\siii}{[S\,{\scriptsize III}]}

\graphicspath{{./}{figures/}}

\begin{document}

\title{Class I/II Jets with JWST: Mass loss rates, Asymmetries, and Binary induced Wigglings}

\correspondingauthor{Naman S. Bajaj}
\email{namanbajaj@arizona.edu}

\author[0000-0003-3401-1704]{Naman S. Bajaj}
\affiliation{Lunar and Planetary Laboratory, The University of Arizona, Tucson, AZ 85721, USA}

\author[0000-0001-7962-1683]{Ilaria Pascucci}
\affiliation{Lunar and Planetary Laboratory, The University of Arizona, Tucson, AZ 85721, USA}

\author[0000-0002-6881-0574]{Tracy L. Beck}
\affiliation{The Space Telescope Science Institute, 3700 San Martin Drive, Baltimore, MD 21218, USA}

\author[0000-0002-3232-665X]{Suzan Edwards}
\affiliation{Five College Astronomy Department, Smith College, Northampton, MA, USA}

\author[0000-0002-1593-3693]{Sylvie Cabrit}
\affiliation{Observatoire de Paris - PSL University, Sorbonne Université, LERMA, CNRS, Paris, France}
\affiliation{Univ. Grenoble Alpes, CNRS, IPAG, Grenoble, France}

\author[0000-0002-5758-150X]{Joan R. Najita}
\affiliation{NSF, NOIRLab, Tucson, AZ, USA}

\author[0000-0002-6429-9457]{Kamber Schwarz}
\affiliation{Max-Planck-Institut für Astronomie, Heidelberg, Germany}

\author[0000-0002-3913-7114]{Dmitry Semenov}
\affiliation{Max-Planck-Institut für Astronomie, Heidelberg, Germany}
\affiliation{Department of Chemistry, Ludwig-Maximilians-Universität, Munich, NY, Germany}

\author[0000-0003-3682-6632]{Colette Salyk}
\affiliation{Department of Physics and Astronomy, Vassar College, Poughkeepsie, NY, USA}

\author[0000-0002-3311-5918]{Uma Gorti}
\affiliation{Carl Sagan Center, SETI Institute, Mountain View, CA, USA}
\affiliation{Ames Research Center, NASA, Moffett Field, CA, USA}

\author[0000-0001-5638-1330]{Sean D. Brittain}
\affiliation{Department of Physics and Astronomy, Clemson University, Clemson, SC, USA}

\author[0000-0002-3291-6887]{Sebastiaan Krijt}
\affiliation{Department of Physics and Astronomy, University of Exeter, Exeter, UK}

\author[0000-0003-0522-5789]{Maxime Ruaud}
\affiliation{Carl Sagan Center, SETI Institute, Mountain View, CA, USA}

\author[0000-0002-5943-1222]{James Muzerolle Page}
\affiliation{Instruments Division, Space Telescope Science Institute, Baltimore, MD, USA}

\begin{abstract}

We present JWST NIRSpec spectro-imaging observations of jets from four edge-on protoplanetary disks that exhibit clear signatures of MHD disk winds. Bipolar jets are detected and spatially resolved in over 30 shock-excited forbidden lines, multiple Paschen and Brackett series lines of atomic hydrogen, and the high-energy excitation line of atomic helium (1.083~\micron{}). This helium line is the brightest jet-tracer towards HH~30 and FS~TauB, which also exhibit asymmetric intensity between their red- and blue-shifted lobes in all tracers including the \feii{} and \hei{} lines. Extinction maps reveal no significant differences across the lobes, suggesting an asymmetric jet-launching mechanism rather than environmental effects. Diagnostic line ratios yield consistent shock speeds of 50–60~km~s$^{-1}$, jet ionization fractions of 0.1–0.2, and pre-shock electron densities of 1000~cm$^{-3}$. Combined with pixel-by-pixel electron density maps and \feii{} line luminosities, we estimate jet mass-loss rates using three independent methods, averaging around a few 10$^{-9}$~M$_{\odot}$~yr$^{-1}$. We estimate the accretion rates for these sources as 10 $\times$ the jet mass loss rates and find them to match well with the independently-derived accretion estimates of other Class~II sources in the Taurus star-forming region. Owing to JWST’s high precision, we also investigate jet wiggling and find Tau~042021 to showcase the perfect case of mirror-symmetric wiggling, which can only be explained by the motion of the jet source around a stellar companion. Modeling this wiggling suggests Tau~042021 to host a 0.33 and 0.07~M$_{\odot}$ binary at the center with binary-separation of 1.35~au and an orbital-period of 2.5~years.

\end{abstract}


\section{Introduction} \label{sec:intro}

Protostellar jets are collimated (semi-opening angles $<20$\textdegree), fast-moving ($\sim$100-300 km~s$^{-1}$) flows of shocked gas emerging from the central region of a protoplanetary disk. They are tightly linked to accretion, as evidenced by a strong correlation between the jet mass loss rate ($\dot{M_j}$) and the mass accretion rate ($\dot{M}_{acc}$), with a ratio of $\sim$0.1 \citep[e.g.,][]{Nisini2018}. Jets appear to be the innermost part of the wide-angled (semi-opening angle $\sim$ 10-40\textdegree), slow-moving ($\sim$10-40 km~s$^{-1}$), radially extended disk winds that are launched from disk radii $\sim$0.1-20 au, with growing evidence suggesting that jet mass loss constitutes only ~10\% of the total mass loss from these winds \citep[e.g.,][for a recent review]{Pascucci2023}.

We observed one Class\,I and three Class\,II edge-on disks with the \texttt{James Webb Space Telescope} (JWST) Near Infrared Spectrograph (NIRSpec) to spatially resolve jet and wind emission in multiple diagnostic lines \cite[proposal ID - 1621,][]{Pascucci2021}. The edge-on configuration was selected as it offers clear visibility of the outflows and less contamination from the disk. Recently, in \citet[][\textbf{hereafter Paper I}]{Pascucci2024} we focused on the morphologies of these outflows, and a distinctive configuration attributed to MHD disk winds was found in all four sources where the narrow \feii{} jet is nested inside the wider, hollow H$_2$ wind cones. Particularly for HH~30, for which the wind was also detected in the rovibrational CO\,(v=1-0) and the pure rotational CO\,(J=2-1) lines, the nested morphology extended further with successively wider cones.

Simulations that include non-ideal MHD effects have found that the radially extended MHD disk winds remove enough angular momentum from the disk to drive accretion \citep[e.g.,][]{Bai2013}. We aim to test this theory observationally by comparing the wind mass-loss rates to the mass accretion rates for our observed systems, which are expected to be similar if the winds indeed drive accretion. The mass accretion rates for these systems would be significantly underestimated if using traditional methods of UV excess or \hi{} recombination lines without accounting for the high extinction from the edge-on disk \citep[e.g.,][]{Arulanantham2024}. Hence, in this study, we estimate the jet mass loss rates (Section \ref{sec:jmlr}) as an indirect tracer of the mass accretion rates using the empirical relation found between the two \citep[e.g.,][]{Hartigan1995,Nisini2018}. We will estimate the wind mass-loss rates using the spatially resolved H$_2$ lines in future work (Beck et al. in prep, \textbf{hereafter Paper III}).

Thanks to the greater sensitivity of JWST compared to existing ground-based facilities, we will also characterize the jet in detail and estimate various jet properties. We find our dataset to be extremely rich, with 34 atomic transitions tracing jets in the wavelength range corresponding to the G140H/F100LP grating-filter ($\sim$0.97-1.82~\micron{}, Sections \ref{sec:linedet}, \ref{sec:new_lines}). We use \feii{} transitions sharing the same upper energy level to create pixel-by-pixel extinction maps (Section \ref{sec:ext_map_corr}) along the jet to assess whether dust extinction is the prime source of observed jet asymmetries (Section \ref{sec:jet_asymmetry}) and to provide accurate fluxes for all lines, yielding valuable diagnostic ratios. We calculate jet mass loss rates from pixel-by-pixel electron density maps (Section \ref{sec:elec_den}) coupled with line fluxes and estimates of the shock speed and ionization fraction provided by various line ratios (Section \ref{sec:shock_velocity_density}). Additionally, owing to the unprecedented spatial resolution and precision of JWST at these near-IR (NIR) wavelengths, we trace these jets very close to the star ($<$0.5\arc{}), which enables us to look for signatures of stellar multiplicity through a comparison of the jet lateral motion on the two sides, often referred as jet wiggling (Section \ref{sec:jet_wigg_obs}). Binaries are the only mechanism known to create mirror-symmetric jet wiggles and the wiggle period and amplitude can well constrain the binary separation and mass ratio \citep[e.g.,][]{Anglada2007,Lee2010}, which are otherwise extremely difficult to obtain for edge-on disks where the central star(s) is blocked by the disk. Lastly, we provide conclusions in Section \ref{sec:conclusion}. 

\section{Targets, Observations and Data Reduction} \label{sec:obs&data}

All four sources observed here (HH~30, FS~TauB, Tau~042021, IRAS~04302) are part of the Taurus star-forming region, which is located at an average distance of 140 pc with an estimated age of $\sim$1-2\,Myr \citep[e.g.,][]{Luhman2004}. These objects were selected based on their edge-on configuration, known dynamical stellar masses \citep{Villenave2020}, and to span a range of stellar masses \citep{Pascucci2021}. In the following, we provide more details about the sources (Section \ref{sec:overview}) and the observations and data reduction (Section \ref{sec:obs&red}).

\subsection{Overview of the Sources} \label{sec:overview}

The main star and disk properties of our sample are summarized in Table~1 of Paper~I. In short, all disks are inclined by more than 70\textdegree{} and are fairly large (dust radii $\ge 130$\,au), with the central stars spanning a range of stellar masses ($\sim 0.4-1.7$\,M$_{\odot}$). Only IRAS~04302 is surrounded by a significant envelope; hence, it is likely a Class~I source.

The jets of FS~TauB (also known as Haro~6-5B) and HH~30 were first imaged by \cite{Mundt1983,Mundt1984} and have been well characterized at multiple wavelengths, including optical/near-IR \citep[e.g.,][]{Mundt1990,Bacciotti1999,Hartigan2007} and radio \citep[mm,][]{Louvet2018,Lopez-Vazquez2024}. They are known to be asymmetric in length and brightness \citep{Eisloffel1998,Bacciotti1999,Liu2012} and our study provides high-resolution visual extinction maps to evaluate if the asymmetries are intrinsic to the launch mechanism (Section \ref{sec:ext_map_corr}). Jet mass loss rates for these sources are reported in the literature using optical data \citep{Mundt1987,Mundt1990,Bacciotti1999}; here we will use the NIR lines observed with JWST and use three different methods to calculate the jet mass loss rates. In contrast, the jets from IRAS~04302 (also known as the `Butterfly star') and Tau~042021 are relatively less studied. Recently, \cite{Arulanantham2024} detected a jet towards Tau~042021 in the forbidden atomic lines of \neii{}, \neiii{}, \niii{}, \feii{}, \arii{}, and \siii{}. They also estimate a mass accretion rate of 2~$\times$~10$^{-11}$~M$_{\odot}$~yr$^{-1}$ with the \hi{} Humphreys $\alpha$ line which, as we will show in Section \ref{sec:Taurus}, is significantly underestimated. All jets have been laterally resolved with our NIRSpec data and the semi-opening angles using the \feii{} 1.644~\micron{} line are provided in Table~2 of Paper~I. 

\subsection{Observations and Data Reduction} \label{sec:obs&red}

\begin{figure*}
    \centering
    \includegraphics[width=7in]{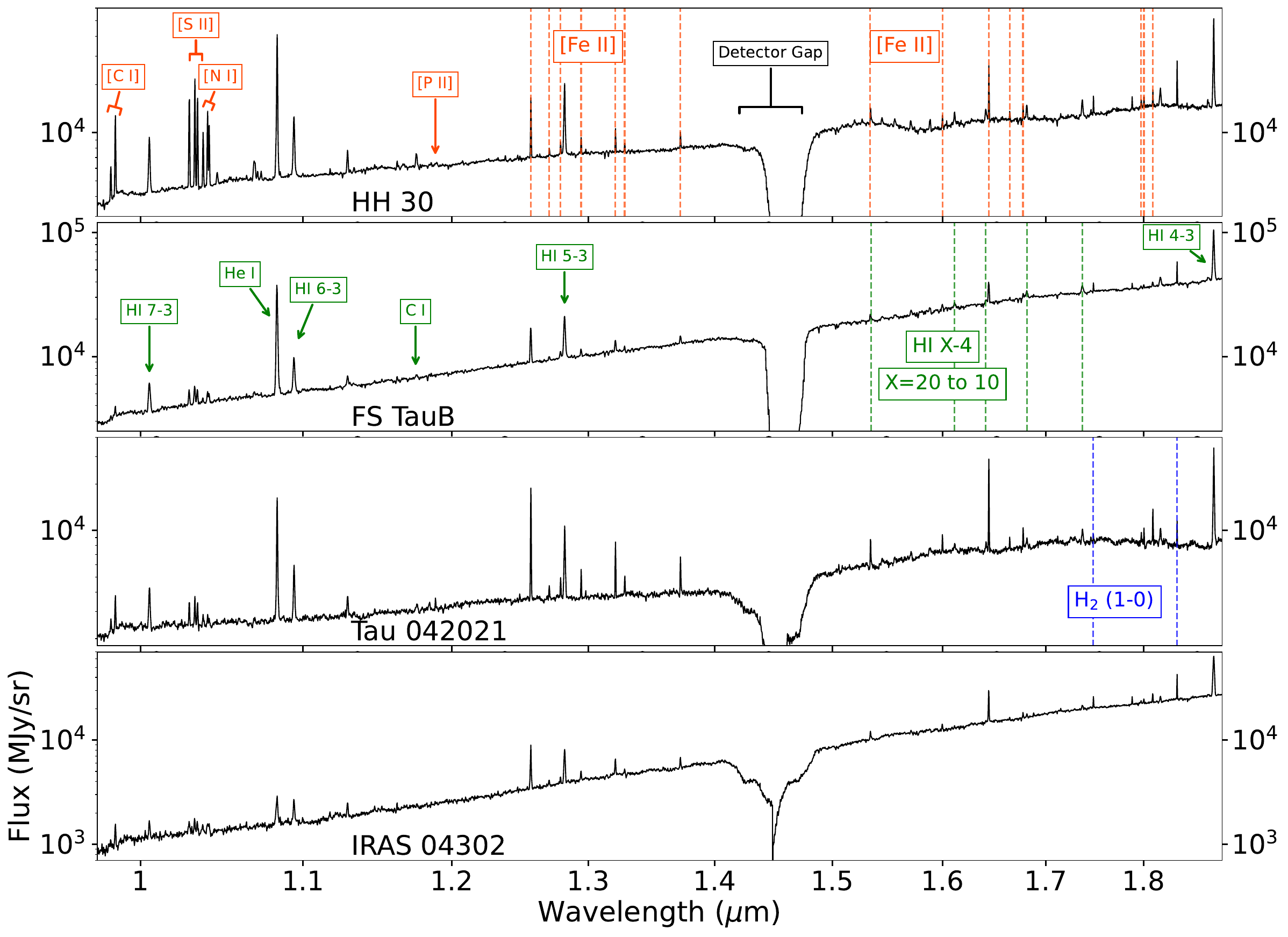}
    \caption{NIRSpec 0.97-1.82 \micron{} (G140H/F100LP grating filter) spectra of all the sources summed over the entire field of view. The forbidden atomic lines are identified in the first panel and other atomic and molecular lines in the second and third panels, respectively, but apply to all the panels. The strong dip in all the spectra between $\sim$\,1.408-1.486 \micron{} corresponds to the detector gap and is highlighted in the first panel.}
    \label{fig:spec_img}
\end{figure*}

All sources were observed with the NIRSpec Integral Field Unit (IFU) between Sep 17 and 26, 2022 using the same settings and exposure times. They were each observed for 1.5 hrs on-source, using a 4-point dither pattern optimized for an extended source. A total of 30 groups were taken in `NRSIRS2RAPID' readout mode and the integration time per grating-filter combination and complete dither was 1809s. The data were taken in three high-resolution grating-filter combinations [G140H/F100LP, G235H/F170LP, and G395H/F290LP] covering the wavelength range of $\sim$ 0.97$-$5.2\,$\micron$. 

We used version `1.13.4.dev19+gbddb39c6' of the JWST calibration pipeline \citep{Bushouse2024}, the latest version made available as of January 25, 2024. We used the Calibration Reference Data System (CRDS) version `11.17.15' and context `jwst\_1188.pmap' for reference files and their selection rules, respectively. The data reduction began with the retrieval of `uncal' files from the Mikulksi Archive for Space Telescopes (MAST\footnote{The Mikulski Archive for Space Telescopes (MAST) is a NASA-funded project to support and provide a variety of astronomical data archives, with the primary focus on the optical, ultraviolet, and near-infrared parts of the spectrum. MAST is located at the Space Telescope Science Institute (STScI).}). We performed ramps-to-slopes processing of these files using the \texttt{Detector1Pipeline} function from jwst\_pipeline. We customized the \texttt{jump} step by switching on \texttt{expand\_large\_events} which detects `snowballs' in the data, and we set \texttt{after\_jump\_flag\_dn1} to 1000 and \texttt{after\_jump\_flag\_time1} to 50 which flags the groups after jump with data number (DN) above 1000 and any groups within the first 50 seconds, respectively. We used multiple cores to perform the jump step to speed up the process. Next, we ran the \texttt{Spec2Pipeline}, which includes important calibration steps such as photometric calibration, flat field correction, wcs assignment, etc. We used all the default parameters except the `NSClean' algorithm, which was switched off by default. This algorithm was developed particularly to remove faint vertical banding and picture frame noise (background) from NIRSpec IFU \citep{Rauscher2024}. For one of our sources, Tau~042021, we found that using the `NSClean' algorithm improves the spectral match in the overlapping wavelength regions.

The next step of the JWST calibration process is to run the \texttt{Spec3Pipeline}, which stitches the dithers together, rejects outliers, and creates IFU cubes and spectra, but we found that the outlier detection missed many outliers. As such, before running the \texttt{Spec3Pipeline}, we performed outlier detection and rejection using a custom script (J. Morrison, personal communication, 2023) where the files are first checked for their data quality (DQ) flags, which were set in the Detector1 step. All the pixels with DQ flags as `NO\_SAT\_CHECK' and `UNRELIABLE\_FLAT' are set to `DO\_NOT\_USE' if they are not already. Also, the pixels with the value `nan' are set to `DO\_NOT\_USE'. A (7,7) kernel is defined with one dimension in the image plane, whereas the other is in the wavelength plane, and any value above 99.6 percentile of the values within the kernel is flagged. As a second check, the process is repeated with a kernel size of (3,3). During the second process, the aim is mainly to flag the large positive and large negative pixels. After completing this process, we run the \texttt{Spec3Pipeline} with these modified `cal' files and create the IFU cubes in the `ifualign' mode to avoid any further data interpolation. To verify that no useful scientific data were flagged, we compare the output cube fluxes with the cubes created following the standard calibration process, i.e., without using the above-described custom outlier detection step.

Finally, to properly compare NIRSpec IFU data across the full wavelength range, we deconvolve the cubes at each wavelength using a corresponding model point spread function (PSF). In doing this, we effectively remove the PSF broadening as a function of wavelength. The PSF datacubes for NIRSpec IFU were generated by deconvolving commissioning observations of a point source \citep[PID 1128\footnote{\url{http://doi.org/10.17909/2c5e-dm80}},][]{Leonardo2023} using a model NIRSpec cube created with the WebbPSF package \citep{Perrin2014}. In this process, we apply the optical path differential (OPD) file relevant to the date of the commissioning observation. Then, we construct a convolution kernel cube to adjust for optical variations between the actual data and PSF model predictions. This kernel enables the alignment of any WebbPSF OPD calculation with an observed NIRSpec IFU dataset. In this study, we generate the PSF model using the WebbPSF OPD model for the time of the HH~30 observations (wavefront error of 75~nm rms) and convolve using our kernel to match the scientific data. Additional details on the PSF deconvolution are provided in the Methods section of Paper~I; full details and an application to wind mass loss rate estimates will be discussed in Paper~III.

\begin{figure*}
    \centering
    \includegraphics[width=6in]{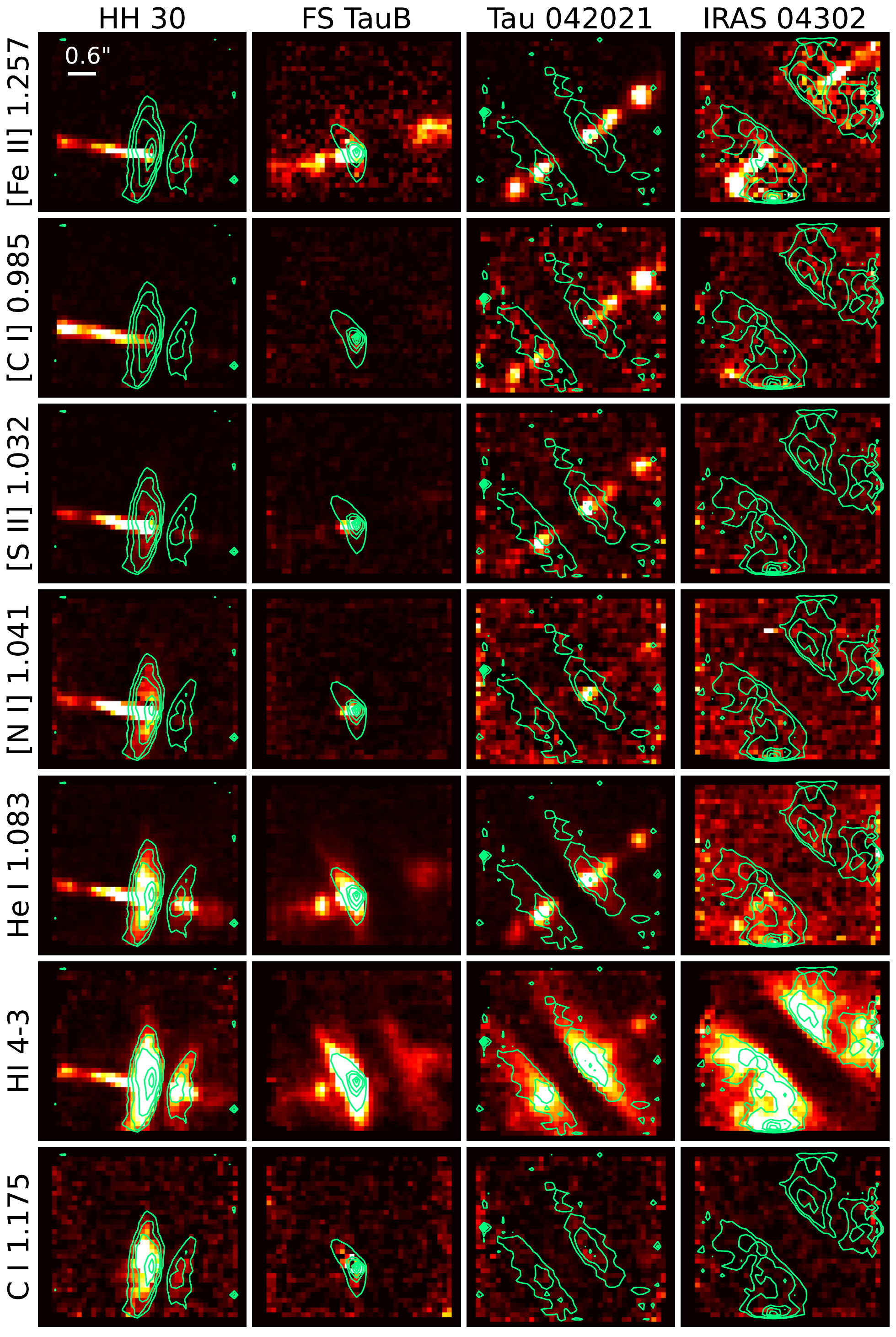}
    \caption{Continuum-subtracted emission in the following lines: \feii{} 1.257 \micron{}, [\ci{}] 0.985 \micron{}, \sii{} 1.032 \micron{}, \ngi{} 1.040+1.041 \micron{}, \hei{} 1.083 \micron{}, \hi{}~4-3, and permitted \ci{} 1.175 \micron{} towards all four sources. For \ngi{}, we integrate the emission over both 1.040 \micron{} and 1.041 \micron{} transitions to increase the S/N. The respective continuum emission is overlaid on the line emissions and is highlighted in green.}
    \label{fig:line_img}
\end{figure*}

\section{Line Morphologies and Extinction} \label{sec:results}

In this section, we first present the integrated spectra in the shortest wavelength cube (G140H) and the method employed to calculate the integrated line fluxes. Then, we detail the method used to create the continuum-subtracted line emission images to investigate the morphologies traced by different lines. Lastly, we estimate extinction towards the jets using the \feii{} 1.257 and 1.644 \micron{} lines, which share the same upper energy level, making their ratio only a function of extinction. 

\subsection{Line Detections} \label{sec:linedet}

To evaluate which gas emission lines are detected, we generate a 1D spectrum, where the flux at each wavelength is estimated by summing all the pixels in the image cube at that wavelength. Extended emission from our targets covers the majority of the NIRSpec FOV (3\arc{}\,x\,3\arc{}) necessitating to sum over all the pixels (see, e.g., Figure \ref{fig:line_img}). These spectra are shown in Figure \ref{fig:spec_img} with several strong transitions highlighted. At these short wavelengths ($\sim$0.97-1.82\,\micron{}), we find several atomic and forbidden atomic lines and a few molecular lines. 

In HH~30, which has the strongest lines and the richest spectrum, we detect a total of nineteen \feii{} transitions, which provide a comprehensive inventory of jet tracers imaged at the same spatial resolution. We also find two [\ci{}], two \pii{}, four \sii{}, and two \ngi{} lines. We find the \ngi{} 1.040/1.041 \micron{} doublet to be deblended when detected, which is seen for the first time towards low-mass young stars \citep[see][for their recent discovery of deblended forbidden N\,I lines towards an intermediate-mass young star]{Katoh2024}. Additionally, we detect two permitted transitions of \ci{}, and \hei{}, each and several hydrogen lines from the Paschen (n=3) and Brackett (n=4) series, namely \hi{} (4 to 7)-3 and \hi{} (10 to 20)-4 in the lowest wavelength range. The remaining Brackett lines in the series (\hi{} (5 to 9)-4) are detected at higher wavelengths (Paper I). Finally, H$_2$ is the only molecule observed at these wavelengths, and we find the (1-0), (2-0), and (3-1) transitions in the spectra of HH~30. Some of the above-mentioned lines are weaker and/or not detected towards other sources. A full list of lines and their integrated fluxes are provided in Table \ref{tab:line_fluxes} along with 3$\sigma$ upper limits when not detected. Note that the fluxes listed in Table \ref{tab:line_fluxes} are total observed fluxes, i.e., they can include contributions from the scattered light close to the star. To investigate the morphology of the observed lines, we subtract the continuum under the line in each spaxel. 

\subsection{Spaxel-by-spaxel Continuum Subtraction} \label{sec:cont_sub}

We generate line-only maps for select lines by removing the continuum emission contribution in each spaxel. To do this, we consider $\sim$ 30 continuum data points on both sides of the line emission (excluding the line itself) and use \texttt{sklearn.linear\_model.RANSACRegressor}\footnote{\url{https://scikit-learn.org/stable/modules/generated/sklearn.linear_model.RANSACRegressor.html}} to fit a straight line. For wavelength points in a spaxel under consideration, we subtract the continuum model flux and create a corresponding continuum-subtracted cube. In this continuum-subtracted cube, to calculate integrated line fluxes in each spaxel, we consider $\sim$ 12 wavelength points on either side of the line center and fit a Gaussian curve using \texttt{scipy.optimize.curve\_fit}\footnote{\url{https://docs.scipy.org/doc/scipy/reference/generated/scipy.optimize.curve_fit.html}}. We provide the fit with sensible initial guesses and parameter lower and upper bounds, which are unique to every spaxel. As \texttt{curve\_fit} is known to be sensitive to the quality of initial guesses, we vary the initial guesses 16 times for each spaxel and select the fit with the lowest reduced-$\chi^2$ value. Most often, more than 10 of these Gaussian fits are the same as the best-fit curve, indicating the robustness of the fit. As a further check, we create residual cubes by subtracting the best-fit curve from the data and find the residuals at the jet locations to be under 2$\sigma$, where $\sigma$ is the standard deviation measured in the residual cube pixels surrounding the jet. Finally, the integrated flux is calculated as the area under the best-fit Gaussian curve using \texttt{scipy.integrate.trapezoid}\footnote{\url{https://docs.scipy.org/doc/scipy/reference/generated/scipy.integrate.trapezoid.html}}, and a single image is created with each pixel representing the integrated line flux in the corresponding spaxel in the cube. In cases where the peak flux is less than 3 $\times$ the standard deviation of the points outside the line, an upper limit is calculated using a Gaussian profile with amplitude 3$\sigma$ and width given by the instrument's spectral resolution at that wavelength. 

Simultaneously, using the continuum model flux for each wavelength point in a spaxel, we create continuum emission cubes \citep[see][for a similar method applied to the forbidden nobel gas lines and H$_2$ rovibrational lines, respectively]{Bajaj2024,Pascucci2024}. We repeat this procedure and re-generate the continuum cubes and the integrated line flux images for each transition individually. To estimate the source position along the jet axis, we adopt the same approach as in Paper~I; for HH~30, IRAS~04302, and Tau~042021 with inclination $>$84\textdegree, we take the midpoint of the `dark' continuum valley between flared disk emissions as the star position. For FS~TauB, with an inclination of $\sim$74\textdegree, the continuum emission is more compact and one-sided (green countors in Figure \ref{fig:spec_img}) likely tracing scattered light emission close to the star. Accordingly, we consider the bottom edge of the slightly-resolved continuum emission to be the star position.

\subsection{Line morphology: jet, wind, disk} \label{sec:line_morphology}

\begin{figure*}
    \centering
    \includegraphics[width=7.5in]{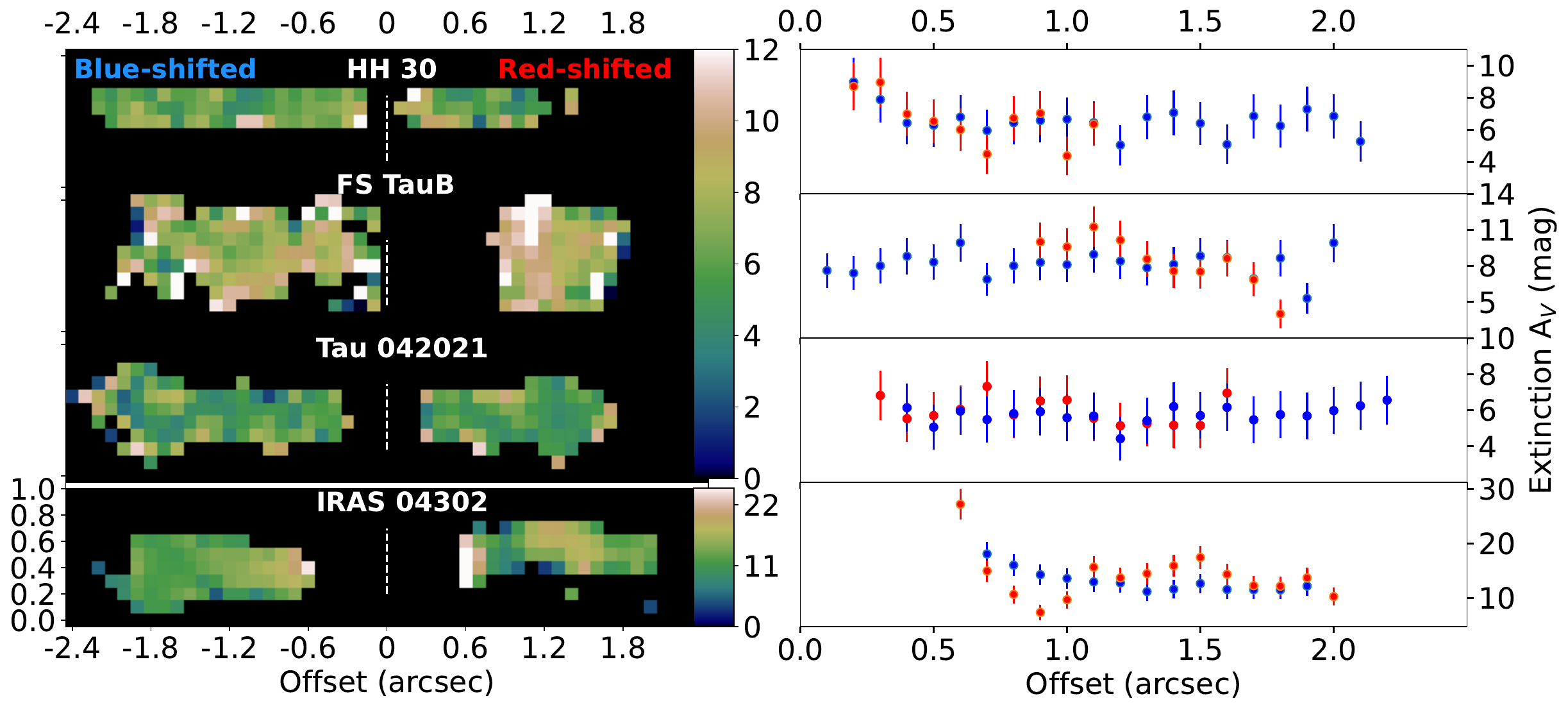}
    \caption{The left panel shows pixel-by-pixel visual extinction maps (A$v$) inferred from the jet emission with the source name displayed above each map and the color bar on the right. There is a single color bar for all sources except IRAS~04302 (the only Class\,I source in the sample), which shows higher extinction values. The maps are arranged so the left side is blue-shifted while the right side is red-shifted (Paper~I). The dashed lines in the maps highlight the estimated source position along the jet. The right panel shows extinction as a function of distance from the source (sources are arranged in the same order as the left panels), averaged across the jet. Blue and red points represent extinction on the blue- and red-shifted sides, respectively.}
    \label{fig:ext_map}
\end{figure*}

We can classify the line morphologies into three categories: jet, disk wind, and disk. Jet is identified as a narrow emission starting near the center of the disk continuum and extending away from the disk almost perpendicularly. Disk wind emission also starts near the center of the disk continuum and broadens as it extends away from the disk in a cone-like morphology (see, e.g., Paper I). Finally, disk emission is identified when no such extended outflow structure is seen, and emission is observed to be coincident with the continuum contours. We present continuum-subtracted images of select atomic and forbidden atomic lines in Figure \ref{fig:line_img}. The corresponding continuum contours are overlaid in green.

In all sources, forbidden atomic lines, when detected, are the cleanest tracers of jets (see e.g., HH~30 which has detections in all \feii{}, [\ci{}], \sii{}, \ngi{} lines shown in Figure \ref{fig:line_img}). \hei{} and \hi{} lines also trace jets, along with scattered light emission from the disk surface. Finally, the permitted \ci{} lines trace the disk when detected, whereas the H$_2$ lines (not shown here, see Paper I) trace disk wind emission.

\subsection{Extinction Maps and Extinction Correction}
\label{sec:ext_map_corr}
Extinction toward jets can be estimated using the ratio of the intensities of lines originating from the same upper energy level which, without any extinction, only depends on the ratio of their Einstein A coefficients and their wavelengths. JWST NIRSpec covers two such bright \feii{} lines, a$^4$D-a$^6$D(7/2 $\rightarrow$ 9/2) at 1.257 \micron{} and a$^4$D-a$^4$F(7/2 $\rightarrow$ 9/2) at 1.644 \micron{}, for which the ratio of Einstein coefficients is 1.5 \citep{Tayal2018} giving the intrinsic intensity ratio  ($[F_{\lambda_1}/F_{\lambda_2}]_{int}$) as 1.96. We can then calculate the difference in the extinction magnitudes at any two wavelengths $\lambda_1$ and $\lambda_2$ as

\begin{equation} \label{eq:1}
     A_{\lambda_1} - A_{\lambda_2} = 2.5 \times \log_{10}\left(\frac{F_{\lambda_1}/F_{\lambda_2}}{[F_{\lambda_1}/F_{\lambda_2}]_{int}}\right)
\end{equation}

Where A$_{\lambda}$ is the extinction magnitude at wavelength $\lambda$, and the numerator is the observed flux ratio. As the two \feii{} lines ($\lambda_1$ = 1.257 \micron{} and $\lambda_2$ = 1.644 \micron{}) fall in the same IFU grating, we calculate the flux ratio in each pixel and retrieve pixel-by-pixel visual extinction maps of the jets, see Figure~\ref{fig:ext_map}. We note that the resulting extinction using $[F_{\lambda_1}/F_{\lambda_2}]_{int}$ = 1.96 is only a lower limit. \cite{Rubinstein2021}, using a completely different methodology determined an intrinsic ratio of 2.6 for the same lines towards HH objects, which will lead to 3 mag higher extinction than those derived here. Further details on calculating the extinction maps are provided in Appendix \ref{sec:Extinction_appendix}. 

We find a visual extinction magnitude ranging $\sim$4-13\,mag towards all the sources except IRAS~04302, which shows much higher extinction ($\sim$7-30\,mag). This is consistent with IRAS~04302 being a more embedded Class~I source than the other sources. In all cases, we see similar extinction on both sides of the jets. In the case of FS~TauB, we find a large gap between the star and the red-shifted extinction map. This is due to the lower disk inclination angle of FS~TauB (74\textdegree) compared to the other sources ($>$84\textdegree) due to which the disk of FS~TauB occults the red-shifted jet lobe such that we do not see any emission in that region, i.e., very high extinction. In the red-shifted lobe of IRAS~04302, extinction increases around 1-1.5 arcsec from the star where the emission is offset across the jet, as can be seen in the corresponding extinction map.

Next, we correct the integrated flux maps for extinction by using Equations \ref{eq:1}, \ref{eq:2}, and \ref{eq:3} and the pixel-by-pixel extinction maps. The extinction-corrected integrated flux F$_{corr}$ can be written as

\begin{equation} \label{eq:5}
\begin{split}
    \log{\frac{F_{corr, \lambda 1}}{F_{\lambda_1}}} = \left(\frac{0.282 \times A_V (\lambda_1/1.25~\micron{})^{-1.7}}{2.5}\right); \\  \hfill \\ \log{\frac{F_{corr, \lambda 2}}{F_{\lambda_2}}} = \left(\frac{0.176 \times A_V (\lambda_2/1.65~\micron{})^{-1.7}}{2.5}\right)
\end{split}
\end{equation}

To reduce the impact of uncertainty in the extinction law index (-1.7) on extinction corrected flux, we perform the correction relative to 1.25 \micron{} (Equation \ref{eq:5}, top) for lines that center closer to this wavelength, i.e. \hei{} 1.083 \micron{}, Pa$\beta$ 1.282 \micron{}, and \feii{} lines at 1.257 \micron{}, 1.271 \micron{}, 1.279 \micron{}, and 1.295 \micron{}. Similarly, for lines that center close to 1.65 \micron{}, extinction correction is performed relative to this wavelength (Equation \ref{eq:5}, bottom). These include the \feii{} 1.644 \micron{}, 1.533 \micron{}, 1.600 \micron{}, 1.664 \micron{}, and 1.677 \micron{} lines. Many of these lines are useful diagnostics of the jet's physical properties and are discussed next.

\section{Line ratios along the jet}

\begin{figure}
    \centering
    \includegraphics[width=\columnwidth]{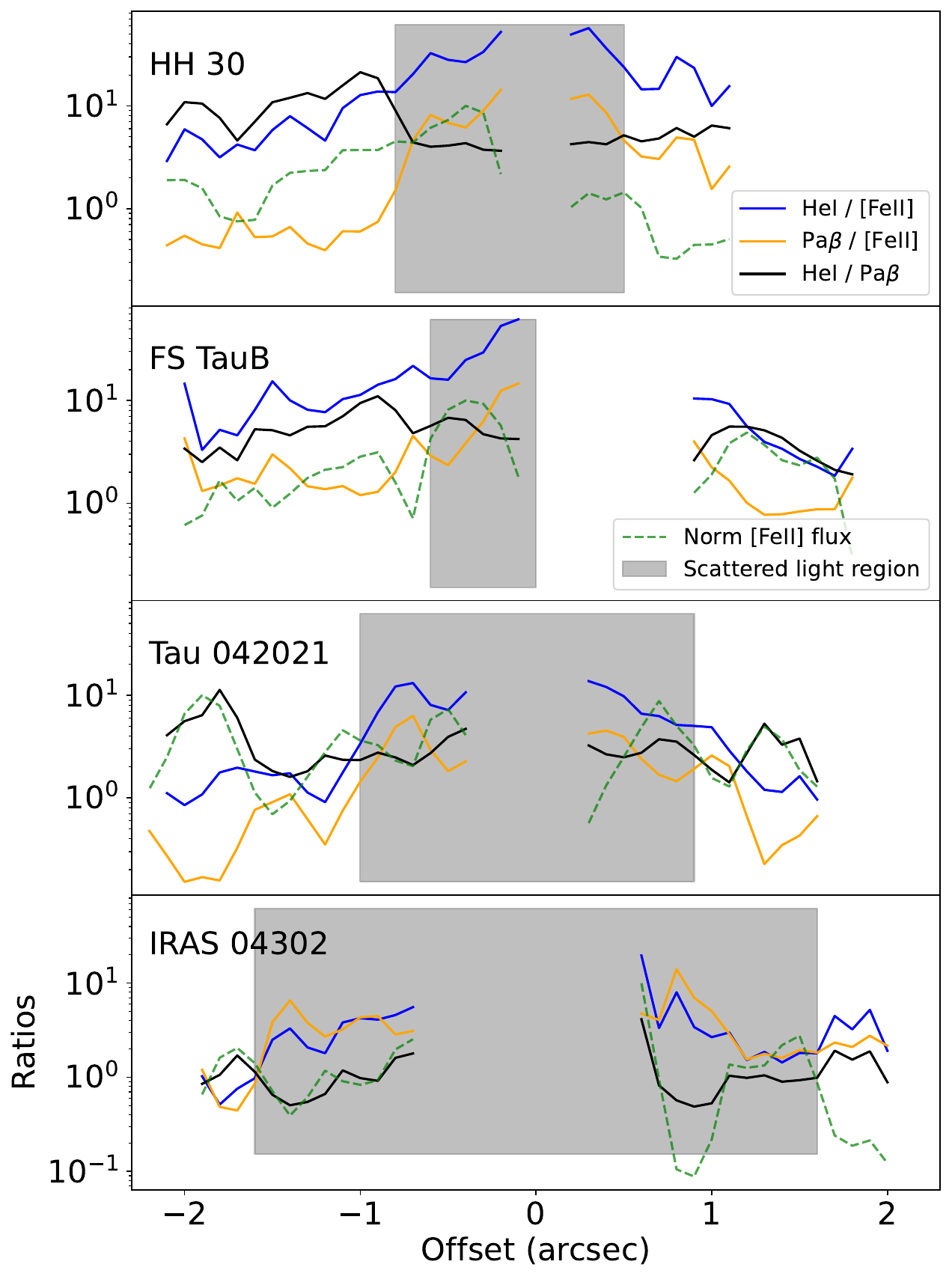}
    \caption{Extinction-corrected line ratios as a function of distance from the source. The green dashed lines show the normalized \feii{} 1.644 \micron{} line flux. The gray bands highlight the scattered-light dominated region and since the \hei{} and \hi{} lines have a scattered emission component too (Figure \ref{fig:line_img}), the ratios there are not representative of the jet. However, outside the gray bands are the jet-only ratios. For the red-shifted side of HH~30, the Pa$\beta$/\feii{} ratio should be considered as an upper limit and the \hei{}/Pa$\beta$ ratio as a lower limit as we could only estimate an upper limit for Pa$\beta$ in this region (see Section \ref{sec:shock_velocity_density}).}
    \label{fig:line_ratios}
\end{figure}

As discussed in Section \ref{sec:line_morphology}, we find several lines tracing jets. Of particular interest are the \hi{} and \hei{} lines whose emission is often attributed mostly to accretion \citep[e.g.,][]{Gatti2008,Alcala2014}. Our maps clearly show a strong jet component, too. The fact that these lines also trace the jet is not well known and only a few evidence (spectro-astrometric) exist that show weak \hei{} emission in a jet \citep{Takami2002b,Takami2003,Podio2008}. To further investigate these lines, we plot their extinction-corrected integrated fluxes across the jet with respect to the well-known \feii{} 1.644 \micron{} jet tracer as a function of distance from the star. In the scattered-light dominated regions in \hi{} and \hei{}, we define the jet from the scattering free \feii{} 1.644 \micron{} line. These ratios are plotted in Figure \ref{fig:line_ratios}, where the gray bands highlight the scattered light-dominated regions. These regions were established by analyzing the vertical profiles of the continuum emission to find the location where the profiles change from being narrow (jet) to broad (continuum).

While the \hi{}/\feii{} line ratio is often less than unity in the jet-dominated region, we find that the \hei{}/\feii{} can be up to $\sim 10$ in HH~30 and FS~TauB. This indicates that \hei{} can be used to probe much fainter jet emission, hence ejection toward lower accretors. \hei{} is not easily excited since it has a high excitation energy ($\sim$20~eV), compared to other forbidden lines ($\sim$2~eV). Hence, the near-IR line ratios presented here could be useful diagnostics of various jet parameters including jet temperature. In the following subsection, we use the ratios of some of these lines to estimate the parameters necessary for calculating jet mass loss rates.

\subsection{Shock Velocities, Pre-shock Densities, and Ionization Fraction} \label{sec:shock_velocity_density}

\begin{figure}
    \centering
    \includegraphics[width=\columnwidth]{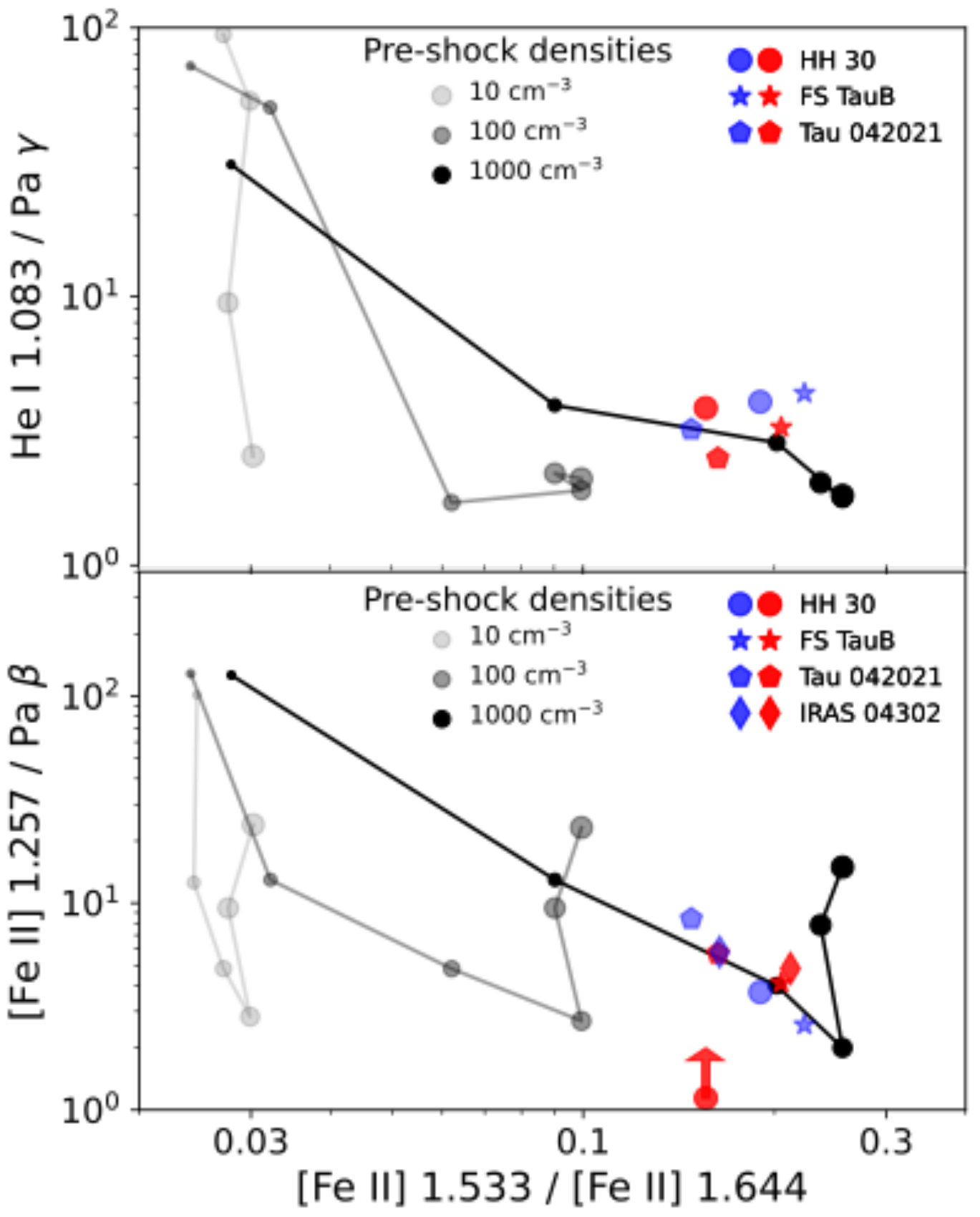}\\ 
    \caption{\textit{Top:} \hei{}, Pa$\gamma$ and \feii{} line ratios as diagnostics of the shock speed and pre-shock electron densities. Blue and red markers represent the blue- and red-shifted jets, respectively, for each source. The \hei{}/Pa$\gamma$ ratios are generated using the publicly available shock code \texttt{MAPPINGS~V} \citep{Dopita2017,Sutherland2017,Sutherland2018}. The curves represent increasing pre-shock densities from light gray to black. Growing circle sizes along each curve represent the following shock speeds: 20, 40, 60, 80, 100, 120 km~s$^{-1}$. Each line flux was calculated in the scattered-light-free region of the jet and was corrected for extinction before taking the ratio. \textit{Bottom:} \feii{} and Pa$\beta$ line ratios as diagnostics of the shock speed and pre-shock electron densities. The curves are digitized and recreated from \cite{Koo2016}.}
    \label{fig:shock_speed_density}
\end{figure}

Shocks are produced when faster-moving gas collides with slower-moving gas ahead with the velocity difference referred to as the shock velocity. We use the \hei{} 1.083~\micron{}/Pa$\gamma$, and \feii{} 1.257~\micron{}/Pa$\beta$ flux ratios calculated in the jet to estimate the shock velocities and the \feii{} 1.533~\micron{}/\feii{} 1.644~\micron{} ratio to estimate the pre-shock densities. While \feii{}~1.257/Pa$\beta$ ratio is often used to estimate the shock velocities \citep[e.g.,][]{Koo2016}, it is restricted by the uncertainties in \feii{} rate coefficients \citep[e.g.,][]{Nussbaumer1988,Pesenti2003,Bautista2015,Tayal2018} as well as potential Fe depletion in the jet. Hence, we use the extinction-independent, abundance-free, and shock-sensitive ratio of \hei{} 1.083~\micron{}/Pa$\gamma$ calculated using the \texttt{MAPPINGS V} shock code \citep{Dopita2017,Sutherland2017,Sutherland2018} to estimate the shock velocities. Simultaneously, we also use the \feii{} 1.257~\micron{}/Pa$\beta$ ratio using shock model predictions from \cite{Koo2016} to compare the results. \cite{Koo2016} updated the shock code, initially written by \cite{Raymond1979} and \cite{Cox1985}, using the improved \feii{} atomic parameters from \cite{Bautista2015}. These models assume solar abundances in the jet.

We plot line ratios of both the blue- and red-shifted jets for all the sources in Figure \ref{fig:shock_speed_density} where model predictions are shown in grey circles (growing in size for increasing shock speeds and darker colors for higher pre-shock electron densities). The observed \hei{} 1.083~\micron{}/Pa$\gamma$ flux ratios shown in the top figure are consistent with a pre-shock electron density of $\sim$1000~cm$^{-3}$ and shock speeds in the range 50-60~km~s$^{-1}$. Similarly, in the bottom figure that includes the \feii{} 1.257~\micron{}/Pa$\beta$ flux ratios, all points are consistent with a pre-shock electron density of 1000~cm$^{-3}$ including the red-shifted jet of HH~30, which is a lower limit\footnote{In most pixels that correspond to the red-shifted jet of HH~30, we could only estimate a 3~$\sigma$ upper limit of the Pa$\beta$ flux, leading to a lower limit in \feii{}/Pa$\beta$.} and a shock velocity of $\sim$60~km~s$^{-1}$ towards HH~30, FS~TauB, and IRAS~04302 and $\sim$50~km~s$^{-1}$ towards Tau~042021. Our estimates of 50-60~km~s$^{-1}$ shock speeds are consistent with other literature estimates \cite[e.g.,][]{Hartigan1994,Lavalley-Fouquet2000,Garcia-Lopez2008,Dopita2017}. The fact that the shock speeds are lower than the jet velocity suggests that these shocks are generated internally within the jet, likely caused by variations in the jet velocity. The good match found between the two ratios supports the latest atomic parameters for Iron. Additionally, the match found between models assuming solar abundance and the observed points in the bottom plot (Figure \ref{fig:shock_speed_density}) suggests solar abundance of Iron in the jet, i.e., nearly complete liberation of Iron in these shocks and/or jet launch within the dust sublimation radius.

\begin{figure}
    \centering
    \includegraphics[width=\columnwidth]{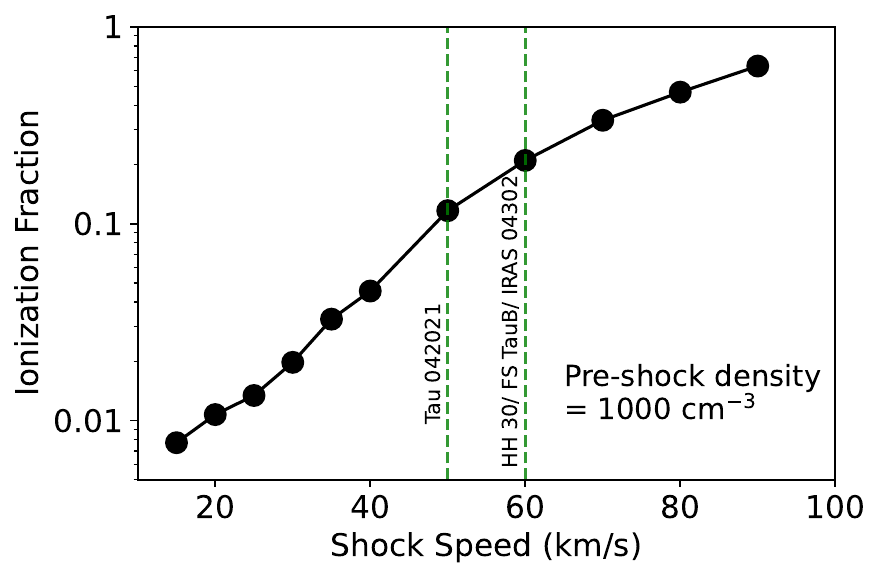}
    \caption{Relation between the ionization fraction and shock speed for a pre-shock density of 1000~cm$^{-3}$ and magnetic field strength of 100~$\mu$G. The relation is digitized and recreated from \cite{Hartigan1994} and was derived using the shock model predictions for \sii{} line ratios.}
    \label{fig:ionization_fraction}
\end{figure}

Using \sii{} diagnostic ratios common to both the shock velocity and ionization fraction, \cite{Hartigan1994} found a relation between the two for different pre-shock electron densities \citep[see Figure 16 of][]{Hartigan1994}. Since \feii{} is formed in a similar cooling zone as \sii{} \citep[e.g.,][]{Reiter2015}, we use the same relation to estimate the post-shock ionization fraction. In Figure \ref{fig:ionization_fraction}, we plot one such relation predicted for pre-shock electron density = 1000~cm$^{-3}$, as found for our targets, and pre-shock magnetic field strength = 100~$\mu$G. Using Figure \ref{fig:ionization_fraction}, we infer a post-shock ionization fraction of 0.12 towards Tau~042021 and 0.21 towards other sources. These values are consistent with the previous estimates for HH~30 \citep{Bacciotti1999,Hartigan2007} and other jets in the literature \citep[e.g.,][]{Hamann1994,Shinn2013}. We note that the variation in the inferred ionization fraction for pre-shock magnetic field strengths $\leq$100~$\mu$G is negligible. We use these estimates of the shock speeds and ionization fractions to calculate the jet mass loss rates (Section \ref{sec:jmlr}).

\begin{figure*}
    \centering
    \includegraphics[width=7.5in]{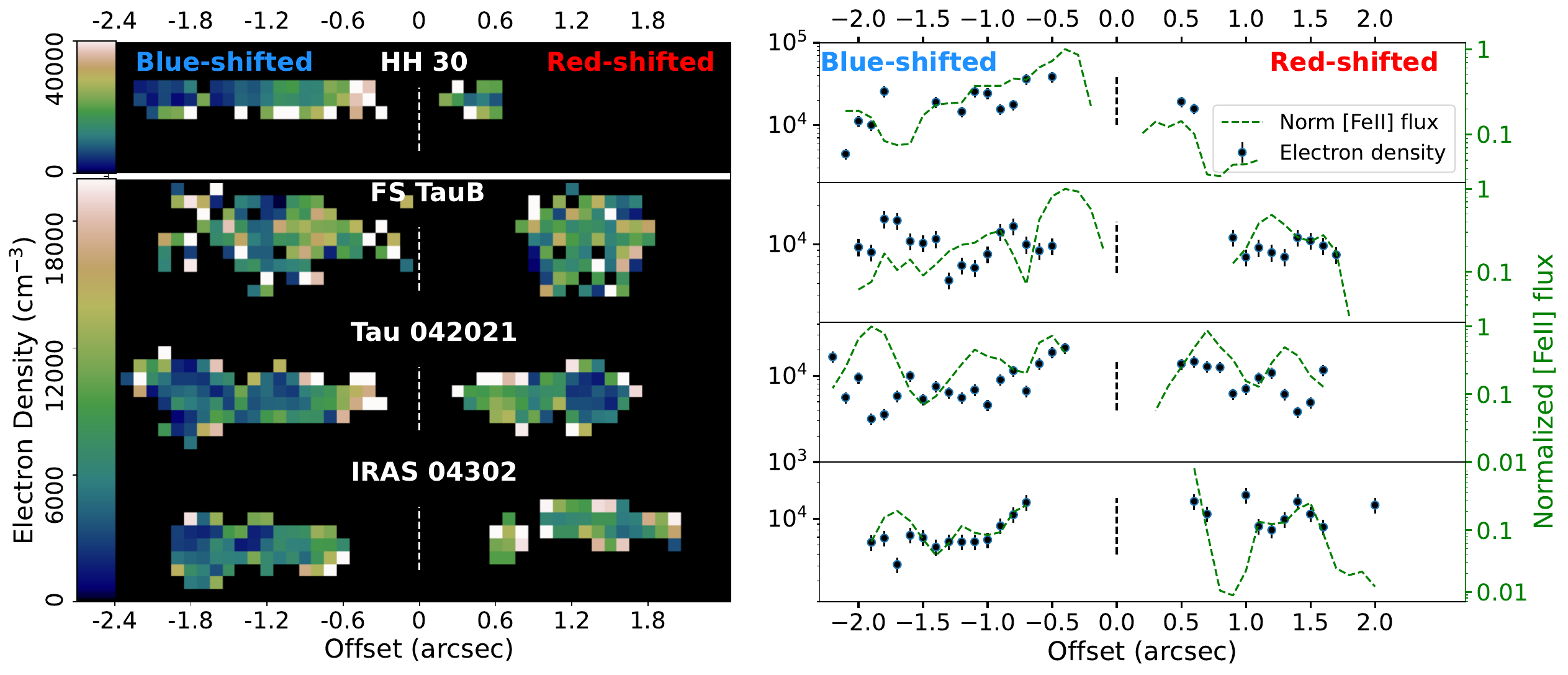}
    \caption{The left panel shows pixel-by-pixel electron density maps (n$_e$) inferred from the jet emission with the source name displayed above each map and the color bar on the right. There is a single color bar for all sources except HH~30, which shows higher electron densities. The right panel shows the average electron density as a function of distance from the source (sources are arranged in the same order as the left panels). The vertical bars represent a 15\% error as described in the text. Green dashed lines represent the extinction corrected normalized \feii{} 1.644~\micron{} line fluxes. The maps are arranged so the left side is blue-shifted while the right side is red-shifted. The dashed lines in the maps and the plots highlight the estimated source position along the jet.}
    \label{fig:elecden_map}
\end{figure*}

\section{Jet mass loss rates from forbidden Fe II lines}

Jet mass loss rates towards edge-on disks can be used to estimate the mass accretion rate onto the star as well as compare with the wind mass-loss rates to test the disk-wind-driven evolution scenario. Most methods of calculating the jet mass loss rate rely on either the line luminosity or the post-shock electron density or both \citep[e.g.,][]{Hartigan1995,Dougados2010,Agra-Amboage2011}. Accordingly, we first estimate the post-shock electron density in Section \ref{sec:elec_den} and then calculate the mass loss rates in Section \ref{sec:jmlr}.

\subsection{Electron Density Maps} \label{sec:both_elec_den}

We first estimate the electron densities using diagnostic \feii{} line ratios in Section \ref{sec:elec_den} and then compare them with those derived from the \sii{} 1.03~\micron{} lines in Section \ref{sec:S_elec_den}.

\subsubsection{\feii{}-derived Electron Densities} \label{sec:elec_den}

NIR \feii{} line ratios can be used as a diagnostic of electron density of the emitting material \citep[e.g.,][]{Takami2006,Dougados2010,Koo2016}. In the NIRSpec wavelength range, several \feii{} lines have similar excitation energies, the ratios of which mainly depend on the electron density and weakly depend on the temperature. The most commonly used ratios to estimate jet densities are \feii{} 1.533/1.644 \micron{}, 1.600/1.644 \micron{}, 1.664/1.644 \micron{}, and 1.677/1.644 \micron{}, as they are bright and have critical densities (10$^{4-5}$\,cm$^{-3}$, see Section \ref{sec:fe_s_ratio_ne}) close to the expected jet densities \citep[e.g.,][]{Pradhan1993,Pesenti2003,Bautista2015,Tayal2018}. 

To obtain the post-shock jet electron density maps, we first create extinction-corrected pixel-by-pixel integrated flux maps for all the lines as discussed in Section \ref{sec:ext_map_corr}. Then, we take pixel-to-pixel ratios of the respective lines and use \texttt{PYTHON PyNeb}\footnote{\url{http://research.iac.es/proyecto/PyNeb/}} package \citep{Luridiana2015} with atomic data from \cite{Tayal2018} to estimate the electron density in each pixel by providing it with the line ratio in that pixel and a temperature value. \texttt{PyNeb} solves the full system of equilibrium equations for an n-level atom. Since we do not know the exact post-shock jet temperature, we collect three density estimates per pixel corresponding to three different temperatures typical of jets: 5000~K, 12500~K, and 20000~K, \citep[][]{Bacciotti1999,Podio2011,Whelan2014}. We find that in most cases, the electron density varies by 10\% within the given temperature range and at most 30\%. We take an average of the three density values. To remove any outliers from the map, we perform sigma clipping using \texttt{astropy} function \texttt{sigma\_clip}\footnote{\url{https://docs.astropy.org/en/stable/api/astropy.stats.sigma_clip.html}} with $\sigma$ = 3. Following this procedure for all four line ratios, we get four corresponding post-shock electron density maps for each source. For robustness, we consider the average of these four maps as the final post-shock electron density map for a source. These maps are shown in Figure \ref{fig:elecden_map} for all the sources. We estimate an uncertainty of $\sim$15\% corresponding to each point, by summing in quadrature 10\% flux ratio uncertainty and 10\% due to the variation of electron density with temperature. 

Comparing our post-shock electron density estimates with the derived pre-shock electron density of 1000~cm$^{-3}$, we get a compression factor in the range 5-80. This is similar to the compression factor found in typical jet shocks (1-100) for a broad range of pre-shock magnetic fields (B=0-3000~$\mu$G), pre-shock electron densities (10-1000~cm$^{-3}$) and shock speeds upto 100~km~s$^{-1}$ using the optical \sii{} diagnostics \citep{Hartigan1994,Raga1996,Heathcote1998}. More recently, \cite{Dopita2017} found the same result for nearly 30 jets using the same \sii{} diagnostics with \texttt{MAPPINGS V} and since the optical \sii{} lines are formed in a similar cooling zone as NIR \feii{}, we expect the compression factors to be similar. We also over plot the extinction-corrected normalized \feii{} 1.644 \micron{} line fluxes for comparison and find them to be, in some cases, correlated with the post-shock electron densities. 

\subsubsection{\sii{}-derived Electron Densities} \label{sec:S_elec_den}

\begin{figure}
    \centering
    \includegraphics[width=\linewidth]{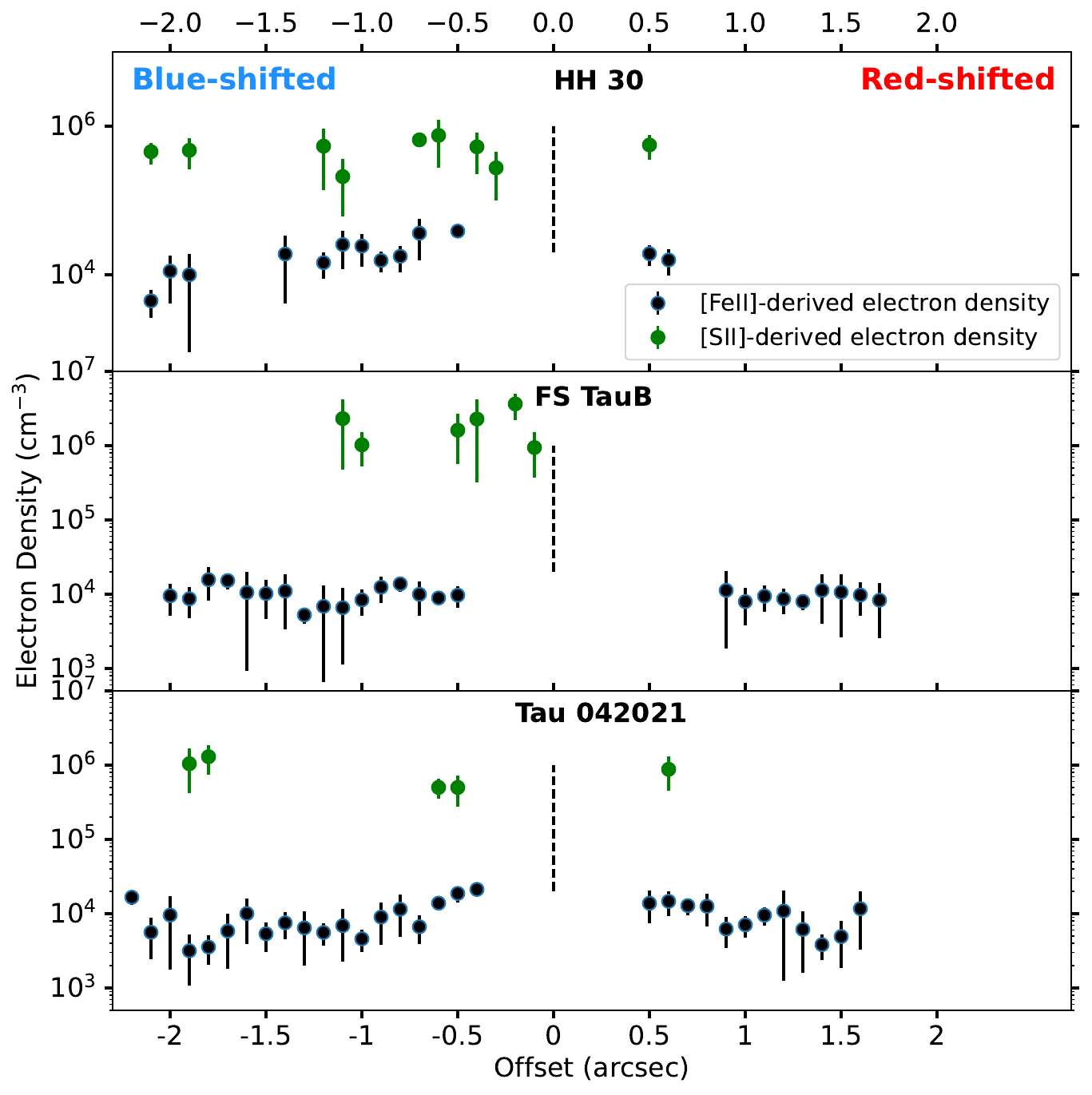}
    \caption{All panels show the average electron density as a function of distance from the source (with the source name at the top of each panel). In black are the electron densities derived using the \feii{} line ratios, and in green, those using the \sii{} line ratios. Each point has an error bar that corresponds to the standard deviation of the electron densities averaged at that offset. The maps are arranged so the left side is blue-shifted while the right side is red-shifted. The dashed lines highlight the estimated source position along the jet.}
    \label{fig:elecden_map_SII}
\end{figure}

Like \feii{}, the NIRSpec wavelength range also contains multiple \sii{} lines that have similar excitation energies and can be used to estimate the electron densities. While the \sii{} 1.03~\micron{} lines have been used in combination with the optical \sii{} lines to estimate electron densities in other sources in the past \citep[e.g.,][]{Nisini2005}, the higher sensitivity and spectral resolution of NIRSpec allows us to calculate line ratios within the 1.03~\micron{} lines. Accordingly, we use the \sii{} 1.0286~\micron{}/1.0336~\micron{} and 1.032~\micron{}/1.037~\micron{} flux ratios. We find that these ratios, while available in a more limited region due to the overall lower fluxes, trace nearly two orders of magnitude higher density than the \feii{} line ratios. Note that our observed ratios are consistent with n$_e$~$<$~n$_c$ (critical density) for both the lines (see Section \ref{sec:fe_s_ratio_ne}). To calculate the electron density, we follow the same methodology as before: for each pixel, we derive six estimates of electron density corresponding to three temperatures and two line ratios. The mean of these is our final estimate for that pixel. We plot the average \sii{}-derived and \feii{}-derived electron densities across the jet in Figure \ref{fig:elecden_map_SII}. It can be seen that the \sii{}-derived estimates are nearly 50-100 times that derived from \feii{} in all sources except HH~30, where it is $\sim$2-50 times. IRAS~04302 has a very weak detection of the \sii{} lines and, hence, is not included in the plot.

\subsection{Methods to estimate mass loss rates} \label{sec:jmlr}

Here, we present jet mass loss rate calculations using three different methods; one based on the post-shock electron density and jet cross-section, and the other two primarily based on the \feii{} line luminosities. Each method has its advantages and limitations. For instance, the first method (Section \ref{sec:jmlr_method1}) is free of reddening/extinction errors; however, it involves knowing the ionization fraction and jet radius. We do not resolve the jet radii very close to the source and, for HH~30, out to about 1.5\arc{} (Paper~I). The latter two methods do not require any knowledge of the jet cross-section but suffer from uncertainties in the extinction and \feii{} flux--electron density relation. Hence, combining these three methods helps to better constrain jet mass loss rates.

\subsubsection{Method 1: Cross-section} \label{sec:jmlr_method1}

In this first method, we calculate the jet mass loss rate as simply the mass density flowing through a cross-sectional area of radius $r_J$ and velocity ($v_J$):

\begin{equation}
    \dot M_J = (\mu m_H \frac{n_e}{x_e}) \times \pi r_J^2 \times v_J
\end{equation}

Where the term in parenthesis is the mass density calculated from the average gas mass $\mu$ (1.24) in units of the hydrogen mass ($m_H$), and the ratio between the post-shock electron density n$_e$ and the ionization fraction x$_e$ gives the hydrogen number density. The post-shock electron density is calculated in Section \ref{sec:elec_den} and x$_e$ in Section \ref{sec:shock_velocity_density} (also listed in Table \ref{tab:mass_loss_accretion}). For HH~30 and FS~TauB, we can use the known deprojected velocities of 120\, and 270\,km~s$^{-1}$ for the blue-shifted lobes, respectively, and 200\,km~s$^{-1}$ for the red-shifted lobes \citep[listed in Table \ref{tab:mass_loss_accretion},][]{Hartigan2007,Estalella2012,Eisloffel1998,Liu2012}.
For Tau~042021 and IRAS~04302, since literature estimates of the jet velocities do not exist, we assume $v_J$ to be 200\,km\,s$^{-1}$, as found through knots close to the launch region for several other sources \citep[e.g.,][]{Mundt1990}. 

We calculate the jet mass loss rate as a function of distance from the source by averaging the electron density in the y-direction (across the jet) at equal distances (pixel size$\sim$14 au) from the source. To calculate the jet radius at these distances, we fit a 1D Gaussian curve to the intensity profile across the jet and estimate the half width at half maximum (HWHM). For HH~30, since the jet is unresolved within 1.5\arc{} from the source, we use the jet size derived by \cite{Hartigan2007} in the optically resolved \sii{} line images. For the other sources, we find that the jet width is less reliable only within 0.5\arc{} from the source (see Figure~3 in Paper~I). We show the mass loss estimates from this method, as a function of distance, in Section \ref{sec:jmlr_comparison_appendix}.

\subsubsection{Method 2: Line Luminosity} \label{sec:jmlr_method2}

In this second method, we use the \feii{} 1.644 \micron{} line luminosity to calculate the jet mass loss rates. Since forbidden lines are optically thin, their luminosity is proportional to the number of atoms that radiate along the line of sight, and hence the mass of the emitting gas \citep[see, e.g.,][]{Hartigan1984,Hartigan1995,Nisini2005,Dougados2010,Agra-Amboage2011,Shinn2013,Fang2018ApJ...868...28F}. Detailed description of this method can be found in the appendix of \cite{Hartigan1995}. Using the atomic data and excitation model for [Fe\,II] from \cite{Tayal2018}, we calculated the emissivity of the [Fe\,II] 1.644\micron{} line at temperature=10,000~K and for an electron density ranging from 5~$\times$~10$^2$ to 5~$\times$~10$^5$~cm$^{-3}$(corresponding to the observed range, see Figure \ref{fig:elecden_map}) as

\begin{equation}
    \frac{j(1.644~\micron{})}{(erg~s^{-1}~sr^{-1}~ion^{-1})} = 1.46 \times 10^{-17} \left(1+\frac{49000}{n_e(cm^{-3})}\right)^{-0.8} 
\end{equation}

where 49,000~cm$^{-3}$ is the critical density of [Fe\,II]~1.644 \micron{} line \citep{Tayal2018}. Following this, we can write the mass loss rate as emissivity divided by the pixel crossing time.

\begin{equation}
    \begin{split}
        \frac{\dot M_J}{(M_{\odot}~yr^{-1})} = 0.98 \times 10^{-8} \times \left(1+\frac{49000}{n_e(cm^{-3})}\right)^{0.8} \\ \times \frac{L_J}{(10^{-4}L_{\odot})} 
        \frac{v_J}{(km~s^{-1})} \times \left({\frac{l_J}{au}}\right)^{-1} \\ \times \left(\frac{[Fe]/[H]}{[Fe]/[H]_{\odot}}\right)^{-1}
        \end{split}
\end{equation}

\begin{figure*}[t]
    \centering
    \includegraphics[width=6in]{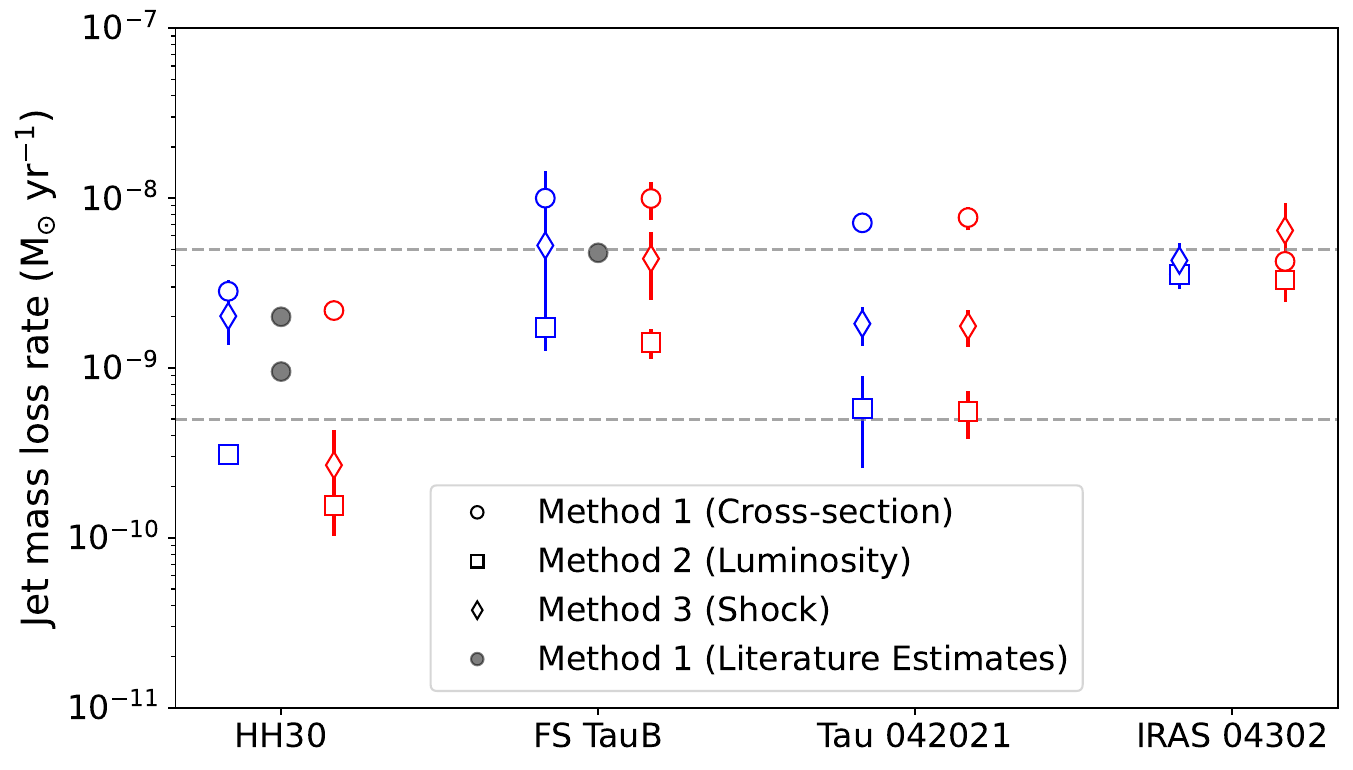}
    \caption{Comparison of jet mass loss rates between (i) red- and blue-shifted sides of the jet, (ii) three methods of calculating the jet mass loss rates, (iii) different sources, and (iv) their literature estimates, when available. The estimates shown here are the median of the values plotted in Figure \ref{fig:jmlr} along the jet. The blue and red points correspond to the blue- and red-shifted lobes. The error bars correspond to the standard error of the distribution in Figure \ref{fig:jmlr}, and an additional error of at least 30\% for HH~30 and FS~TauB and 50\% for Tau~042021 and IRAS~04302 needs to be considered. The dotted lines highlight 5 $\times$ 10$^{-10}$ and 5 $\times$ 10$^{-9}$~M$_{\odot}$~yr$^{-1}$, between which majority of the points lie.}
    \label{fig:jmlr_comp}
\end{figure*}

Here, $L_J$ is the \feii{} 1.644 \micron{} line luminosity, $l_J$ is the length of the jet, $v_J$ is the jet velocity, and $[Fe]/[H]$ is the abundance of Iron relative to Hydrogen normalized by the solar abundance \citep[3.2 $\times$ 10$^{-5}$,][]{Asplund2009}. We consider distance step sizes of one-pixel width ($l_J=14$\,au) and, at any distance, we sum the luminosity in the y-direction (across the jet) in the rotated maps to estimate $L_J$. Jet velocities are as in Table \ref{tab:mass_loss_accretion}. Finally, we consider a solar abundance, i.e. ([Fe]/[H])/([Fe]/[H])$_{\odot}$ = 1, since jets are thought to be launched from the region between the disk inner edge and the dust evaporation radius where Iron is not depleted \citep{Beck-Winchatz1994,Lee2017,Tabone2017}. This is corroborated by recent observations of four protostellar jets where the Iron abundance is found to be solar \citep{Giannini2019} and additionally, our nearly perfect match of the line ratios with the pre-shock density curve of 1000~cm$^{-3}$ that was produced assuming solar abundance of Iron (Figure \ref{fig:shock_speed_density}). We show the mass loss estimates from this method too, as a function of distance, in Section \ref{sec:jmlr_comparison_appendix}.
 
\subsubsection{Method 3: Line Emission from Shock Fronts} \label{sec:jmlr_method3}

In this third method, we again use the \feii{} 1.644 \micron{} line luminosity, assuming that all the line emission originates in the shock fronts. A detailed description of this method is provided in Section \ref{sec:jmlr_method3_appendix}. Following that, we write the mass loss rate as 

\begin{equation} \label{eq:13}
    \frac{\dot M_J}{(M_{\odot}~yr^{-1})} = 7.2 \times 10^{-8}~\frac{L_J}{(10^{-4}L_{\odot})}~\frac{v_J}{v_S}
\end{equation}

Here, the jet velocities are again the same as in Table \ref{tab:mass_loss_accretion}. $v_S$ represents the shock velocity which we estimated for our targets using the Fe/H and He/H line ratios in Section \ref{sec:shock_velocity_density} and listed in Table \ref{tab:mass_loss_accretion}. Following Method 2 (Section \ref{sec:jmlr_method2}), we adopt ([Fe]/[H])/([Fe]/[H])$_{\odot}$ = 1. The luminosity in Equation \ref{eq:13} corresponds to the total luminosity emerging from a jet shock. Since shocks produce the brightest emissions along the jet, we identify them as the region between two adjacent local minima in the extinction-corrected \feii{} 1.644 \micron{} flux profiles (as a function of distance, see Section \ref{sec:shock_id}). These regions are well defined for FS~TauB and Tau~042021 with an average shock length of $\sim$0.6\arc{} (or 84~au); however, for HH~30 and IRAS~04302, they are less clear. Hence, as an alternate, we estimate the luminosity assuming one shock front every 0.6\arc{}. We plot and discuss the mass loss estimates using both techniques (along with Method 1 and 2) in Section \ref{sec:jmlr_comparison_appendix} and find similar values for all the sources.

\subsection{Comparison of the three methods and literature estimates}

\begin{sidewaystable*} 
    \centering
    \caption{Range of Jet Mass Loss Rates} \label{tab:mass_loss_accretion}
    \begin{tabular}{l l c c c | c c | c c | c c c}
        \tableline
        Source & Simbad Name & Stellar Mass & \multicolumn{2}{c|}{Jet Velocity} & \multicolumn{2}{c|}{\feii{} 1.644 \micron{} Flux$^b$} & Ionization Fraction & Shock Speed & \multicolumn{3}{c}{Jet Mass Loss Rate} \\ 
        &  &  & Blue & Red & Blue & Red &  &  & Blue$^c$ & Red$^c$ & Literature$^d$ \\
        &  & (M$_{\odot}$) &  \multicolumn{2}{c|}{(km~s$^{-1}$)}  &  \multicolumn{2}{c|}{(10$^{-18}$ W~m$^{-2}$)}  &  & (km~s$^{-1}$) & \multicolumn{3}{c}{(10$^{-9}$ M$_{\odot}$~yr$^{-1}$)}  \\
        \tableline
        HH~30 & V1213~Tau & 0.5 & 120$^1$ & 200$^1$ & 4.90 & 0.55 & 0.21 & 60 & 0.3--2.8 & 0.1--2.2 & 0.6$^3$ \& 2.0$^4$ \\
        FS~TauB & 2MASS J04220069+2657324 & 0.7 & 270$^2$ & 200$^2$ & 10.3 & 4.25 & 0.21 & 60 & 1.3--14 & 1.1--12 & 3.0$^5$ \\
        Tau~042021$^a$ & SSTtau 042021.4+281349 & 0.4 & \nodata & \nodata & 4.49 & 2.57 & 0.12 & 50 & 0.3--6.9 & 0.4--7.5 & \nodata \\
        IRAS~04302$^a$ & IRAS 04302+2247 & 1.3--1.7 & \nodata & \nodata & 13.16 & 22.41 & 0.21 & 60 & 2.8--5.1 & 2.5--9 & \nodata \\
        \tableline
    \end{tabular}
    \tablecomments{
        $^a$ The blue- and red-shifted velocities for these sources are assumed to be 200 km~s$^{-1}$. \\
        $^b$ These \feii{} 1.644 \micron{} flux estimates are corrected for extinction. Estimated error is 10\%. \\
        $^c$ Jet mass loss rates are minimum and maximum values derived from methods in Figure~\ref{fig:jmlr_comp}. \\
        $^d$ Literature estimates are updated for an ionization fraction of 0.21. \\
        $^1$\cite{Hartigan2007}, $^2$\cite{Liu2012}, $^3$\cite{Mundt1990}, $^4$\cite{Bacciotti1999}, $^5$\cite{Mundt1987}
    }
\end{sidewaystable*}

We plot median jet mass loss rate values in Figure \ref{fig:jmlr_comp} and list their range in Table \ref{tab:mass_loss_accretion}. We find the jet velocity to be the dominant source of uncertainty for all three methods and, hence, suggest an uncertainty of $\sim$30\% for HH~30 and FS~TauB \citep{Hartigan2007,Estalella2012,Eisloffel1998,Liu2012} and $\sim$50\% for Tau~042021 and IRAS~04302. We find that the cross-section method (Method 1, Section \ref{sec:jmlr_method1}) consistently provides a higher estimate than the other two methods except for IRAS~04302. This could be because the telescope beam contains a mix of higher- and lower-density regions, and $n_H$ is dominated by the higher-density regions, which emit more, overestimating the jet mass loss rates. This argument is consistent with the finding in HH~30 that the fraction of the volume of the jet occupied by the emitting material is less than unity \citep{Bacciotti1999}. It is also consistent with the high jet mass loss rates found using the same method (cross-section based) towards four other jet sources \citep{Dougados2010}. IRAS~04302 is the only Class~I source in our study, and it could be that its local density varies less. We also find that the luminosity-based methods (Methods 2 and 3, Sections \ref{sec:jmlr_method2} and \ref{sec:jmlr_method3}, respectively) agree within a factor of 5 for all the sources, except the blue-shifted side of HH~30. Both methods could underestimate mass loss rates as they ignore any contribution from gas that is too cool or diffuse to emit.

Literature estimates of the jet mass loss rates exist only for HH~30 and FS~TauB using the cross-section method. For FS~TauB and HH~30, \cite{Mundt1987} and \cite{Mundt1990} calculated an average mass loss rates of $\sim$3~$\times$~10$^{-9}$~M$_{\odot}$~yr$^{-1}$ and $\sim$0.9~$\times$~10$^{-9}$~M$_{\odot}$~yr$^{-1}$, respectively, corrected for an ionization fraction of 0.21 that we found. They used very low post-shock electron densities of $\sim$700~cm$^{-3}$ calculated at a distance of $>$5\arc{} from the sources. Following Figure \ref{fig:elecden_map}, our post-shock electron densities are at least an order of magnitude higher; however, since they measure the jet at a larger distance from the source, their jet radius is larger by a factor of few, leading to the jet mass loss rate being only $\sim$3 times smaller than our estimate using Method 1 (see Table \ref{tab:mass_loss_accretion}). More recently, \cite{Bacciotti1999} estimated an average jet mass loss rate of ~2~$\times$~10$^{-9}$~M$_{\odot}$~yr$^{-1}$ for HH~30, similar to our estimate using the cross-section method. While they estimated three times lower post-shock electron density using the optical \sii{} lines, they also obtained two times lower ionization fraction (0.1), making the hydrogen densities similar and, hence, the mass loss rates as well. 

\section{Jet wiggling: Tau~042021 is likely a binary} \label{sec:jet_wigg_obs}

In this section, we search for mirror-symmetric jet wiggling and when identified, we model it using a binary circular orbit framework. Wiggles are defined as the lateral movement of the jet axis along the direction of jet propagation. To investigate jet wiggles in our sources, we use the rotated intensity map of \feii{} 1.644 \micron{} described in Section \ref{sec:ext_map_corr}. At every pixel distance along the jet, we use the same Gaussian fits performed to calculate the HWHM/radius of the jet, to now use the emission centroids (see Section \ref{sec:jmlr_method1}). The centroids in the x-direction (along the jet) are measured as the center of the pixels used to fit the Gaussian curve. To estimate the uncertainty, we turn to calculations for the standard deviation of the peak centroid derived in detail by e.g., \cite{Garnir1987,Porter2004} and used for a similar application by \cite{Murphy2021}.

\begin{equation}
    Error = 0.8~\frac{HWHM}{S/N}
\end{equation}

\begin{figure*}
    \centering
    \includegraphics[width=6in]{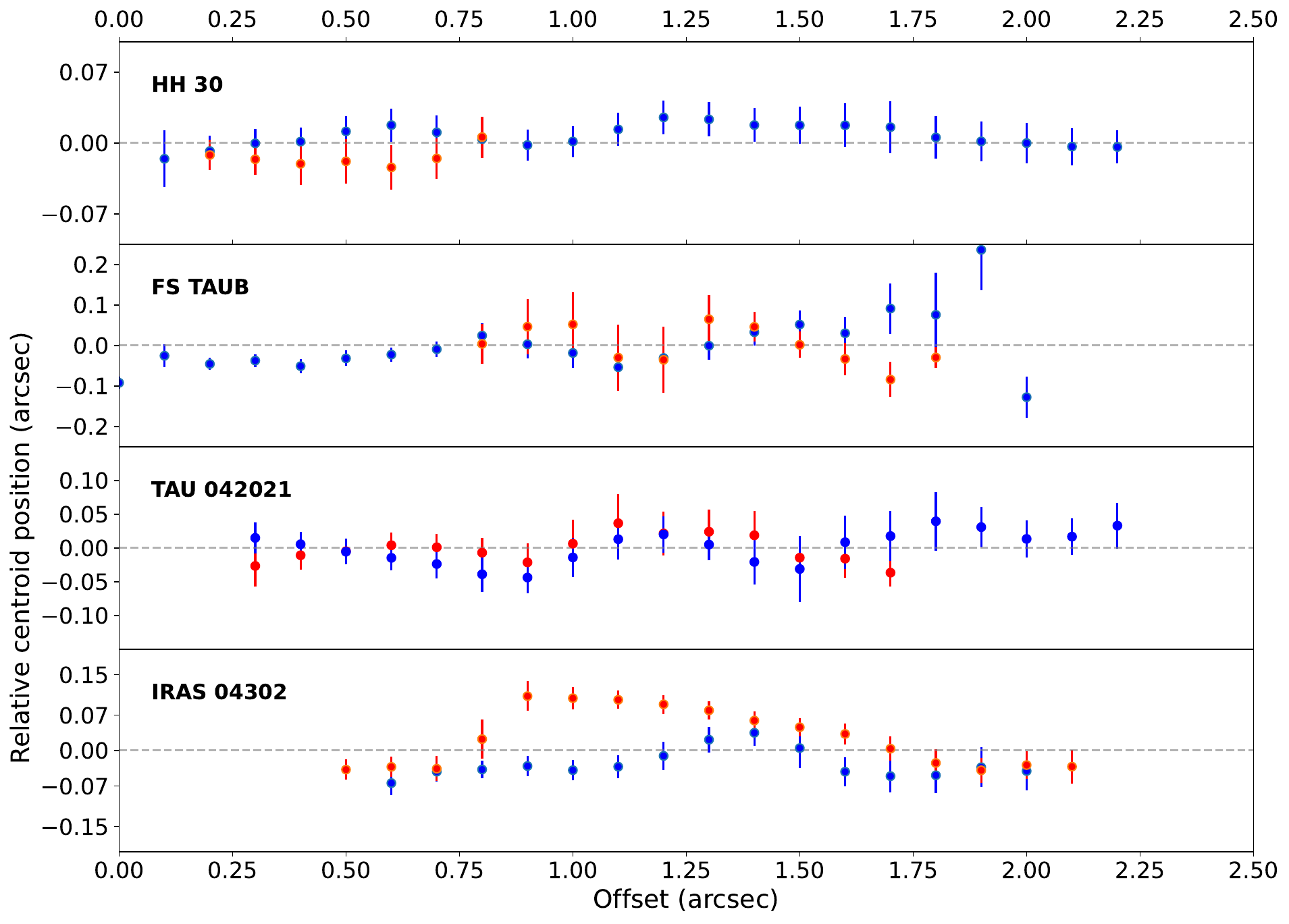}
    \caption{Each panel shows the emission centroids across the jet as a function of distance from the source. The name of the corresponding source is given at the top left. Blue points trace emission centers in the blue-shifted jet lobe, whereas red points trace emission centroids in the red-shifted lobe. Both the red and blue-shifted emissions are placed on the same side of the source for easy comparison. The vertical position of the source (i.e., 0 position on the y-axis) is determined as the average of the highest and lowest data points in the plot.}
    \label{fig:wigg_redblue}
\end{figure*}

We get the HWHM values from the Gaussian fit, and S/N is the signal-to-noise ratio calculated as the ratio of peak intensity and standard deviation of the intensity points outside the jet emission. We plot the blue-shifted and red-shifted jet lobe centroids and errors relative to the position of the star in the x direction (defined in Section \ref{sec:cont_sub}) in Figure \ref{fig:wigg_redblue}. Wiggling structures can be seen for all the sources.

\begin{figure}
    \centering
    \includegraphics[width=\columnwidth]{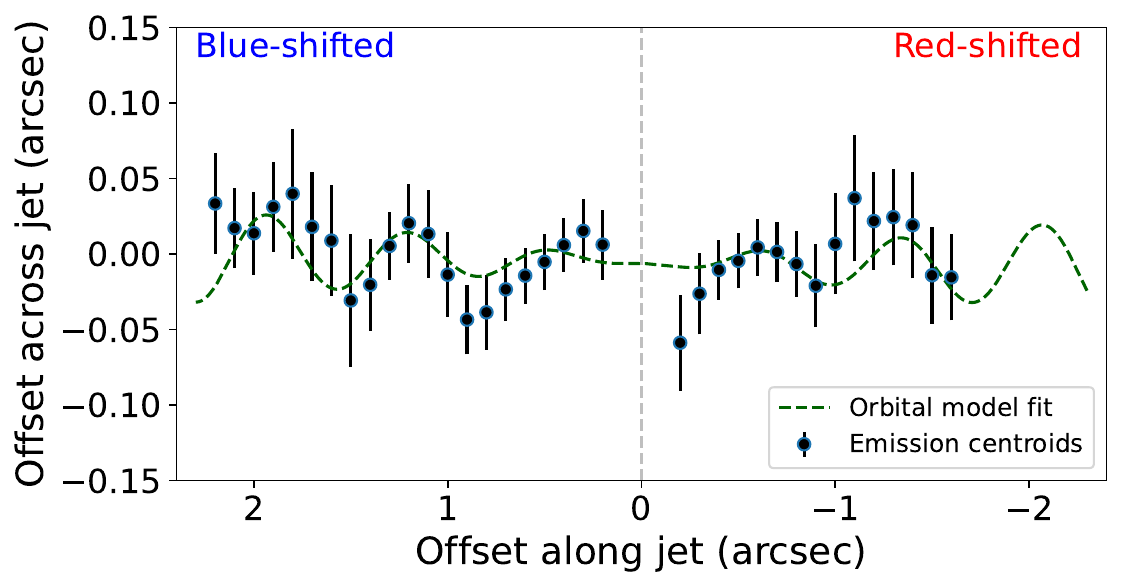}
    \caption{Emission centroids and best-fit binary orbit model for the red-shifted and the blue-shifted jets of Tau~042021, fit simultaneously. The emission centroids and errors in these plots are the same as those shown in Figure \ref{fig:wigg_redblue} for Tau~042021. The negative x-axis is to indicate that it is tracing the red-shifted jet.}
    \label{fig:wiggling_fit}
\end{figure}

Wigglings where the blue and red sides follow each other, also called mirror-symmetric wigglings, can be best explained by a binary system at the center \citep[e.g.,][]{Zinnecker1998}. Of all the sources in our sample, Tau~042021 is the only one with a clear case of mirror-symmetric wiggling which we model assuming a circular orbit for the jet source around a stellar companion \citep[e.g.,][]{Masciadri2002}. In this circular orbit model, the orbital period is calculated as a ratio of the periodic length of the wiggle and the (assumed) jet velocity, whereas, the binary separation is half the observed transverse wiggle of the jet. We refer the reader to \cite{Lee2010} for further details and equations pertaining to this orbital model. Here, we assume that only one of the binary stars is driving the jet, as only a single jet is observed. The jet wiggles of Tau~042021, along with the best-fit model, are shown in Figure \ref{fig:wiggling_fit} with the best-fit parameters given in Table \ref{tab:wiggling_fits}. We find that the orbital parameters derived from the individual fits of the red-shifted lobe and the blue-shifted lobe match with the combined fit within errors. Accordingly, we propose Tau~042021 to be a close-in binary with a separation of $\sim$1.35 au, an orbital period of $\sim$2.5 years, and a mass ratio of $\sim$5:1, i.e., stars with masses $\sim$0.33 M$_{\odot}$ (jet source) and $\sim$0.07 M$_{\odot}$ (stellar companion) considering its total mass of 0.4 M$_{\odot}$. The relatively high 0.89~mm flux of Tau~042021 \citep[124.2 $\pm$ 12.4 mJy,][]{Villenave2020} is consistent with the Taurus population of close-in binaries that host a circumbinary disk \citep[see Figure 6 of][]{Harris2012}. A close-in binary with $\sim$1.35 au separation and binary mass ratio q = 0.2 would lead to an inner tidally truncated gap of $\sim$4 au (assuming binary eccentricity = 0, e.g. Figure~9 from \citealt{Hirsh2020}). In Paper\,I, we found an average upper limit of $\sim$14\,au for the molecular wind launch base traced in near-IR H$_2$, which is still possible with this tidally truncated inner gap. However, it raises the question of whether the innermost portion of a wind launched at $\gtrsim 4$\,au can collimate over only one of the binary stars to create the observed mirror-symmetric wiggle and otherwise properties similar to those of jets from single stars. If not, it may point toward an origin of the jet from Tau~042021 closer to the star than the disk's inner edge.

\begin{table*}
    \centering
    \caption{Best-fit orbital model parameters for the blue and red-shifted jets.}
    \begin{tabular}{c c c c c | c}
        \hline
        Orbital Model Parameters & Relation$^{a}$ & Units & Red-shifted lobe & Blue-shifted lobe & Combined fit \\
        \hline
        Wiggle opening angle & $\sim$2V$_O$/V$_J$$^b$ & (rad) & 0.04 $\pm$ 0.01 & 0.03 $\pm$ 0.01 & 0.03 $\pm$ 0.01 \\
        Periodic length ($\Delta\lambda$) & V$_J$ $\times$ P$_O$ & (au) & 103 $\pm$ 10 & 110 $\pm$ 10 & 102 $\pm$ 3 \\
        Current phase angle & & (rad) & 6.2 $\pm$ 1.7 & 5.2 $\pm$ 1.2 & 5.7 $\pm$ 0.4 \\
        Stellar position vertical offset & & (au) & -2.1 $\pm$ 1.0 & -2.5 $\pm$ 1.3 & 0.7 $\pm$ 0.5 \\
        Reduced $\chi^2$ & & & 0.22 & 0.48 & 0.53 \\
        \hline
        Orbital Period (P$_O$) & $\Delta\lambda$/V$_J$ & (yr) & 2.45 $\pm$ 1.22 & 2.61 $\pm$ 1.30 & 2.42 $\pm$ 1.21 \\
        Binary mass ratio (m) & $^c$ & & 3 $\pm$ 1.5 & 5 $\pm$ 2.5 & 5 $\pm$ 2.5 \\
        Binary separation & (1+m) V$_O$~P$_O$/$\pi$ & (au) & 1.36 $\pm$ 0.40 & 1.38 $\pm$ 0.60 & 1.35 $\pm$ 0.5\\
        \hline
    \end{tabular}
    \tablecomments{The uncertainties attached to the orbital period and binary mass ratio are dominated by the uncertainty in jet velocity, which is assumed to be 200 $\pm$ 100 km~s$^{-1}$ here.}
    \tablenotetext{a}{The relations are taken from the derivation of \cite{Lee2010}}
    \tablenotetext{b}{Here V$_O$ represents the orbital velocity whereas V$_J$ represents the jet velocity}
    \tablenotetext{c}{The relation used to estimate mass ratio from model parameters is Equation 7 of \cite{Lee2010}}
    \label{tab:wiggling_fits}
\end{table*}

Jet wiggling towards HH~30 has been studied previously, and similar orbital models suggested a central binary with $\sim$18~au separation \citep[corresponding to a wiggle period of 16\arc{},][]{Anglada2007,Estalella2012}. We tried to replicate their best-fit model on our data, however, their wiggle period of 16\arc{} is much larger than our field of view ($\sim$3\arc{} in one jet direction). If HH~30 is indeed a binary with $\sim$18~au separation as suggested, then the small-scale wiggles we observe likely have a different origin than a binary, e.g., disk/jet precession.

\section{Discussion} \label{sec:discussion}

Many jet-tracing lines were detected towards our sources, including some that were previously not well-established as jet diagnostics, e.g., \hi{} and \hei{}. We discuss the importance of these lines in relation to the jet's physical properties in Section \ref{sec:new_lines}. We estimated jet mass loss rates towards all the sources using three different methods (Section \ref{sec:jmlr}). Here, we present them as mass accretion rate tracers and compare the $\dot{M}_{acc}$ estimates, whenever possible, with literature wind mass-loss rates to conduct a preliminary test of the disk-wind driven accretion scenario (Section \ref{sec:Taurus}). We also found similar extinctions towards the blue and red lobes of HH~30 and FS~TauB (Section \ref{sec:ext_map_corr}) and discuss here the implications on the jet asymmetry (Section \ref{sec:jet_asymmetry}).

\subsection{New jet tracers and excitation mechanism} \label{sec:new_lines}

We found many forbidden and permitted atomic jet-tracing lines in this work (Figure \ref{fig:line_img}). The forbidden lines of Fe, N, and S are well known to originate in the jet \citep[see][for a review]{Ray2021}; however, the lines of [\ci{}], \hei{} and \hi{} are less commonly known to trace jets and should be studied in detail using shock models. 

The [\ci{}] transitions at 0.982 and 0.985~\micron{} can result from collisional excitation, where atoms are raised to an excited state and emit a line upon returning to the ground state, or from a recombination cascade \citep{Escalante1990}. In the latter case, the permitted \ci{} lines at 1.07 and 1.17~\micron{} are expected to be only 1 to 10 times weaker than the [\ci{}] 0.982+0.985~\micron{} lines \citep{Walmsley2000}. The non-detection of the permitted \ci{} 1.07+1.17~\micron{} lines in the jet, for HH~30 and Tau~042021 where strong [\ci{}] lines are detected, suggests collisional excitation of [\ci{}]. For instance, the total [\ci{}] 0.982+0.985~\micron{} flux in the jet towards HH~30 is 440~$\times$~10$^{-20}$~W~m$^{-2}$ (Table \ref{tab:line_fluxes}) and the sensitivity for permitted \ci{} 1.07+1.17~\micron{} lines is 10-20~~$\times$~10$^{-20}$~W~m$^{-2}$. So, the non-detection of permitted \ci{} 1.07+1.17~\micron{} lines in the jet suggests that these lines are $\gtrsim$20 times weaker than the [\ci{}] lines (in the jet). This result is also consistent with those derived by \cite{Nisini2005} for spatially unresolved HH~1 jet. 

Recently, \cite{Aru2024} proposed the [\ci{}] 0.8727~\micron{} line as a tracer of externally irradiated proplyds based on its detection and non-detection in high and low UV environments, respectively. Since the [\ci{}] 0.982+0.985~\micron{} lines have excitation energies (1.26~eV) similar to that of [\ci{}] 0.8727~\micron{} (2.68~eV), we expect the latter to be detected towards HH~30 and Tau~042021 (e.g., with \texttt{VLT/MUSE}) and also trace jets. 
According to \cite{Aru2024}, the detection of [\ci{}] 0.8727~\micron{} suggests strong external irradiation by massive stars, which are absent in Taurus \citep{Luhman2004,Gudel2007}. Instead, the [\ci{}] 0.982+0.985~\micron{} lines in our Taurus sample trace jets rather than the disk surface,  highlighting the need to spatially resolve the [\ci{}] 0.8727~\micron{} to pin down its origin.

The \hei{} 1.083~\micron{} arises from the 2$^{3}$P-2$^{3}$S transition, at just $\sim 1$\,eV above the orthohelium metastable  2$^{3}$S state, which is at $\sim 20$\,eV above the ground. The 1.083~\micron{} line has been used to probe outflowing material from young accreting stars \citep[e.g.,][]{Edwards2003} and, more recently, from exoplanet atmospheres \citep[e.g.,][]{Ninan2020,Levine2024}. High-resolution spectroscopy revealed complex \hei{} 1.083~\micron{} profiles toward young accreting stars with emission likely due to magnetospheric accretion and jets \citep[e.g.,][]{Takami2002b} along with broad blueshifted absorption (up to 400\,km~s$^{-1}$) caused by a fast wind emerging from the star or its immediate surroundings \citep{Edwards2003}. Our results expand upon the high-resolution spectroscopy of young stars by spatially resolving the jet contribution. We detect \hei{} 1.083~\micron{} emission out to $\sim 200$\,au from the star, with knotty structures clearly visible in FS~TauB and Tau~042021. This spatial extent and morphology rule out stellar ionization of helium as the dominant mechanism. Instead, the results support the suggestion by \cite{Takami2002b} that shock velocities in the jet are sufficient to excite the \hei{} line. This interpretation is further corroborated by the detection of the jet in the \neii{} line at 12.8~\micron{} towards Tau~042021 \citep{Arulanantham2024}, as the ionization of Neon also requires high energy ($\sim 22$\,eV). 

Importantly, in combination with the Pa$\gamma$ line, it provides a critical test of the diagnostic potential of Iron to estimate shock speeds since the Iron-based diagnostic can be affected by uncertainties in extinction, Iron abundance in the jet, and rate coefficients. We used both \feii{}/\hi{} and \hei{}/\hi{} ratios to estimate the shock speeds in Section \ref{sec:shock_velocity_density} and found similar results suggesting (i) The latest rate coefficients for Iron are well constrained, making Iron a reliable diagnostic of jets and (ii) The abundance of Iron in the jet is solar, either due to complete liberation of iron in the shocks or due to jet launching from inside the dust sublimation radius \citep[e.g.,][]{Lee2017,Tabone2017}. Another significant finding from our study is that the \hei{} 1.083~\micron{} flux can be up to an order of magnitude higher in the jet than the \feii{} at 1.644\,\micron\ (Figure \ref{fig:line_ratios}), pointing to the \hei{} 1.083~\micron{} transition as a potentially more sensitive tracer of jets in lower accreting stars.

The fact that we found some of the hydrogen recombination lines, e.g., Pa$\alpha$/\,$\beta$/$\gamma$ tracing the jets, in addition to the scattered light emission is also an exciting result (Section \ref{sec:line_morphology}). The NIR hydrogen recombination lines are commonly used as tracers of accretion based on the empirical relations derived between the accretion luminosity and hydrogen line intensities \citep[e.g.,][]{Natta2006,Alcala2014,Rigliaco2015,Rogers2024}. If the hydrogen recombination line emission had a fixed percentage contribution from the jet relative to the accreting region, then the empirical relations would not be affected by it, however, the Pa$\beta$/\feii{} ratio (Figure \ref{fig:line_ratios}) showed that the strength of Pa$\beta$ in the jet varies for every source. This can explain some of the spread observed in the accretion rate-\hi{} luminosity relation and suggests that the \hi{} should be used with caution for calculating the mass accretion rates.  

\subsection{The Origin of the Discrepancy Between \feii{}-derived and \sii{}-derived Electron Densities}

We found a significant difference in the electron densities estimated using the NIR \feii{} and \sii{} diagnostics (see Section \ref{sec:both_elec_den} and Figure \ref{fig:elecden_map_SII}). In Section \ref{sec:fe_s_ratio_ne}, we show that our observed line ratios fall on the linear portion of the ratio-density curves, meaning n$_e$~$<$~n$_c$ and that our electron density estimates are valid. There are a few possible ways to explain this discrepancy. (i) Since the rate coefficients for Iron are difficult to calculate, it could be that the \feii{}-derived electron densities are systematically underestimated. In this case, a constant difference between the \feii{}-derived and \sii{}-derived electron densities would be expected. However, we find the \sii{}/\feii{} density ratio to vary from $\sim$2-50 in HH~30 to $\sim$50-100 in other sources (Figure \ref{fig:elecden_map_SII}).

(ii) Another possibility is that the \sii{} NIR lines could be tracing a denser region of the jet, such as the head of the bow shock or a very narrow jet spine. In the former case, the intensity maps would be expected to peak at different locations as a function of distance between the two species. However, we find their intensity curves to be quite similar. We test the latter case by comparing the radius of the jet (FWHM) in the \feii{} 1.644~\micron{} line and the \sii{} 1.032~\micron{} line. We find that the low SNR detection of \sii{} further out from the star makes it difficult to reach any conclusion. Simultaneous high SNR detection of both species in a broad jet will be needed to test this possibility. (iii) Another possibility could be unresolved clumpiness in the jet material. This is likely as the jet is intermittent. In this case, \sii{} would be tracing denser knots of material ejecta, leading to a much smaller filling factor in \sii{} than \feii{}. If cases (i) or (ii) are true, then the jet mass loss rates derived in this work using the \feii{} lines could potentially be underestimated. However, if case (iii) is true, then \feii{} is likely a better diagnostic of the jet mass loss rate than the NIR \sii{} lines. Hence, a detailed investigation of the origin of this density discrepancy is crucial in future works.

\subsection{Mass Accretion Rates and Comparison with the Winds} \label{sec:Taurus}

\begin{figure*}
    \centering
    \includegraphics[width=1.5\columnwidth]{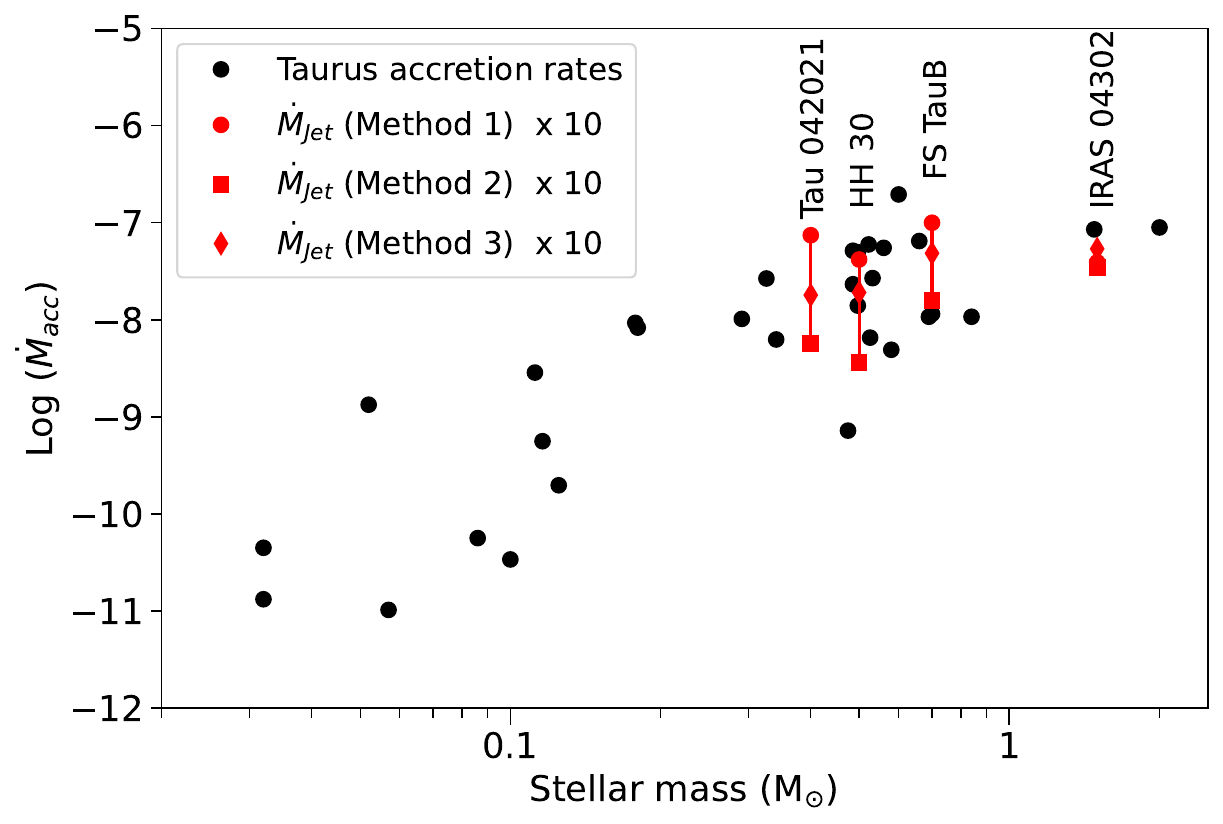}
    \caption{In red are the mass accretion estimates from this work calculated using the jet mass loss rates acquired using three different methods. These estimates are placed alongside literature values of Taurus sources (in black) calculated using Hydrogen recombination lines and/or UV excess by \cite{Herczeg2008,Herczeg2014,Testi2022} and compiled by \cite{Manara2023}.}
    \label{fig:jmlr_taurus}
\end{figure*}

Mass accretion rates are available for hundreds of stars in different star-forming regions and are obtained either from emission lines probing accreting gas or UV excess emission from the accretion shocks \citep[see][for a recent review]{Manara2023}. Figure \ref{fig:jmlr_taurus} shows $\dot{M}_{acc}$ against stellar mass for sources that belong to the Taurus star-forming region like our targets. As noted in other star-forming regions \citep[e.g.,][]{Testi2022}, $\dot{M}_{acc}$ is positively correlated with $M_{\star}$.

Several studies have shown that, for edge-on disks, these methods underestimate the mass accretion rate by orders of magnitude \citep[e.g.,][]{Manara2016,Arulanantham2024}. This is due to the accretion region being heavily obscured by the disk, leading to reduced line and UV continuum flux reaching the telescope. For instance, using the extinction-corrected Pa$\alpha$ line luminosity of Tau~042021, we find log(L$_{acc}$/L$_{\odot}$) = -4.1 \citep[following the empirical relation between Pa$\alpha$ and L$_{acc}$ derived by][]{Rogers2024}. Assuming a stellar radius of 1.25~R$_{\odot}$ for a 0.4~M$_{\odot}$ star \citep{Lin2023}, the resulting accretion rate is 10$^{-11}$~M$_{\odot}$~yr$^{-1}$, nearly three orders of magnitude lower than expected (see Figure \ref{fig:jmlr_taurus}). Jet mass loss rates provide an alternative way to estimate mass accretion rates for such edge-on systems. As mentioned earlier, jet mass loss rates correlate with mass accretion rates, and the $\dot{M}_{jet}$/$\dot{M}_{acc}$ ratio ranges from 0.01-0.5 with an average value of 0.07 \citep[e.g.,][]{Nisini2018}. We over plot our accretion estimates calculated as 10 $\times$ the jet mass loss rates in Figure \ref{fig:jmlr_taurus} and find them to agree well with the Taurus population. This demonstrates that our selected sources are neither high nor low accretors but are rather typical of this region, meaning that our small sample of edge-on disks is representative for testing the disk-wind-driven accretion scenario for stars with masses 0.4-1.5~M$_{\odot}$.

We can conduct a preliminary test of the disk-wind-driven accretion scenario by comparing our $\dot{M}_{acc}$ with available $\dot{M}_{wind}$. More specifically, in the disk-wind-driven accretion scenario, we expect $\dot{M}_{wind}$ to be 1-4 times $\dot{M}_{acc}$ \citep[see][Section 3.2.2]{Pascucci2023}. For Tau~042021, \cite{Arulanantham2024} estimated the wind mass loss rate using mid-IR spatially resolved warm H$_2$ emission as 6.5 $\times$ 10$^{-9}$~M$_{\odot}$~yr$^{-1}$. Comparison to our mass accretion rate estimates (see Table \ref{tab:mass_loss_accretion}) gives a ratio of $\sim$0.1-2. This partially agrees with the expected ratio, considering that Method 1 gives an upper limit on the jet mass loss rate and, correspondingly, an upper limit on the accretion rate and a lower limit on the ratio. On the other hand, for HH~30, the wind mass loss rate has been estimated to be 9 $\times$ 10$^{-8}$ M$_{\odot}$~yr$^{-1}$ using spatially resolved $^{12}$CO(2-1) molecular emission \citep{Louvet2018,Lopez-Vazquez2024}. Our estimate of the mass accretion rate is nearly an order of magnitude smaller than the wind mass loss rate, giving a ratio of $\sim$3.2-30. So, in both cases, we find there are $\dot{M}_{acc}$ estimates consistent with the disk-wind-driven accretion scenario. However, some values lie outside the expected range. We will further test this scenario by comparing the mass accretion rates found in this work, with the wind mass-loss rates contemporaneously estimated using the spatially resolved H$_2$ lines for all the sources in our data (Paper~III).

\subsection{The Origin of the Asymmetric Jets} \label{sec:jet_asymmetry}

Nearly 50\% of the jets are found to be asymmetric in intensity, morphology, velocity, temperature, density, or a combination of these parameters \citep{Hirth1994,Mundt1987,Woitas2005,Podio2011}. These asymmetries are seen on either side of the jet. Two such sources known to have asymmetric lengths and velocities are HH~30 and FS~TauB. In terms of length, both sources have shorter red-shifted components than the blue-shifted: for HH~30 the red-shifted lobe extends out to 35\arc{} and 85\arc{} and the blue out to 2$^{\prime{}}$ \citep{Mundt1990,Ray1996,Bacciotti1999,Hartigan2007}; for FS~TauB the red-shifted lobe extends out to $\sim$38\arc{} and the blue-shifted lobe to $\sim$2$^{\prime{}}$ \citep{Mundt1991,Eisloffel1998,Liu2012}. In terms of deprojected velocity, \cite{Eisloffel1998} found the blue-shifted lobe of FS~TauB to be 20\% faster than the red-shifted lobe in the optical \sii{} line and \cite{Liu2012} found it to be 34\% faster in the \feii{} line. On the contrary, for HH~30, \cite{Hartigan2007} and \cite{Estalella2012} found the red-shifted lobe to be 70\% faster than the blue-shifted lobe using the proper motion of the knots within 3.2\arc{}. We observed this length asymmetry in the near-IR \feii{} lines, too, albeit at scales of a few arcsec (see Figure \ref{fig:line_img}).

There have been several indiscernible explanations for jet asymmetries, primarily due to the lack of theoretical investigation into the matter as well as the lack of observed jet properties for both lobes. Our new estimates of the extinction (Section \ref{sec:ext_map_corr}), electron density (Section \ref{sec:elec_den}), shock speed, ionization fraction (Section \ref{sec:shock_velocity_density}), the jet mass loss rates (Section \ref{sec:jmlr}) and literature estimates of the velocity, along with the observed wind asymmetry (see Paper~I) provide a fresh view to this problem.

The asymmetric jets of HH~30 and FS~TauB can be produced either by an intrinsically asymmetric jet launch mechanism that operates differently on the two sides or by interaction with the immediate surrounding material, such that the ambient material is denser on the weaker side of the jet. The latter case can be tested by mapping the visual extinction on both sides of the jet. If the asymmetry is caused by external factors such as the surrounding material, the weaker jet lobe will face higher extinction. Our extinction maps (Section \ref{sec:ext_map_corr} and Figure \ref{fig:ext_map}) show that there isn't any significant extinction difference between the two sides that would make the counter-jet more obscured than the jet, suggesting that their jet asymmetry is intrinsic to the jet launch mechanism.

\begin{enumerate}
    \item One possible explanation is given by the X-wind scenario in which the low intensity of a jet lobe is explained by less ambient material available to entrain on that side. In this case, the high velocity of the other jet lobe could be due to the atomic emission partially tracing the entrained layers \citep{Louvet2018}. Since HH~30 and FS~TauB are Class~II sources that have shed their envelope and do not show any significant differences in the extinction (Figure \ref{fig:ext_map}), we do not prefer this explanation. This is also consistent with the interpretation in Paper~I.

    \item Another explanation comes from conserving the linear momentum of the jet lobes where higher velocity will lead to lower mass loss on that side of the jet, as no recoil effect has been observed towards such young star systems. Considering the standard errors on the mass loss rates shown in Figure \ref{fig:jmlr_comp} along with the 30\% uncertainty propagated through velocity, Method 1 for HH~30 and all methods for FS~TauB give similar mass loss rates on the two sides. However, Method 2 and 3 for HH~30 suggest that the mass loss rate is higher in the blue-shifted lobe of the jet ($\dot{M}_{j(blue)}$/$\dot{M}_{j(red)}$$\sim$1.6-2.8 and $\dot{M}_{j(blue)}$/$\dot{M}_{j(red)}$$\sim$1.7-20, respectively), which is partially comparable to the velocity difference, where the red-shifted lobe is observed to be faster than the blue-shifted lobe (v$_{j(red)}$/v$_{j(blue)}$=1.7). Such difference in jet velocities could originate from a difference in the launch radius, where the higher velocity component is launched closer to the source as opposed to the lower velocity component, which is launched further away \citep{Ferreira2006}. Non-ideal MHD effects, such as the Hall effect can also lead to a velocity difference with the direction of the higher velocity component governed by the sense of disk rotation. If the jets and the winds are driven by the same physical process, then such an asymmetry would also lead to an asymmetric wind, which has now been observed towards both sources \citep{Louvet2018,Pascucci2024}. 
\end{enumerate}
 
Considering the above scenarios, we suggest the origin of the HH~30 and FS~TauB jet in a collimated inner MHD disk wind and that the apparent brightness/velocity asymmetry is due to an intrinsic difference in the launch mechanism on the two sides. IRAS~04302 and Tau~042021 do not seem to be asymmetric in our data based on their length or [Fe\,II] intensity, however, high-res spectra and/or proper motion studies are needed to investigate any velocity asymmetry.

\section{Summary and Conclusions} \label{sec:conclusion}

We calibrated, deconvolved, and analyzed JWST NIRSpec IFU observations of four edge-on disks focusing on the wavelength range $\sim$1-1.9\,\micron{} which is rich in jet diagnostics. We analyzed 1D spectra integrated on the entire IFU to identify gas emission lines, performed spaxel-by-spaxel continuum subtraction, and integrated the line emission in every spaxel for selected lines to create jet intensity maps. Further, we used: i) the ratio of \feii{} 1.257 and 1.644 \micron{} lines to estimate extinction towards the jet in every pixel; ii) \feii{}, \hei{} and hydrogen line ratios to estimate the shock speeds, pre-shock densities, and ionization fractions; and iii) pairs of \feii{} and \sii{} lines that share similar excitation energies to estimate the post-shock electron density. Using these quantities, we estimated jet mass loss rates with 3 different methods. Finally, to investigate jet precession, we plotted emission centroids perpendicular to the jet axis as a function of distance for both the blue-shifted and the red-shifted jet. The main results from this work are listed below:

\begin{itemize}
    \item We detect more than 40 emission lines towards each source. The forbidden lines of \feii{}, [\ci{}], \pii{}, \sii{}, and \ngi{} all trace narrow jets as expected. We consistently find that \hi{}~X-3 (where X=4,5,6,7) and \hei{} trace a jet (in addition to scatted light from the disk). For HH~30 and FS~TauB, we also find that the \hei{} line can be an order of magnitude stronger than the \feii{} in the jet, thus revealing a potentially more sensitive tracer for weaker jets from lower accretors. Finally, we also find two permitted \ci{} lines that seem to trace emission from the disk itself when detected.

    \item We find an excellent match of our data with the pre-shock density curve of 1000~cm$^{-3}$ and shock speeds of 50-60 km$s^{-1}$ for all the sources \citep[using shock model from][]{Koo2016}. These values are consistent with those found for other jets from low-mass stars. We also find the jets to be $\sim$21\% ionized except for Tau~042021, which is likely less ionized ($\sim$12\%).

    \item The shock speeds estimated using the \hei{}/Pa$\gamma$ ratio in the jet (using the \texttt{MAPPINGS V} shock model) match well with those derived using the \feii{}/Pa$\beta$ ratio assuming solar abundance of Iron. This supports the complete liberation of Iron in the jet and/or jet launch within the dust sublimation radius. 

    \item Using \feii{} line ratios, we find the post-shock jet electron densities to be the largest close to the source and that they decline gradually outwards, except for FS~TauB, where the densities remain similar. Densities are in the range 10$^3$-10$^5$~cm$^{-3}$ with HH~30, the narrowest jet, showing the largest electron density. We also use the NIR \sii{} line ratios for the first time and find 2-100 times larger electron densities than those found using the \feii{} line ratios.

    \item We constrain the jet mass loss rates between $\sim$0.1-10 $\times$ 10$^{-9}$ M$_{\odot}$~yr$^{-1}$. We find the post-shock electron density and jet cross-section-based method to consistently provide a higher estimate than the luminosity-based methods. Our estimated mass accretion rates (10 $\times$ $\dot{M}_{jet}$) are consistent with the $\dot{M}_{acc}$-M$_{\odot}$ spread of the Taurus star-forming region. Comparison with literature wind mass loss rate estimates gives $\dot{M}_{wind}$/$\dot{M}_{acc}$ $\sim$ 0.1-2 for Tau~042021, and $\sim$ 3.2-30 for HH~30.

    \item We also find all the jets to wiggle along their flow axes. Of particular interest is Tau~042021, which shows mirror-symmetric wiggles between the red and blue-shifted jets. A simple circular orbit model suggests Tau~042021 to be a close binary with a separation of $\sim$1.35 au, orbital period of $\sim$2.5 yr, and a mass ratio of $\sim$4:1.
\end{itemize}

With a strong detection of 17 \feii{} lines tracing jets in the NIRSpec IFU mode between 0.97 and 1.82 \micron{}, we present one of the richest datasets of Class~I/II sources to characterize jets. The lack of extinction difference on the two sides of the jet suggests that the interstellar material surrounding these sources has similar densities and that the length and/or brightness asymmetries of HH~30 and FS~TauB are intrinsic to the jet launch mechanism. Owing to the higher mass loss rates observed on the side with lower velocity (for HH~30), non-ideal MHD effects or an asymmetry in the launch radius are better suited to explain the observed jet intensity and velocity asymmetry. Such an asymmetry would also lead to an asymmetric wind emission, which has already been reported towards both sources \citep{Louvet2018,Pascucci2024}. This would then suggest the origin of the HH~30 and FS~TauB jet in an inner, more collimated, and atomic component of the wider MHD disk winds, rather than X-winds. We conducted a preliminary test of whether these MHD disk winds drive accretion in the disk by comparing the literature wind mass-loss rates to the mass accretion rates from this work and found the $\dot{M}_{wind}$/$\dot{M}_{acc}$ ratio for Tau~042021 and HH~30 to partially agree with the expected range (1-4). We will conduct a more detailed test of this scenario by contemporaneously measuring the wind mass-loss rates using the spatially resolved H$_2$ lines for all the sources in Paper~III.

\section{acknowledgments}

This work is based on observations made with the NASA/ESA/CSA James Webb Space Telescope. The data were obtained from the Mikulski Archive for Space Telescopes at the Space Telescope Science Institute, which is operated by the Association of Universities for Research in Astronomy, Inc., under NASA contract NAS 5-03127 for JWST. The specific observations analyzed can be accessed via \dataset[doi:10.17909/s62y-vt90]{https://doi.org/10.17909/s62y-vt90}. These observations are associated with the GO Cycle 1 program 1621. N.~S.~B. and I.~P. acknowledge partial support from NASA/STScI GO grant JWST-GO-01621.001. D.~S. acknowledges support from the European Research Council under the Horizon 2020 Framework Program via the ERC Advanced Grant Origins 83 24 28 (PI: Th. Henning).


\vspace{5mm}
\facilities{James Webb Space Telescope}

\software{Astropy \citep{astropy:2013,astropy:2018,astropy:2022}, JWST \citep{Bushouse2024}, Matplotlib \citep{Hunter:2007}, NumPy \citep{harris2020array}, PyNeb \citep{Luridiana2015}, Scikit-learn \citep{scikit-learn}, SciPy \citep{2020SciPy-NMeth}}

\bibliography{citations.bib}{}

\begin{thebibliography}{}
\expandafter\ifx\csname natexlab\endcsname\relax\def\natexlab#1{#1}\fi
\providecommand{\url}[1]{\href{#1}{#1}}
\providecommand{\dodoi}[1]{doi:~\href{http://doi.org/#1}{\nolinkurl{#1}}}
\providecommand{\doeprint}[1]{\href{http://ascl.net/#1}{\nolinkurl{http://ascl.net/#1}}}
\providecommand{\doarXiv}[1]{\href{https://arxiv.org/abs/#1}{\nolinkurl{https://arxiv.org/abs/#1}}}

\bibitem[{{Agra-Amboage} {et~al.}(2011){Agra-Amboage}, {Dougados}, {Cabrit}, \& {Reunanen}}]{Agra-Amboage2011}
{Agra-Amboage}, V., {Dougados}, C., {Cabrit}, S., \& {Reunanen}, J. 2011, \aap, 532, A59, \dodoi{10.1051/0004-6361/201015886}

\bibitem[{{Alcal{\'a}} {et~al.}(2014){Alcal{\'a}}, {Natta}, {Manara}, {Spezzi}, {Stelzer}, {Frasca}, {Biazzo}, {Covino}, {Randich}, {Rigliaco}, {Testi}, {Comer{\'o}n}, {Cupani}, \& {D'Elia}}]{Alcala2014}
{Alcal{\'a}}, J.~M., {Natta}, A., {Manara}, C.~F., {et~al.} 2014, \aap, 561, A2, \dodoi{10.1051/0004-6361/201322254}

\bibitem[{{Anglada} {et~al.}(2007){Anglada}, {L{\'o}pez}, {Estalella}, {Masegosa}, {Riera}, \& {Raga}}]{Anglada2007}
{Anglada}, G., {L{\'o}pez}, R., {Estalella}, R., {et~al.} 2007, \aj, 133, 2799, \dodoi{10.1086/517493}

\bibitem[{{Aru} {et~al.}(2024){Aru}, {Mauco}, {Manara}, {Haworth}, {Ballering}, {Boyden}, {Campbell-White}, {Facchini}, {Rosotti}, {Winter}, {Miotello}, {McLeod}, {Robberto}, {Petr-Gotzens}, {Ballabio}, {Vicente}, {Ansdell}, \& {Cleeves}}]{Aru2024}
{Aru}, M.-L., {Mauco}, K., {Manara}, C.~F., {et~al.} 2024, arXiv e-prints, arXiv:2410.21018, \dodoi{10.48550/arXiv.2410.21018}

\bibitem[{{Arulanantham} {et~al.}(2024){Arulanantham}, {McClure}, {Pontoppidan}, {Beck}, {Sturm}, {Harsono}, {Boogert}, {Cordiner}, {Dartois}, {Drozdovskaya}, {Espaillat}, {Melnick}, {Noble}, {Palumbo}, {Pendleton}, {Terada}, \& {van Dishoeck}}]{Arulanantham2024}
{Arulanantham}, N., {McClure}, M.~K., {Pontoppidan}, K., {et~al.} 2024, \apjl, 965, L13, \dodoi{10.3847/2041-8213/ad35c9}

\bibitem[{{Asplund} {et~al.}(2009){Asplund}, {Grevesse}, {Sauval}, \& {Scott}}]{Asplund2009}
{Asplund}, M., {Grevesse}, N., {Sauval}, A.~J., \& {Scott}, P. 2009, \araa, 47, 481, \dodoi{10.1146/annurev.astro.46.060407.145222}

\bibitem[{{Astropy Collaboration} {et~al.}(2013){Astropy Collaboration}, {Robitaille}, {Tollerud}, {Greenfield}, {Droettboom}, {Bray}, {Aldcroft}, {Davis}, {Ginsburg}, {Price-Whelan}, {Kerzendorf}, {Conley}, {Crighton}, {Barbary}, {Muna}, {Ferguson}, {Grollier}, {Parikh}, {Nair}, {Unther}, {Deil}, {Woillez}, {Conseil}, {Kramer}, {Turner}, {Singer}, {Fox}, {Weaver}, {Zabalza}, {Edwards}, {Azalee Bostroem}, {Burke}, {Casey}, {Crawford}, {Dencheva}, {Ely}, {Jenness}, {Labrie}, {Lim}, {Pierfederici}, {Pontzen}, {Ptak}, {Refsdal}, {Servillat}, \& {Streicher}}]{astropy:2013}
{Astropy Collaboration}, {Robitaille}, T.~P., {Tollerud}, E.~J., {et~al.} 2013, \aap, 558, A33, \dodoi{10.1051/0004-6361/201322068}

\bibitem[{{Astropy Collaboration} {et~al.}(2018){Astropy Collaboration}, {Price-Whelan}, {Sip{\H{o}}cz}, {G{\"u}nther}, {Lim}, {Crawford}, {Conseil}, {Shupe}, {Craig}, {Dencheva}, {Ginsburg}, {Vand erPlas}, {Bradley}, {P{\'e}rez-Su{\'a}rez}, {de Val-Borro}, {Aldcroft}, {Cruz}, {Robitaille}, {Tollerud}, {Ardelean}, {Babej}, {Bach}, {Bachetti}, {Bakanov}, {Bamford}, {Barentsen}, {Barmby}, {Baumbach}, {Berry}, {Biscani}, {Boquien}, {Bostroem}, {Bouma}, {Brammer}, {Bray}, {Breytenbach}, {Buddelmeijer}, {Burke}, {Calderone}, {Cano Rodr{\'\i}guez}, {Cara}, {Cardoso}, {Cheedella}, {Copin}, {Corrales}, {Crichton}, {D'Avella}, {Deil}, {Depagne}, {Dietrich}, {Donath}, {Droettboom}, {Earl}, {Erben}, {Fabbro}, {Ferreira}, {Finethy}, {Fox}, {Garrison}, {Gibbons}, {Goldstein}, {Gommers}, {Greco}, {Greenfield}, {Groener}, {Grollier}, {Hagen}, {Hirst}, {Homeier}, {Horton}, {Hosseinzadeh}, {Hu}, {Hunkeler}, {Ivezi{\'c}}, {Jain}, {Jenness}, {Kanarek}, {Kendrew}, {Kern}, {Kerzendorf}, {Khvalko}, {King}, {Kirkby}, {Kulkarni},
  {Kumar}, {Lee}, {Lenz}, {Littlefair}, {Ma}, {Macleod}, {Mastropietro}, {McCully}, {Montagnac}, {Morris}, {Mueller}, {Mumford}, {Muna}, {Murphy}, {Nelson}, {Nguyen}, {Ninan}, {N{\"o}the}, {Ogaz}, {Oh}, {Parejko}, {Parley}, {Pascual}, {Patil}, {Patil}, {Plunkett}, {Prochaska}, {Rastogi}, {Reddy Janga}, {Sabater}, {Sakurikar}, {Seifert}, {Sherbert}, {Sherwood-Taylor}, {Shih}, {Sick}, {Silbiger}, {Singanamalla}, {Singer}, {Sladen}, {Sooley}, {Sornarajah}, {Streicher}, {Teuben}, {Thomas}, {Tremblay}, {Turner}, {Terr{\'o}n}, {van Kerkwijk}, {de la Vega}, {Watkins}, {Weaver}, {Whitmore}, {Woillez}, {Zabalza}, \& {Astropy Contributors}}]{astropy:2018}
{Astropy Collaboration}, {Price-Whelan}, A.~M., {Sip{\H{o}}cz}, B.~M., {et~al.} 2018, \aj, 156, 123, \dodoi{10.3847/1538-3881/aabc4f}

\bibitem[{{Astropy Collaboration} {et~al.}(2022){Astropy Collaboration}, {Price-Whelan}, {Lim}, {Earl}, {Starkman}, {Bradley}, {Shupe}, {Patil}, {Corrales}, {Brasseur}, {N{"o}the}, {Donath}, {Tollerud}, {Morris}, {Ginsburg}, {Vaher}, {Weaver}, {Tocknell}, {Jamieson}, {van Kerkwijk}, {Robitaille}, {Merry}, {Bachetti}, {G{"u}nther}, {Aldcroft}, {Alvarado-Montes}, {Archibald}, {B{'o}di}, {Bapat}, {Barentsen}, {Baz{'a}n}, {Biswas}, {Boquien}, {Burke}, {Cara}, {Cara}, {Conroy}, {Conseil}, {Craig}, {Cross}, {Cruz}, {D'Eugenio}, {Dencheva}, {Devillepoix}, {Dietrich}, {Eigenbrot}, {Erben}, {Ferreira}, {Foreman-Mackey}, {Fox}, {Freij}, {Garg}, {Geda}, {Glattly}, {Gondhalekar}, {Gordon}, {Grant}, {Greenfield}, {Groener}, {Guest}, {Gurovich}, {Handberg}, {Hart}, {Hatfield-Dodds}, {Homeier}, {Hosseinzadeh}, {Jenness}, {Jones}, {Joseph}, {Kalmbach}, {Karamehmetoglu}, {Ka{l}uszy{'n}ski}, {Kelley}, {Kern}, {Kerzendorf}, {Koch}, {Kulumani}, {Lee}, {Ly}, {Ma}, {MacBride}, {Maljaars}, {Muna}, {Murphy}, {Norman}, {O'Steen},
  {Oman}, {Pacifici}, {Pascual}, {Pascual-Granado}, {Patil}, {Perren}, {Pickering}, {Rastogi}, {Roulston}, {Ryan}, {Rykoff}, {Sabater}, {Sakurikar}, {Salgado}, {Sanghi}, {Saunders}, {Savchenko}, {Schwardt}, {Seifert-Eckert}, {Shih}, {Jain}, {Shukla}, {Sick}, {Simpson}, {Singanamalla}, {Singer}, {Singhal}, {Sinha}, {Sip{H{o}}cz}, {Spitler}, {Stansby}, {Streicher}, {{{S}}umak}, {Swinbank}, {Taranu}, {Tewary}, {Tremblay}, {Val-Borro}, {Van Kooten}, {Vasovi{'c}}, {Verma}, {de Miranda Cardoso}, {Williams}, {Wilson}, {Winkel}, {Wood-Vasey}, {Xue}, {Yoachim}, {Zhang}, {Zonca}, \& {Astropy Project Contributors}}]{astropy:2022}
{Astropy Collaboration}, {Price-Whelan}, A.~M., {Lim}, P.~L., {et~al.} 2022, \apj, 935, 167, \dodoi{10.3847/1538-4357/ac7c74}

\bibitem[{{Bacciotti} {et~al.}(1999){Bacciotti}, {Eisl{\"o}ffel}, \& {Ray}}]{Bacciotti1999}
{Bacciotti}, F., {Eisl{\"o}ffel}, J., \& {Ray}, T.~P. 1999, \aap, 350, 917

\bibitem[{{Bai} \& {Stone}(2013)}]{Bai2013}
{Bai}, X.-N., \& {Stone}, J.~M. 2013, \apj, 769, 76, \dodoi{10.1088/0004-637X/769/1/76}

\bibitem[{{Bajaj} {et~al.}(2024){Bajaj}, {Pascucci}, {Gorti}, {Alexander}, {Sellek}, {Morrison}, {Gaspar}, {Clarke}, {Xie}, {Ballabio}, \& {Deng}}]{Bajaj2024}
{Bajaj}, N.~S., {Pascucci}, I., {Gorti}, U., {et~al.} 2024, \aj, 167, 127, \dodoi{10.3847/1538-3881/ad22e1}

\bibitem[{{Bautista} {et~al.}(2015){Bautista}, {Fivet}, {Ballance}, {Quinet}, {Ferland}, {Mendoza}, \& {Kallman}}]{Bautista2015}
{Bautista}, M.~A., {Fivet}, V., {Ballance}, C., {et~al.} 2015, \apj, 808, 174, \dodoi{10.1088/0004-637X/808/2/174}

\bibitem[{{Beck-Winchatz} {et~al.}(1994){Beck-Winchatz}, {Bohm}, \& {Noriega-Crespo}}]{Beck-Winchatz1994}
{Beck-Winchatz}, B., {Bohm}, K.~H., \& {Noriega-Crespo}, A. 1994, \pasp, 106, 1271, \dodoi{10.1086/133504}

\bibitem[{{Bushouse} {et~al.}(2024){Bushouse}, {Eisenhamer}, {Dencheva}, {Davies}, {Greenfield}, {Morrison}, {Hodge}, {Simon}, {Grumm}, {Droettboom}, {Slavich}, {Sosey}, {Pauly}, {Miller}, {Jedrzejewski}, {Hack}, {Davis}, {Crawford}, {Law}, {Gordon}, {Regan}, {Cara}, {MacDonald}, {Bradley}, {Shanahan}, {Jamieson}, {Teodoro}, {Williams}, \& {Pena-Guerrero}}]{Bushouse2024}
{Bushouse}, H., {Eisenhamer}, J., {Dencheva}, N., {et~al.} 2024, {JWST Calibration Pipeline}, 1.13.4,  Zenodo, \dodoi{10.5281/zenodo.6984365}

\bibitem[{{Cabrit}(2002)}]{Cabrit2002}
{Cabrit}, S. 2002, in EAS Publications Series, Vol.~3, EAS Publications Series, ed. J.~{Bouvier} \& J.-P. {Zahn}, 147--182, \dodoi{10.1051/eas:2002049}

\bibitem[{{Cox} \& {Raymond}(1985)}]{Cox1985}
{Cox}, D.~P., \& {Raymond}, J.~C. 1985, \apj, 298, 651, \dodoi{10.1086/163649}

\bibitem[{{Dopita} \& {Sutherland}(2017)}]{Dopita2017}
{Dopita}, M.~A., \& {Sutherland}, R.~S. 2017, \apjs, 229, 35, \dodoi{10.3847/1538-4365/aa6542}

\bibitem[{{Dos Santos, Leonardo}(2023)}]{Leonardo2023}
{Dos Santos, Leonardo}. 2023, Commissioning data for NIRSpec/G140H,  STScI/MAST, \dodoi{10.17909/2C5E-DM80}

\bibitem[{{Dougados} {et~al.}(2010){Dougados}, {Bacciotti}, {Cabrit}, \& {Nisini}}]{Dougados2010}
{Dougados}, C., {Bacciotti}, F., {Cabrit}, S., \& {Nisini}, B. 2010, in Lecture Notes in Physics, Berlin Springer Verlag, ed. P.~J.~V. {Garcia} \& J.~M. {Ferreira}, Vol. 793, 213, \dodoi{10.1007/978-3-642-02289-0_7}

\bibitem[{{Edwards} {et~al.}(2003){Edwards}, {Fischer}, {Kwan}, {Hillenbrand}, \& {Dupree}}]{Edwards2003}
{Edwards}, S., {Fischer}, W., {Kwan}, J., {Hillenbrand}, L., \& {Dupree}, A.~K. 2003, \apjl, 599, L41, \dodoi{10.1086/381077}

\bibitem[{{Eisl{\"o}ffel} \& {Mundt}(1998)}]{Eisloffel1998}
{Eisl{\"o}ffel}, J., \& {Mundt}, R. 1998, \aj, 115, 1554, \dodoi{10.1086/300282}

\bibitem[{{Escalante} \& {Victor}(1990)}]{Escalante1990}
{Escalante}, V., \& {Victor}, G.~A. 1990, \apjs, 73, 513, \dodoi{10.1086/191479}

\bibitem[{{Estalella} {et~al.}(2012){Estalella}, {L{\'o}pez}, {Anglada}, {G{\'o}mez}, {Riera}, \& {Carrasco-Gonz{\'a}lez}}]{Estalella2012}
{Estalella}, R., {L{\'o}pez}, R., {Anglada}, G., {et~al.} 2012, \aj, 144, 61, \dodoi{10.1088/0004-6256/144/2/61}

\bibitem[{{Fang} {et~al.}(2018){Fang}, {Pascucci}, {Edwards}, {Gorti}, {Banzatti}, {Flock}, {Hartigan}, {Herczeg}, \& {Dupree}}]{Fang2018ApJ...868...28F}
{Fang}, M., {Pascucci}, I., {Edwards}, S., {et~al.} 2018, \apj, 868, 28, \dodoi{10.3847/1538-4357/aae780}

\bibitem[{{Ferreira} {et~al.}(2006){Ferreira}, {Dougados}, \& {Cabrit}}]{Ferreira2006}
{Ferreira}, J., {Dougados}, C., \& {Cabrit}, S. 2006, \aap, 453, 785, \dodoi{10.1051/0004-6361:20054231}

\bibitem[{{Garcia Lopez} {et~al.}(2008){Garcia Lopez}, {Nisini}, {Giannini}, {Eisl{\"o}ffel}, {Bacciotti}, \& {Podio}}]{Garcia-Lopez2008}
{Garcia Lopez}, R., {Nisini}, B., {Giannini}, T., {et~al.} 2008, \aap, 487, 1019, \dodoi{10.1051/0004-6361:20079045}

\bibitem[{{Garnir} {et~al.}(1987){Garnir}, {Baudinet-Robinet}, \& {Dumont}}]{Garnir1987}
{Garnir}, H.-P., {Baudinet-Robinet}, Y., \& {Dumont}, P.-D. 1987, Nuclear Instruments and Methods in Physics Research B, 28, 146, \dodoi{10.1016/0168-583X(87)90051-6}

\bibitem[{{Gatti} {et~al.}(2008){Gatti}, {Natta}, {Randich}, {Testi}, \& {Sacco}}]{Gatti2008}
{Gatti}, T., {Natta}, A., {Randich}, S., {Testi}, L., \& {Sacco}, G. 2008, \aap, 481, 423, \dodoi{10.1051/0004-6361:20078971}

\bibitem[{{Giannini} {et~al.}(2019){Giannini}, {Nisini}, {Antoniucci}, {Biazzo}, {Alcal{\'a}}, {Bacciotti}, {Fedele}, {Frasca}, {Harutyunyan}, {Munari}, {Rigliaco}, \& {Vitali}}]{Giannini2019}
{Giannini}, T., {Nisini}, B., {Antoniucci}, S., {et~al.} 2019, \aap, 631, A44, \dodoi{10.1051/0004-6361/201936085}

\bibitem[{{G{\"u}del} {et~al.}(2007){G{\"u}del}, {Briggs}, {Arzner}, {Audard}, {Bouvier}, {Feigelson}, {Franciosini}, {Glauser}, {Grosso}, {Micela}, {Monin}, {Montmerle}, {Padgett}, {Palla}, {Pillitteri}, {Rebull}, {Scelsi}, {Silva}, {Skinner}, {Stelzer}, \& {Telleschi}}]{Gudel2007}
{G{\"u}del}, M., {Briggs}, K.~R., {Arzner}, K., {et~al.} 2007, \aap, 468, 353, \dodoi{10.1051/0004-6361:20065724}

\bibitem[{{Hamann} {et~al.}(1994){Hamann}, {Simon}, {Carr}, \& {Prato}}]{Hamann1994}
{Hamann}, F., {Simon}, M., {Carr}, J.~S., \& {Prato}, L. 1994, \apj, 436, 292, \dodoi{10.1086/174904}

\bibitem[{Harris {et~al.}(2020)Harris, Millman, van~der Walt, Gommers, Virtanen, Cournapeau, Wieser, Taylor, Berg, Smith, Kern, Picus, Hoyer, van Kerkwijk, Brett, Haldane, del R{\'{i}}o, Wiebe, Peterson, G{\'{e}}rard-Marchant, Sheppard, Reddy, Weckesser, Abbasi, Gohlke, \& Oliphant}]{harris2020array}
Harris, C.~R., Millman, K.~J., van~der Walt, S.~J., {et~al.} 2020, Nature, 585, 357, \dodoi{10.1038/s41586-020-2649-2}

\bibitem[{{Harris} {et~al.}(2012){Harris}, {Andrews}, {Wilner}, \& {Kraus}}]{Harris2012}
{Harris}, R.~J., {Andrews}, S.~M., {Wilner}, D.~J., \& {Kraus}, A.~L. 2012, \apj, 751, 115, \dodoi{10.1088/0004-637X/751/2/115}

\bibitem[{{Hartigan} {et~al.}(1995){Hartigan}, {Edwards}, \& {Ghandour}}]{Hartigan1995}
{Hartigan}, P., {Edwards}, S., \& {Ghandour}, L. 1995, \apj, 452, 736, \dodoi{10.1086/176344}

\bibitem[{{Hartigan} {et~al.}(1984){Hartigan}, {Lada}, {Stocke}, \& {Tapia}}]{Hartigan1984}
{Hartigan}, P., {Lada}, C.~J., {Stocke}, J., \& {Tapia}, S. 1984, in Bulletin of the American Astronomical Society, Vol.~16, 998

\bibitem[{{Hartigan} \& {Morse}(2007)}]{Hartigan2007}
{Hartigan}, P., \& {Morse}, J. 2007, \apj, 660, 426, \dodoi{10.1086/513015}

\bibitem[{{Hartigan} {et~al.}(1994){Hartigan}, {Morse}, \& {Raymond}}]{Hartigan1994}
{Hartigan}, P., {Morse}, J.~A., \& {Raymond}, J. 1994, \apj, 436, 125, \dodoi{10.1086/174887}

\bibitem[{{Heathcote} {et~al.}(1998){Heathcote}, {Reipurth}, \& {Raga}}]{Heathcote1998}
{Heathcote}, S., {Reipurth}, B., \& {Raga}, A.~C. 1998, \aj, 116, 1940, \dodoi{10.1086/300548}

\bibitem[{{Hensley} \& {Draine}(2020)}]{Hensley2020}
{Hensley}, B.~S., \& {Draine}, B.~T. 2020, \apj, 895, 38, \dodoi{10.3847/1538-4357/ab8cc3}

\bibitem[{{Herczeg} \& {Hillenbrand}(2008)}]{Herczeg2008}
{Herczeg}, G.~J., \& {Hillenbrand}, L.~A. 2008, \apj, 681, 594, \dodoi{10.1086/586728}

\bibitem[{{Herczeg} \& {Hillenbrand}(2014)}]{Herczeg2014}
---. 2014, \apj, 786, 97, \dodoi{10.1088/0004-637X/786/2/97}

\bibitem[{{Hirsh} {et~al.}(2020){Hirsh}, {Price}, {Gonzalez}, {Ubeira-Gabellini}, \& {Ragusa}}]{Hirsh2020}
{Hirsh}, K., {Price}, D.~J., {Gonzalez}, J.-F., {Ubeira-Gabellini}, M.~G., \& {Ragusa}, E. 2020, \mnras, 498, 2936, \dodoi{10.1093/mnras/staa2536}

\bibitem[{{Hirth} {et~al.}(1994){Hirth}, {Mundt}, {Solf}, \& {Ray}}]{Hirth1994}
{Hirth}, G.~A., {Mundt}, R., {Solf}, J., \& {Ray}, T.~P. 1994, \apjl, 427, L99, \dodoi{10.1086/187374}

\bibitem[{Hunter(2007)}]{Hunter:2007}
Hunter, J.~D. 2007, Computing in Science \& Engineering, 9, 90, \dodoi{10.1109/MCSE.2007.55}

\bibitem[{{Katoh} {et~al.}(2024){Katoh}, {Yasui}, {Ikeda}, {Kobayashi}, {Matsunaga}, {Kondo}, {Sameshima}, {Hamano}, {Mizumoto}, {Kawakita}, {Fukue}, {Otsubo}, \& {Takenaka}}]{Katoh2024}
{Katoh}, H., {Yasui}, C., {Ikeda}, Y., {et~al.} 2024, \apj, 965, 70, \dodoi{10.3847/1538-4357/ad2842}

\bibitem[{{Koo} {et~al.}(2016){Koo}, {Raymond}, \& {Kim}}]{Koo2016}
{Koo}, B.-C., {Raymond}, J.~C., \& {Kim}, H.-J. 2016, Journal of Korean Astronomical Society, 49, 109, \dodoi{10.5303/JKAS.2016.49.3.109}

\bibitem[{{Lavalley-Fouquet} {et~al.}(2000){Lavalley-Fouquet}, {Cabrit}, \& {Dougados}}]{Lavalley-Fouquet2000}
{Lavalley-Fouquet}, C., {Cabrit}, S., \& {Dougados}, C. 2000, \aap, 356, L41

\bibitem[{{Lee} {et~al.}(2010){Lee}, {Hasegawa}, {Hirano}, {Palau}, {Shang}, {Ho}, \& {Zhang}}]{Lee2010}
{Lee}, C.-F., {Hasegawa}, T.~I., {Hirano}, N., {et~al.} 2010, \apj, 713, 731, \dodoi{10.1088/0004-637X/713/2/731}

\bibitem[{{Lee} {et~al.}(2017){Lee}, {Ho}, {Li}, {Hirano}, {Zhang}, \& {Shang}}]{Lee2017}
{Lee}, C.-F., {Ho}, P. T.~P., {Li}, Z.-Y., {et~al.} 2017, Nature Astronomy, 1, 0152, \dodoi{10.1038/s41550-017-0152}

\bibitem[{{Levine} {et~al.}(2024){Levine}, {Vissapragada}, {Feinstein}, {King}, {Hernandez}, {Corrales}, {Greklek-McKeon}, \& {Knutson}}]{Levine2024}
{Levine}, W.~G., {Vissapragada}, S., {Feinstein}, A.~D., {et~al.} 2024, \aj, 168, 65, \dodoi{10.3847/1538-3881/ad5354}

\bibitem[{{Lin} {et~al.}(2023){Lin}, {Li}, {Tobin}, {Ohashi}, {J{\o}rgensen}, {Looney}, {Aso}, {Takakuwa}, {Aikawa}, {van't Hoff}, {de Gregorio-Monsalvo}, {Encalada}, {Flores}, {Gavino}, {Han}, {Kido}, {Koch}, {Kwon}, {Lai}, {Lee}, {Lee}, {Phuong}, {Sai}, {Sharma}, {Sheehan}, {Thieme}, {Williams}, {Yamato}, \& {Yen}}]{Lin2023}
{Lin}, Z.-Y.~D., {Li}, Z.-Y., {Tobin}, J.~J., {et~al.} 2023, \apj, 951, 9, \dodoi{10.3847/1538-4357/acd5c9}

\bibitem[{{Liu} {et~al.}(2012){Liu}, {Shang}, {Pyo}, {Takami}, {Walter}, {Yan}, {Wang}, {Ohashi}, \& {Hayashi}}]{Liu2012}
{Liu}, C.-F., {Shang}, H., {Pyo}, T.-S., {et~al.} 2012, \apj, 749, 62, \dodoi{10.1088/0004-637X/749/1/62}

\bibitem[{{L{\'o}pez-V{\'a}zquez} {et~al.}(2024){L{\'o}pez-V{\'a}zquez}, {Lee}, {Fern{\'a}ndez-L{\'o}pez}, {Louvet}, {Guerra-Alvarado}, \& {Zapata}}]{Lopez-Vazquez2024}
{L{\'o}pez-V{\'a}zquez}, J.~A., {Lee}, C.-F., {Fern{\'a}ndez-L{\'o}pez}, M., {et~al.} 2024, \apj, 962, 28, \dodoi{10.3847/1538-4357/ad132a}

\bibitem[{{Louvet} {et~al.}(2018){Louvet}, {Dougados}, {Cabrit}, {Mardones}, {M{\'e}nard}, {Tabone}, {Pinte}, \& {Dent}}]{Louvet2018}
{Louvet}, F., {Dougados}, C., {Cabrit}, S., {et~al.} 2018, \aap, 618, A120, \dodoi{10.1051/0004-6361/201731733}

\bibitem[{{Luhman}(2004)}]{Luhman2004}
{Luhman}, K.~L. 2004, \apj, 617, 1216, \dodoi{10.1086/425647}

\bibitem[{{Luridiana} {et~al.}(2015){Luridiana}, {Morisset}, \& {Shaw}}]{Luridiana2015}
{Luridiana}, V., {Morisset}, C., \& {Shaw}, R.~A. 2015, \aap, 573, A42, \dodoi{10.1051/0004-6361/201323152}

\bibitem[{{Manara} {et~al.}(2023){Manara}, {Ansdell}, {Rosotti}, {Hughes}, {Armitage}, {Lodato}, \& {Williams}}]{Manara2023}
{Manara}, C.~F., {Ansdell}, M., {Rosotti}, G.~P., {et~al.} 2023, in Astronomical Society of the Pacific Conference Series, Vol. 534, Protostars and Planets VII, ed. S.~{Inutsuka}, Y.~{Aikawa}, T.~{Muto}, K.~{Tomida}, \& M.~{Tamura}, 539, \dodoi{10.48550/arXiv.2203.09930}

\bibitem[{{Manara} {et~al.}(2016){Manara}, {Rosotti}, {Testi}, {Natta}, {Alcal{\'a}}, {Williams}, {Ansdell}, {Miotello}, {van der Marel}, {Tazzari}, {Carpenter}, {Guidi}, {Mathews}, {Oliveira}, {Prusti}, \& {van Dishoeck}}]{Manara2016}
{Manara}, C.~F., {Rosotti}, G., {Testi}, L., {et~al.} 2016, \aap, 591, L3, \dodoi{10.1051/0004-6361/201628549}

\bibitem[{{Masciadri} \& {Raga}(2002)}]{Masciadri2002}
{Masciadri}, E., \& {Raga}, A.~C. 2002, \apj, 568, 733, \dodoi{10.1086/338767}

\bibitem[{{Mathis}(1990)}]{Mathis1990}
{Mathis}, J.~S. 1990, \araa, 28, 37, \dodoi{10.1146/annurev.aa.28.090190.000345}

\bibitem[{{Mundt} {et~al.}(1987){Mundt}, {Brugel}, \& {Buehrke}}]{Mundt1987}
{Mundt}, R., {Brugel}, E.~W., \& {Buehrke}, T. 1987, \apj, 319, 275, \dodoi{10.1086/165453}

\bibitem[{{Mundt} {et~al.}(1984){Mundt}, {Buehrke}, {Fried}, {Neckel}, {Sarcander}, \& {Stocke}}]{Mundt1984}
{Mundt}, R., {Buehrke}, T., {Fried}, J.~W., {et~al.} 1984, \aap, 140, 17

\bibitem[{{Mundt} {et~al.}(1990){Mundt}, {Buehrke}, {Solf}, {Ray}, \& {Raga}}]{Mundt1990}
{Mundt}, R., {Buehrke}, T., {Solf}, J., {Ray}, T.~P., \& {Raga}, A.~C. 1990, \aap, 232, 37

\bibitem[{{Mundt} \& {Fried}(1983)}]{Mundt1983}
{Mundt}, R., \& {Fried}, J.~W. 1983, \apjl, 274, L83, \dodoi{10.1086/184155}

\bibitem[{{Mundt} {et~al.}(1991){Mundt}, {Ray}, \& {Raga}}]{Mundt1991}
{Mundt}, R., {Ray}, T.~P., \& {Raga}, A.~C. 1991, \aap, 252, 740

\bibitem[{{Murphy} {et~al.}(2021){Murphy}, {Dougados}, {Whelan}, {Bacciotti}, {Coffey}, {Comer{\'o}n}, {Eisl{\"o}ffel}, \& {Ray}}]{Murphy2021}
{Murphy}, A., {Dougados}, C., {Whelan}, E.~T., {et~al.} 2021, \aap, 652, A119, \dodoi{10.1051/0004-6361/202141315}

\bibitem[{{Natta} {et~al.}(2006){Natta}, {Testi}, \& {Randich}}]{Natta2006}
{Natta}, A., {Testi}, L., \& {Randich}, S. 2006, \aap, 452, 245, \dodoi{10.1051/0004-6361:20054706}

\bibitem[{{Ninan} {et~al.}(2020){Ninan}, {Stefansson}, {Mahadevan}, {Bender}, {Robertson}, {Ramsey}, {Terrien}, {Wright}, {Diddams}, {Kanodia}, {Cochran}, {Endl}, {Ford}, {Fredrick}, {Halverson}, {Hearty}, {Jennings}, {Kaplan}, {Lubar}, {Metcalf}, {Monson}, {Nitroy}, {Roy}, \& {Schwab}}]{Ninan2020}
{Ninan}, J.~P., {Stefansson}, G., {Mahadevan}, S., {et~al.} 2020, \apj, 894, 97, \dodoi{10.3847/1538-4357/ab8559}

\bibitem[{{Nisini} {et~al.}(2018){Nisini}, {Antoniucci}, {Alcal{\'a}}, {Giannini}, {Manara}, {Natta}, {Fedele}, \& {Biazzo}}]{Nisini2018}
{Nisini}, B., {Antoniucci}, S., {Alcal{\'a}}, J.~M., {et~al.} 2018, \aap, 609, A87, \dodoi{10.1051/0004-6361/201730834}

\bibitem[{{Nisini} {et~al.}(2005){Nisini}, {Bacciotti}, {Giannini}, {Massi}, {Eisl{\"o}ffel}, {Podio}, \& {Ray}}]{Nisini2005}
{Nisini}, B., {Bacciotti}, F., {Giannini}, T., {et~al.} 2005, \aap, 441, 159, \dodoi{10.1051/0004-6361:20053097}

\bibitem[{{Nussbaumer} \& {Storey}(1988)}]{Nussbaumer1988}
{Nussbaumer}, H., \& {Storey}, P.~J. 1988, \aap, 193, 327

\bibitem[{{Pascucci} {et~al.}(2024){Pascucci}, {Beck}, {Cabrit}, {Bajaj}, {Edwards}, {Fabien}, \& {Najita}}]{Pascucci2024}
{Pascucci}, I., {Beck}, T., {Cabrit}, S., {et~al.} 2024, Nature Astronomy, \dodoi{10.1038/s41550-024-02385-7}

\bibitem[{{Pascucci} {et~al.}(2023){Pascucci}, {Cabrit}, {Edwards}, {Gorti}, {Gressel}, \& {Suzuki}}]{Pascucci2023}
{Pascucci}, I., {Cabrit}, S., {Edwards}, S., {et~al.} 2023, in Astronomical Society of the Pacific Conference Series, Vol. 534, Protostars and Planets VII, ed. S.~{Inutsuka}, Y.~{Aikawa}, T.~{Muto}, K.~{Tomida}, \& M.~{Tamura}, 567, \dodoi{10.48550/arXiv.2203.10068}

\bibitem[{{Pascucci} {et~al.}(2021){Pascucci}, {Beck}, {Brittain}, {Cabrit}, {Edwards}, {Gorti}, {Krijt}, {Muzerolle}, {Najita}, {Ruaud}, {Salyk}, {Schwarz}, {Semenov}, \& {Testi}}]{Pascucci2021}
{Pascucci}, I., {Beck}, T., {Brittain}, S., {et~al.} 2021, {Testing the emerging paradigm of wind-driven accretion with NIRSpec spectro-imaging}, JWST Proposal. Cycle 1, ID. \#1621

\bibitem[{Pedregosa {et~al.}(2011)Pedregosa, Varoquaux, Gramfort, Michel, Thirion, Grisel, Blondel, Prettenhofer, Weiss, Dubourg, Vanderplas, Passos, Cournapeau, Brucher, Perrot, \& Duchesnay}]{scikit-learn}
Pedregosa, F., Varoquaux, G., Gramfort, A., {et~al.} 2011, Journal of Machine Learning Research, 12, 2825

\bibitem[{{Peeters} {et~al.}(2024){Peeters}, {Habart}, {Bern{\'e}}, {Sidhu}, {Chown}, {Van De Putte}, {Trahin}, {Schroetter}, {Canin}, {Alarc{\'o}n}, {Schefter}, {Khan}, {Pasquini}, {Tielens}, {Wolfire}, {Dartois}, {Goicoechea}, {Maragkoudakis}, {Onaka}, {Pound}, {Vicente}, {Abergel}, {Bergin}, {Bernard-Salas}, {Boersma}, {Bron}, {Cami}, {Cuadrado}, {Dicken}, {Elyajouri}, {Fuente}, {Gordon}, {Issa}, {Joblin}, {Kannavou}, {Lacinbala}, {Languignon}, {Le Gal}, {Meshaka}, {Okada}, {Robberto}, {R{\"o}llig}, {Schirmer}, {Tabone}, {Zannese}, {Aleman}, {Allamandola}, {Auchettl}, {Baratta}, {Bejaoui}, {Bera}, {Black}, {Boulanger}, {Bouwman}, {Brandl}, {Brechignac}, {Br{\"u}nken}, {Buragohain}, {Burkhardt}, {Candian}, {Cazaux}, {Cernicharo}, {Chabot}, {Chakraborty}, {Champion}, {Colgan}, {Cooke}, {Coutens}, {Cox}, {Demyk}, {Meyer}, {Foschino}, {Garc{\'\i}a-Lario}, {Gerin}, {Gottlieb}, {Guillard}, {Gusdorf}, {Hartigan}, {He}, {Herbst}, {Hornekaer}, {J{\"a}ger}, {Janot-Pacheco}, {Kaufman}, {Kendrew}, {Kirsanova},
  {Klaassen}, {Kwok}, {Labiano}, {Lai}, {Lee}, {Lefloch}, {Le Petit}, {Li}, {Linz}, {Mackie}, {Madden}, {Mascetti}, {McGuire}, {Merino}, {Micelotta}, {Misselt}, {Morse}, {Mulas}, {Neelamkodan}, {Ohsawa}, {Paladini}, {Palumbo}, {Pathak}, {Pendleton}, {Petrignani}, {Pino}, {Puga}, {Rangwala}, {Rapacioli}, {Ricca}, {Roman-Duval}, {Roser}, {Roueff}, {Rouill{\'e}}, {Salama}, {Sales}, {Sandstrom}, {Sarre}, {Sciamma-O'Brien}, {Sellgren}, {Shenoy}, {Teyssier}, {Thomas}, {Togi}, {Verstraete}, {Witt}, {Wootten}, {Ysard}, {Zettergren}, {Zhang}, {Zhang}, \& {Zhen}}]{Peeters2024}
{Peeters}, E., {Habart}, E., {Bern{\'e}}, O., {et~al.} 2024, \aap, 685, A74, \dodoi{10.1051/0004-6361/202348244}

\bibitem[{{Perrin} {et~al.}(2014){Perrin}, {Sivaramakrishnan}, {Lajoie}, {Elliott}, {Pueyo}, {Ravindranath}, \& {Albert}}]{Perrin2014}
{Perrin}, M.~D., {Sivaramakrishnan}, A., {Lajoie}, C.-P., {et~al.} 2014, in Society of Photo-Optical Instrumentation Engineers (SPIE) Conference Series, Vol. 9143, Space Telescopes and Instrumentation 2014: Optical, Infrared, and Millimeter Wave, ed. J.~{Oschmann}, Jacobus~M., M.~{Clampin}, G.~G. {Fazio}, \& H.~A. {MacEwen}, 91433X, \dodoi{10.1117/12.2056689}

\bibitem[{{Pesenti} {et~al.}(2003){Pesenti}, {Dougados}, {Cabrit}, {O'Brien}, {Garcia}, \& {Ferreira}}]{Pesenti2003}
{Pesenti}, N., {Dougados}, C., {Cabrit}, S., {et~al.} 2003, \aap, 410, 155, \dodoi{10.1051/0004-6361:20031131}

\bibitem[{{Podio} {et~al.}(2006){Podio}, {Bacciotti}, {Nisini}, {Eisl{\"o}ffel}, {Massi}, {Giannini}, \& {Ray}}]{Podio2006}
{Podio}, L., {Bacciotti}, F., {Nisini}, B., {et~al.} 2006, \aap, 456, 189, \dodoi{10.1051/0004-6361:20054156}

\bibitem[{{Podio} {et~al.}(2011){Podio}, {Eisl{\"o}ffel}, {Melnikov}, {Hodapp}, \& {Bacciotti}}]{Podio2011}
{Podio}, L., {Eisl{\"o}ffel}, J., {Melnikov}, S., {Hodapp}, K.~W., \& {Bacciotti}, F. 2011, \aap, 527, A13, \dodoi{10.1051/0004-6361/201016049}

\bibitem[{{Podio} {et~al.}(2008){Podio}, {Garcia}, {Bacciotti}, {Antoniucci}, {Nisini}, {Dougados}, \& {Takami}}]{Podio2008}
{Podio}, L., {Garcia}, P.~J.~V., {Bacciotti}, F., {et~al.} 2008, \aap, 480, 421, \dodoi{10.1051/0004-6361:20078694}

\bibitem[{{Porter} {et~al.}(2004){Porter}, {Oudmaijer}, \& {Baines}}]{Porter2004}
{Porter}, J.~M., {Oudmaijer}, R.~D., \& {Baines}, D. 2004, \aap, 428, 327, \dodoi{10.1051/0004-6361:20035686}

\bibitem[{{Pradhan} \& {Zhang}(1993)}]{Pradhan1993}
{Pradhan}, A.~K., \& {Zhang}, H.~L. 1993, \apjl, 409, L77, \dodoi{10.1086/186864}

\bibitem[{{Raga} {et~al.}(1996){Raga}, {B{\"o}hm}, \& {Cant{\'o}}}]{Raga1996}
{Raga}, A.~C., {B{\"o}hm}, K.~H., \& {Cant{\'o}}, J. 1996, \rmxaa, 32, 161

\bibitem[{{Rauscher}(2024)}]{Rauscher2024}
{Rauscher}, B.~J. 2024, \pasp, 136, 015001, \dodoi{10.1088/1538-3873/ad1b36}

\bibitem[{{Ray} \& {Ferreira}(2021)}]{Ray2021}
{Ray}, T.~P., \& {Ferreira}, J. 2021, \nar, 93, 101615, \dodoi{10.1016/j.newar.2021.101615}

\bibitem[{{Ray} {et~al.}(1996){Ray}, {Mundt}, {Dyson}, {Falle}, \& {Raga}}]{Ray1996}
{Ray}, T.~P., {Mundt}, R., {Dyson}, J.~E., {Falle}, S. A.~E.~G., \& {Raga}, A.~C. 1996, \apjl, 468, L103, \dodoi{10.1086/310239}

\bibitem[{{Raymond}(1979)}]{Raymond1979}
{Raymond}, J.~C. 1979, \apjs, 39, 1, \dodoi{10.1086/190562}

\bibitem[{{Reiter} {et~al.}(2015){Reiter}, {Smith}, {Kiminki}, \& {Bally}}]{Reiter2015}
{Reiter}, M., {Smith}, N., {Kiminki}, M.~M., \& {Bally}, J. 2015, \mnras, 450, 564, \dodoi{10.1093/mnras/stv634}

\bibitem[{{Rieke} \& {Lebofsky}(1985)}]{Rieke1985}
{Rieke}, G.~H., \& {Lebofsky}, M.~J. 1985, \apj, 288, 618, \dodoi{10.1086/162827}

\bibitem[{{Rigliaco} {et~al.}(2015){Rigliaco}, {Pascucci}, {Duchene}, {Edwards}, {Ardila}, {Grady}, {Mendigut{\'\i}a}, {Montesinos}, {Mulders}, {Najita}, {Carpenter}, {Furlan}, {Gorti}, {Meijerink}, \& {Meyer}}]{Rigliaco2015}
{Rigliaco}, E., {Pascucci}, I., {Duchene}, G., {et~al.} 2015, \apj, 801, 31, \dodoi{10.1088/0004-637X/801/1/31}

\bibitem[{{Rogers} {et~al.}(2024){Rogers}, {de Marchi}, \& {Brandl}}]{Rogers2024}
{Rogers}, C., {de Marchi}, G., \& {Brandl}, B. 2024, \aap, 684, L8, \dodoi{10.1051/0004-6361/202449282}

\bibitem[{{Rubinstein}(2021)}]{Rubinstein2021}
{Rubinstein}, A.~E. 2021, Research Notes of the American Astronomical Society, 5, 214, \dodoi{10.3847/2515-5172/ac283f}

\bibitem[{{Rynkun} {et~al.}(2019){Rynkun}, {Gaigalas}, \& {J{\"o}nsson}}]{Rynkun2019}
{Rynkun}, P., {Gaigalas}, G., \& {J{\"o}nsson}, P. 2019, \aap, 623, A155, \dodoi{10.1051/0004-6361/201834931}

\bibitem[{{Shinn} {et~al.}(2013){Shinn}, {Pyo}, {Lee}, {Lee}, {Kim}, {Koo}, {Sung}, {Chun}, {Lyo}, {Moon}, {Kyeong}, {Park}, {Hur}, \& {Lee}}]{Shinn2013}
{Shinn}, J.-H., {Pyo}, T.-S., {Lee}, J.-J., {et~al.} 2013, \apj, 777, 45, \dodoi{10.1088/0004-637X/777/1/45}

\bibitem[{{Sutherland} {et~al.}(2018){Sutherland}, {Dopita}, {Binette}, \& {Groves}}]{Sutherland2018}
{Sutherland}, R., {Dopita}, M., {Binette}, L., \& {Groves}, B. 2018, {MAPPINGS V: Astrophysical plasma modeling code}, Astrophysics Source Code Library, record ascl:1807.005

\bibitem[{{Sutherland} \& {Dopita}(2017)}]{Sutherland2017}
{Sutherland}, R.~S., \& {Dopita}, M.~A. 2017, \apjs, 229, 34, \dodoi{10.3847/1538-4365/aa6541}

\bibitem[{{Tabone} {et~al.}(2017){Tabone}, {Cabrit}, {Bianchi}, {Ferreira}, {Pineau des For{\^e}ts}, {Codella}, {Gusdorf}, {Gueth}, {Podio}, \& {Chapillon}}]{Tabone2017}
{Tabone}, B., {Cabrit}, S., {Bianchi}, E., {et~al.} 2017, \aap, 607, L6, \dodoi{10.1051/0004-6361/201731691}

\bibitem[{{Takami} {et~al.}(2003){Takami}, {Bailey}, \& {Chrysostomou}}]{Takami2003}
{Takami}, M., {Bailey}, J., \& {Chrysostomou}, A. 2003, \aap, 397, 675, \dodoi{10.1051/0004-6361:20021544}

\bibitem[{{Takami} {et~al.}(2002){Takami}, {Chrysostomou}, {Bailey}, {Gledhill}, {Tamura}, \& {Terada}}]{Takami2002b}
{Takami}, M., {Chrysostomou}, A., {Bailey}, J., {et~al.} 2002, \apjl, 568, L53, \dodoi{10.1086/340277}

\bibitem[{{Takami} {et~al.}(2006){Takami}, {Chrysostomou}, {Ray}, {Davis}, {Dent}, {Bailey}, {Tamura}, {Terada}, \& {Pyo}}]{Takami2006}
{Takami}, M., {Chrysostomou}, A., {Ray}, T.~P., {et~al.} 2006, \apj, 641, 357, \dodoi{10.1086/500352}

\bibitem[{{Tayal} \& {Zatsarinny}(2010)}]{Tayal2010}
{Tayal}, S.~S., \& {Zatsarinny}, O. 2010, \apjs, 188, 32, \dodoi{10.1088/0067-0049/188/1/32}

\bibitem[{Tayal \& Zatsarinny(2018)}]{Tayal2018}
Tayal, S.~S., \& Zatsarinny, O. 2018, Phys. Rev. A, 98, 012706, \dodoi{10.1103/PhysRevA.98.012706}

\bibitem[{{Testi} {et~al.}(2022){Testi}, {Natta}, {Manara}, {de Gregorio Monsalvo}, {Lodato}, {Lopez}, {Muzic}, {Pascucci}, {Sanchis}, {Miranda}, {Scholz}, {De Simone}, \& {Williams}}]{Testi2022}
{Testi}, L., {Natta}, A., {Manara}, C.~F., {et~al.} 2022, \aap, 663, A98, \dodoi{10.1051/0004-6361/202141380}

\bibitem[{{Villenave} {et~al.}(2020){Villenave}, {M{\'e}nard}, {Dent}, {Duch{\^e}ne}, {Stapelfeldt}, {Benisty}, {Boehler}, {van der Plas}, {Pinte}, {Telkamp}, {Wolff}, {Flores}, {Lesur}, {Louvet}, {Riols}, {Dougados}, {Williams}, \& {Padgett}}]{Villenave2020}
{Villenave}, M., {M{\'e}nard}, F., {Dent}, W.~R.~F., {et~al.} 2020, \aap, 642, A164, \dodoi{10.1051/0004-6361/202038087}

\bibitem[{Virtanen {et~al.}(2020)Virtanen, Gommers, Oliphant, Haberland, Reddy, Cournapeau, Burovski, Peterson, Weckesser, Bright, {van der Walt}, Brett, Wilson, Millman, Mayorov, Nelson, Jones, Kern, Larson, Carey, Polat, Feng, Moore, {VanderPlas}, Laxalde, Perktold, Cimrman, Henriksen, Quintero, Harris, Archibald, Ribeiro, Pedregosa, {van Mulbregt}, \& {SciPy 1.0 Contributors}}]{2020SciPy-NMeth}
Virtanen, P., Gommers, R., Oliphant, T.~E., {et~al.} 2020, Nature Methods, 17, 261, \dodoi{10.1038/s41592-019-0686-2}

\bibitem[{{Walmsley} {et~al.}(2000){Walmsley}, {Natta}, {Oliva}, \& {Testi}}]{Walmsley2000}
{Walmsley}, C.~M., {Natta}, A., {Oliva}, E., \& {Testi}, L. 2000, \aap, 364, 301

\bibitem[{{Whelan} {et~al.}(2014){Whelan}, {Bonito}, {Antoniucci}, {Alcal{\'a}}, {Giannini}, {Nisini}, {Bacciotti}, {Podio}, {Stelzer}, \& {Comer{\'o}n}}]{Whelan2014}
{Whelan}, E.~T., {Bonito}, R., {Antoniucci}, S., {et~al.} 2014, \aap, 565, A80, \dodoi{10.1051/0004-6361/201322037}

\bibitem[{{Woitas} {et~al.}(2005){Woitas}, {Bacciotti}, {Ray}, {Marconi}, {Coffey}, \& {Eisl{\"o}ffel}}]{Woitas2005}
{Woitas}, J., {Bacciotti}, F., {Ray}, T.~P., {et~al.} 2005, \aap, 432, 149, \dodoi{10.1051/0004-6361:20034439}

\bibitem[{{Zinnecker} {et~al.}(1998){Zinnecker}, {McCaughrean}, \& {Rayner}}]{Zinnecker1998}
{Zinnecker}, H., {McCaughrean}, M.~J., \& {Rayner}, J.~T. 1998, \nat, 394, 862, \dodoi{10.1038/29716}

\end{thebibliography}
\bibliographystyle{aasjournal}

\appendix
\section{Line Fluxes}

\begin{longtable}{p{1cm}p{1cm}p{1cm}p{1.3cm}p{1cm}p{1cm}|p{1cm}p{1cm}|p{1cm}p{1cm}|p{1cm}p{1cm}|p{1cm}p{1cm}}
\caption{Integrated line fluxes}
    \\ \hline
    Species & Upper Level & Lower Level & Einstein Coeff$^a$ & Upper level energy$^a$ & Listed wave & \multicolumn{2}{c}{HH 30}  & \multicolumn{2}{c}{FS TauB} & \multicolumn{2}{c}{Tau 042021} & \multicolumn{2}{c}{IRAS 04302} \\
    & ~ & ~ & ~ & ~  & ~ & Flux  & Center & Flux & Center & Flux  & Center  & Flux & Center   \\
    & ~  &  ~ & (s$^{-1}$)  & (K)  & (\micron{}) & (10$^{-20}$ W~m$^{-2}$)  & (\micron{})  & (10$^{-20}$ W~m$^{-2}$)  & (\micron{})     & (10$^{-20}$ W~m$^{-2}$)   & (\micron{})  & (10$^{-20}$  W~m$^{-2}$)   & (\micron{})   \\ 
    \hline
    {[}Fe II{]} & a4D7/2 & a6D9/2 & 4.74E-03  & 11446 & 1.2571 & 236 & 1.25708  & 384 & 1.25697  & 379 & 1.25708 & 184 & 1.25699 \\
    {[}Fe II{]} & a4D7/2 & a6D7/2 & 1.31E-03 & 11446 & 1.3209 & 64.1 & 1.32097  & 105 & 1.32089  & 103 & 1.32097 & 57.6 & 1.32088    \\
    {[}Fe II{]} & a4D7/2 & a6D5/2  & ~ & 11446 & 1.3722      & 40.6 & 1.37225  & 76  & 1.37225  & 95.9 & 1.37217  & 34.4 & 1.37221    \\
    {[}Fe II{]} & a4D7/2 & a4F9/2 & 6.00E-03 & 11446  & 1.644 & 197  & 1.64409  & 392  & 1.64395  & 332  & 1.64409 & 292  & 1.64396    \\
    {[}Fe II{]} & a4D7/2           & a4F7/2      & 1.32E-03             & 11446              & 1.8099      & 42.4            & 1.80998  & 92               & 1.80975  & 73                & 1.80997    & 79.8              & 1.80984    \\
    {[}Fe II{]} & a4D5/2           & a6D7/2      &                      & 12074              & 1.2489      & 4.94            & 1.24894  & 7.04             & 1.24902  & 11.8              & 1.24884    & \textless{}13.6   & \\
    {[}Fe II{]} & a4D5/2           & a6D5/2      & 1.98E-03             & 12074              & 1.2946      & 44              & 1.29469  & 57.6             & 1.29456  & 47.8              & 1.29465    & 27.5              & 1.29455    \\
    {[}Fe II{]} & a4D5/2           & a6D3/2      & 1.17E-03             & 12074              & 1.3281      & 28.1            & 1.32815  & 33               & 1.328    & 37.6              & 1.32819    & 16.5              & 1.32811    \\
    {[}Fe II{]} & a4D5/2           & a4F9/2      & 3.12E-03             & 163383             & 1.5336      & 57.6            & 1.53401  & 85.8             & 1.53382  & 52.9              & 1.53397    & 45.4              & 1.53387    \\
    {[}Fe II{]} & a4D5/2           & a4F7/2      & 2.49E-03             & 12074              & 1.6773      & 33.9            & 1.67742  & 36.9             & 1.67741  & 37.5              & 1.67744    & 37.8              & 1.67732    \\
    {[}Fe II{]} & a4D5/2           & a4F5/2      & 1.82E-03             & 12074              & 1.8005      & 21.9            & 1.8006   & 48.9             & 1.80032  & 31.6              & 1.80055    & 32.5              & 1.80045    \\
    {[}Fe II{]} & a4D3/2           & a6D3/2      & 2.45E-03             & 12489              & 1.2791      & 32.2            & 1.27916  & 33.7             & 1.27902  & 27                & 1.2792     & 16.1              & 1.27904    \\
    {[}Fe II{]} & a4D3/2           & a6D1/2      & 1.08E-03             & 12489              & 1.2981      & 7.69            & 1.2982   & \textless{}46.1  &          & 9.49              & 1.29826    & 22.6              & 1.2985     \\
    {[}Fe II{]} & a4D3/2           & a4F7/2      & 4.20E-03             & 12489              & 1.5999      & 33              & 1.59999  & 80.6             & 1.59986  & 25.4              & 1.60002    & 46.2              & 1.59981    \\
    {[}Fe II{]} & a4D3/2           & a4F3/2      & 2.12E-03             & 12489              & 1.7976      & 20.9            & 1.79764  & 27.6             & 1.79741  & 11.6              & 1.79775    & 16.8              & 1.79749    \\
    {[}Fe II{]} & a4D1/2           & a6D1/2      & 3.32E-03             & 12728              & 1.2707      & 21.8            & 1.27074  & 36.3             & 1.27067  & 20                & 1.27065    & 17.3              & 1.27055    \\
    {[}Fe II{]} & a4D1/2           & a4F5/2      & 4.70E-03             & 12728              & 1.6642      & 22.7            & 1.66434  & 45.9             & 1.66435  & 28.7              & 1.66437    & 19.2              & 1.66428    \\
    {[}Fe II{]} & a4D1/2           & a4F3/2      & 2.47E-03             & 12728              & 1.7454      & 14.9            & 1.74543  & 17.6             & 1.74519  & 17.4              & 1.74559    & \textless{}143    &            \\
    {[}C I{]}   & 1D2              & 3P2         & 2.26E-04             & 14665              & 0.9853      & 344             & 0.98536  & 26.5             & 0.98532  & 58.2              & 0.98532    & 25.6              & 0.98536    \\
    {[}C I{]}   & 1D2              & 3P1         & 7.56E-05             & 14665              & 0.9827      & 96              & 0.98273  & 12.2             & 0.98282  & 12.1              & 0.983      & 2.29              & 0.98284    \\
    {[}N I{]}   & 2D5/2            & 2Do5/2      & 2.06E+07             & 172869             & 1.041       & 227             & 1.04105  & 51.5             & 1.04094  & 24.4              & 1.04147    & 26.3              & 1.04102    \\
    {[}N I{]}   & 2D3/2            & 2Do5/2      & 2.22E+06             & 151277             & 1.0401      & 349             & 1.04011  & 70.8             & 1.04012  & 25.2              & 1.04012    & 37.2              & 1.04012    \\
    {[}P II{]}  & 1D2              & 3P2         & 1.25E-02             & 12779.6            & 1.1886      & 25.2            & 1.18914  & \textless{}22    &          & 15.6              & 1.18875    & 4.49              & 1.18859    \\
    {[}P II{]}  & 1D2              & 3P1         & 4.54E-03             & 12779.6            & 1.147       & 13.3            & 1.14715  & 9.23             & 1.14709  & 6.29              & 1.1471     & 7.11              & 1.14712    \\
    {[}S II{]}  & 2Po3/2           & 2Do5/2      & 1.56E-01             & 35353              & 1.0323      & 583             & 1.03237  & 78.2             & 1.03225  & 20.6              & 1.03241    & 12.7              & 1.03221    \\
    {[}S II{]}  & 2Po3/2           & 2Do3/2      & 1.15E-01             & 35353              & 1.029       & 480             & 1.02899  & 60.8             & 1.02887  & 45.6              & 1.02901    & 21.7              & 1.0289     \\
    {[}S II{]}  & 2Po1/2           & 2Do5/2      & 6.87E-02             & 35286              & 1.0373      & 198             & 1.03737  & 14.6             & 1.0372   & 21.1              & 1.03728    & 11.8              & 1.037      \\
    {[}S II{]}  & 2Po1/2           & 2Do3/2      & 1.43E-01             & 35286              & 1.0339      & 451             & 1.03396  & 54.4             & 1.03386  & 29                & 1.03394    & 17.1              & 1.03382    \\
    CI          & 3Fo2             & 3D1         & 2.01E+07             & 112584             & 1.1751      & 86.5            & 1.17547  & 38.2             & 1.17557  & 11.9              & 1.17528    & \textless{}10.7   &            \\
    CI          & 3D1              & 3Po0        & 9.29E+06             & 100268             & 1.0688      & 120             & 1.06901  & 22.5             & 1.06891  & \textless{}85.6   &            & 1.99              & 1.06898    \\
    CI          & 3Po1              & 3D1        & 6.92E+06             & 100268             & 1.0710      &        36.8      & 1.07096  &  \textless{21.7}            &   &  \textless{49.3}  &            &        \textless{23.4}       &     \\
    He I        & 3Fo              & 3D          &                      & 275456             & 1.869       & 18.5            & 1.86918  & 124              & m        & \textless{}51.1   &            & 15.2              & 1.86913    \\
    He I        & 3Po0$^a$ & 3S1         & 1.02E+07             & 243278             & 1.083       & 1410            & 1.08337  & 1790             & m        & 539               & 1.08337    & 84.8              & 1.08325    \\
    HI          & 20               & 4           & 1.18E+03             & 157409             & 1.5196      & 30.1            & 1.51964  & 63.9             & m        & \textless{}22.8   &            & \textless{}19.4   &            \\
    HI          & 19               & 4           & 1.53E+03             & 157366             & 1.5265      & 16.2            & 1.52643  & 27.8             & m        & \textless{}28.0   &            & \textless{}17.8   &            \\
    HI          & 18               & 4           & 2.01E+03             & 157316             & 1.5346      & 72              & 1.53412  & 105              & m        & 75.6              & 1.53412    & 74.9              & 1.53412    \\
    HI          & 17               & 4           & 2.69E+03             & 157257             & 1.5443      & 38.9            & 1.54433  & 85.1             & m        & \textless{}20.2   &            & 9.01              & 1.54447    \\ 
    HI          & 16               & 4           & 3.67E+03             & 157187             & 1.5561      & 38.5            & 1.55585  & 66.3             & m        & \textless{}22.3   &            & \textless{}24.5   &            \\ \hline
    HI          & 15               & 4           & 5.11E+03             & 157102             & 1.5705      & 65.3            & 1.5705   & 90.3             & m        & 15.5              & 1.57101    & 29.8              & 1.57086    \\ 
    HI          & 14               & 4           & 7.29E+03             & 156998             & 1.5885      & 68              & 1.58838  & 111              & m        & 54.1              & 1.58822    & 29                & 1.58803    \\
    HI          & 13               & 4           & 1.07E+04             & 156869             & 1.6114      & 62.6            & 1.61136  & 113              & m        & 10.8              & 1.61145    & 18.5              & 1.61137    \\
    HI          & 12               & 4           & 1.62E+04             & 156707             & 1.6412      & 49.2            & 1.64107  & 20.8             & 1.64101  & 15.4              & 1.64131    & \textless{}475    &            \\
    HI          & 11               & 4           & 2.56E+04             & 156499             & 1.6811      & 85.5            & 1.68101  & 157              & 1.6811   & 49.4              & 1.68142    & 31.3              & 1.68118    \\
    HI          & 10               & 4           & 4.24E+04             & 156225             & 1.7367      & 105             & 1.73664  & 244              & 1.73665  & 41.6              & 1.7368     & 59.3              & 1.73668    \\
    HI          & 9                & 4           & 7.46E+04             & 155855             & 1.8181      & 126             & 1.81786  & 276              & 1.81791  & 50.1              & 1.8182     & 89.6              & 1.81797    \\
    HI          & 7                & 3           & 3.36E+05             & 152583             & 1.0052      & 342             & 1.00518  & 222              & 1.00518  & 137               & 1.00523    & 31.3              & 1.00516    \\
    HI          & 6                & 3           & 7.78E+05             & 153420             & 1.0941      & 373             & 1.09412  & 343              & 1.09408  & 173               & 1.09418    & 59.9              & 1.09408    \\
    HI          & 5                & 3           & 2.20E+06             & 151491             & 1.2822      & 505             & 1.28213  & 616              & 1.28211  & 274               & 1.28222    & 199               & 1.28215    \\
    HI          & 4                & 3           & 8.99E+06             & 147940             & 1.8756      & 679             & 1.87569  & 1940             & 1.87558  & 574               & 1.87574    & 1370              & 1.87567    \\
    H$_2$?          & v=6, J=4         & v=3, J=4    & 2.45E-07             & 32132              & 1.0467      &     45.3       &  1.04593  &   \textless{15.2}           &   &     \textless{26}          &     &        \textless{12.4}       &     \\
    H$_2$          & v=3, J=7         & v=1, J=5    & 7.53E-07             & 20856              & 1.1519      & 4.53            & 1.15223  & 7.19             & 1.15192  & 2.62              & 1.15209    & 2.19              & 1.15202    \\
    H$_2$          & v=3, J=5         & v=1, J=3    & 6.59E-07             & 19086              & 1.1857      & \textless{}21.6 &          & \textless{}43.3  &          & \textless{}44.2   &            & 3.59              & 1.18589    \\
    H$_2$          & v=2, J=8         & v=0, J=6   & 3.38E-07             &       16880             & 1.0733      &      33.7      & 1.07324  &  15.6            & 1.07316  &  52.3   &     1.07371       &       23.5        &  1.07285   \\
    H$_2$          & v=2, J=7         & v=0, J=5   & 3.28E-07             &                    & 1.0851      & 6.45            & 1.08522  & 4.43             & 1.08514  & \textless{}640    &            & 4.64              & 1.08533    \\
    H$_2$          & v=2, J=5         & v=0, J=5    & 1.23E-07             & 13890              & 1.2636      & 3.43            & 1.26363  & \textless{}67.1  &          & \textless{}42.1   &            & 6.24              & 1.26375    \\
    H$_2$          & v=2, J=5         & v=0, J=3    & 2.77E-07             & 13890              & 1.1175      & 16.9            & 1.11756  & 9.38             & 1.11745  & 5.16              & 1.11747    & 9.03              & 1.11749    \\
    H$_2$          & v=2, J=3         & v=0, J=3    & 1.29E-07             & 12550              & 1.2473      & 6.27            & 1.24733  & 11.5             & 1.24735  & \textless{}27.4   &            & 6.1               & 1.24728    \\
    H$_2$          & v=2, J=3         & v=0, J=1    & 1.90E-07             & 12550              & 1.162       & 21.6            & 1.16232  & 19.6             & 1.16231  & \textless{}20.6   &            & 7.38              & 1.16224    \\
    H$_2$          & v=2, J=1         & v=0, J=1    & 1.94E-07             & 11789              & 1.238       & 16.3            & 1.23839  & 8.26             & 1.23834  & 6.89              & 1.238      & 2.59              & 1.23838    \\
    H$_2$          & v=1, J=10        & v=0, J=8    & 2.34E-07             & 14220              & 1.7147      & 13.9            & 1.71479  & 24.1             & 1.71479  & 42.7              & 1.71507    & 14.6              & 1.71485    \\
    H$_2$          & v=1, J=9         & v=0, J=7    & 2.98E-07             & 12817              & 1.7479      & 32.5            & 1.74804  & 69.4             & 1.74804  & 10.9              & 1.74803    & 67.4              & 1.74802    \\
    H$_2$          & v=1, J=8         & v=0, J=6    & 3.54E-07             & 11521              & 1.788       & 33.1            & 1.78812  & 43.8             & 1.78815  & \textless{}27.3   &            & 43.9              & 1.78815    \\
    H$_2$          & v=1, J=7         & v=0, J=5    & 3.95E-07             & 10341              & 1.8357      & 149             & 1.83587  & 224              & 1.83585  & 34.6              & 1.83586    & 202               & 1.83585    \\
    OI?         & 3Do2$^a$       & 3P1         & 2.12E+07 & 140264             & 1.1289      & 102             & 1.12899  & 63.9             & 1.12899  & 45.1              & 1.12902    & 26.3              & 1.12899 \\ \hline
    \multicolumn{14}{p{19cm}}{NOTE -- For FS~TauB, some lines required two Gaussian components to fit the spectra. Hence, we do not report the center wavelength for these lines and denote them by `m'. $^a$\citep{Peeters2024}}
\label{tab:line_fluxes}
\end{longtable}

\section{Extinction} \label{sec:Extinction_appendix}

To calculate extinction, we use the \feii{} integrated line flux maps described in Section \ref{sec:cont_sub}. We rotate these maps using the function \texttt{scipy.ndimage.rotate}\footnote{\url{https://docs.scipy.org/doc/scipy/reference/generated/scipy.ndimage.rotate.html}} to make the jet emission horizontal in the image plane. This procedure is followed for both 1.257 \micron{} and 1.644 \micron{} lines, and a simple ratio of their integrated fluxes in each pixel is calculated. Next, using Equation \ref{eq:1}, we retrieve an extinction difference map in each pixel. The two \feii{} line centers, 1.257 \micron{} and 1.644 \micron{} fall close to the J and H band wavelengths, respectively. For $\lambda$$>$0.9 \micron{}, which are the focus of our study, the extinction law is independent of the mean extinction factor R$_V$ \citep{Mathis1990}. Following the extinction law provided by \cite{Rieke1985}
, we can write

\begin{equation} \label{eq:2}
    \frac{A_{\lambda_1}}{A_J} = \left(\frac{\lambda_1}{1.25~\micron{}}\right)^{-1.7}~and~\frac{A_{\lambda_2}}{A_H} = \left(\frac{\lambda_2}{1.65~\micron{}}\right)^{-1.7}
\end{equation}

We adopt R$_V$ = 3.1 to convert the J and H band extinction to the V band extinction:

\begin{equation} \label{eq:3}
    A_J = 0.282 A_V~and~A_H = 0.176 A_V
\end{equation}

By substituting Equations \ref{eq:2} and \ref{eq:3} in Equation \ref{eq:1}, we find a relation between the extinction difference at the two wavelengths and the visual extinction in that region as

\begin{equation} \label{eq:4}
    A_V = \left(\frac{A_{\lambda_1} - A_{\lambda_2}}{0.176\left(\frac{\lambda_2}{1.65~\micron{}}\right)^{-1.7} - 0.282\left(\frac{\lambda_1}{1.25~\micron{}}\right)^{-1.7}}\right)
\end{equation}

Recently, based on the observations of Cyg~OB2-12, which is known to be extinct only by the typical interstellar dust, \cite{Hensley2020} presented updated reddening-to-extinction conversions as A$_J$ = 0.264~A$_V$ and A$_H$ = 0.153~A$_V$. Substituting these in Equation \ref{eq:4} results only in a net 5\% difference in the visible extinction. With the help of Equation \ref{eq:4} and the extinction difference map described earlier, we create pixel-by-pixel visual extinction maps for all the sources. 

\section{Comparison of the \feii{} and \sii{} diagnostic line ratios} \label{sec:fe_s_ratio_ne}

\begin{figure*}
    \centering
    \includegraphics[width=\linewidth]{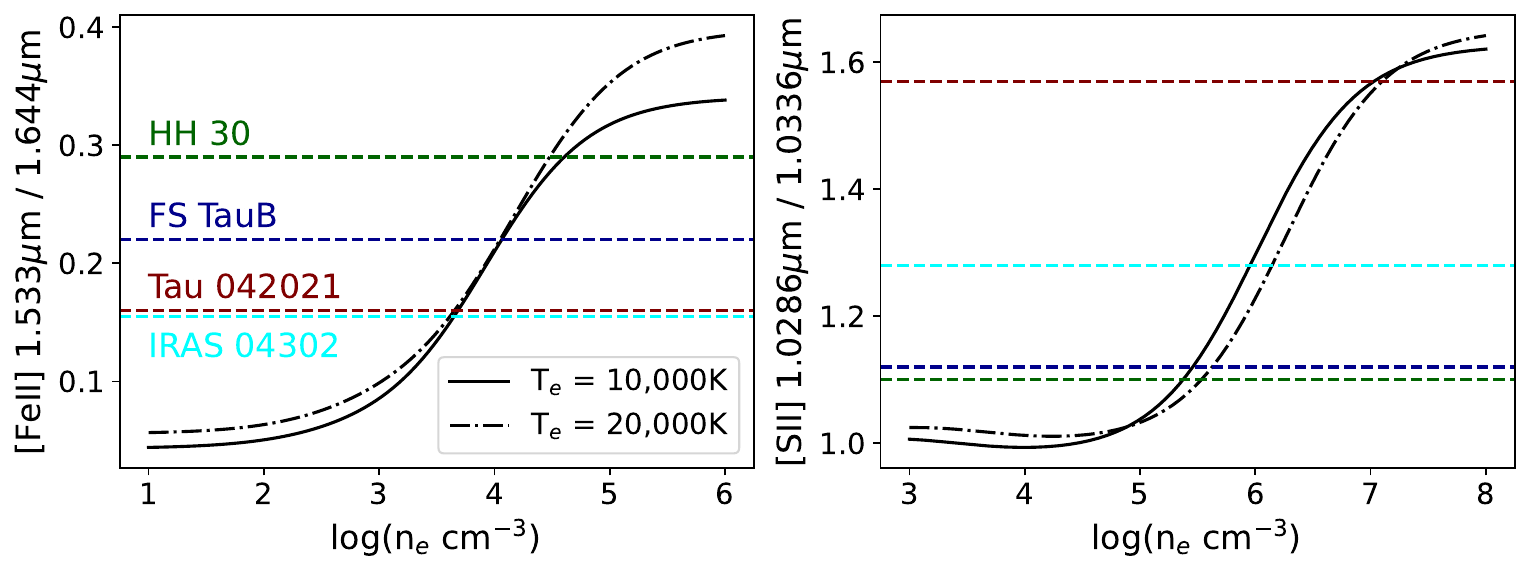}
    \caption{The left and right panels show the variation of \feii{} 1.533~\micron{}/1.644~\micron{} and \sii{} 1.0286~\micron{}/1.0336~\micron{} flux ratios with the electron density, respectively. The solid curves correspond to an electron temperature of 10,000K and the dash-dotted curves to 20,000K. The dashed colored lines show ratios of the total observed fluxes for all sources, as listed in Table \ref{tab:line_fluxes}. The \feii{} curves were created using atomic data and collision strengths from \cite{Tayal2018}. Similarly, the \sii{} curves were made using atomic data from \cite{Rynkun2019} and collision strengths from \cite{Tayal2010}, using \texttt{PyNeb} \citep{Luridiana2015}.}
    \label{fig:Fe_S_ratio_ne}
\end{figure*}

We compare the density diagnosing capability of the \sii{} 1.03 \micron{} and the NIR \feii{} lines by plotting flux ratios as a function of n$_e$ for gas in statistical equilibrium at temperatures 10,000K and 20,000K (Figure \ref{fig:Fe_S_ratio_ne}). These temperatures are representative of those used in Section \ref{sec:both_elec_den} to estimate the electron densities. It can be seen that the \feii{} NIR line ratios are sensitive to gas with electron densities in the range 10$^3$-10$^5$~cm$^{-3}$, whereas, \sii{} NIR line ratios are sensitive to higher electron densities between 10$^5$ and 10$^7$~cm$^{-3}$. This behavior of the NIR \sii{} lines is opposite to the optical \sii{} lines that trace lower densities than NIR \feii{} lines \citep[e.g.,][]{Nisini2005,Podio2006}. In Figure \ref{fig:Fe_S_ratio_ne}, we also show that our observed total line-flux ratios (see Table \ref{tab:line_fluxes}) fall on the linear part of the curves, meaning they are good tracers of the electron density of the regions they are emitted from.

\section{Description of Method 3 for estimating the jet mass loss rate} \label{sec:jmlr_method3_appendix}

Similar to Method 1 (Section \ref{sec:jmlr_method1}), we can write the mass flux entering the shock as 

\begin{equation}
    \dot M_S = \mu m_H n_H v_S A_S
\end{equation}

With $v_S$ and $A_S$ being the shock velocity and the cross-section area, respectively. Using the relation between \feii{} flux and n$_H$v$_S$ \citep[see Figure 7 of][]{Koo2016} for pre-shock densities of 10 and 1000~cm$^{-3}$ and shock speeds in the range 20-100~km~s$^{-1}$ and integrating over the shock area, we derive a direct relation between the mass flux entering the shock and the line luminosity, given as

\begin{equation}
    \frac{\dot M_S}{(M_{\odot}~yr^{-1})} = 7.2 \times 10^{-8}~\left(\frac{L_J}{10^{-4} L_{\odot}}\right) \left(\frac{[Fe]/[H]}{[Fe]/[H]_{\odot}}\right)^{-1}
\end{equation}

The shock model of \cite{Koo2016} uses the Iron rate coefficients and atomic constants derived by \cite{Bautista2015}. At the same time, the jet mass loss rate is related to the mass flux entering the shock by the following equation \citep[see][]{Hartigan1995,Cabrit2002,Agra-Amboage2011}

\begin{equation}
    \dot M_J = \dot M_S \left(\frac{v_J}{v_S}\right) \left(\frac{cos\,\theta}{N_{shock}}\right)
\end{equation}

where $\theta$ = $A_J/A_S$, and $N_{shock}$ is the number of shocks within an aperture. We take $cos\theta$ = 1, i.e. the shock cross-section and the jet cross-section are parallel to each other. We integrate the luminosity over one shock front at a time, leading to $N_{shock}$ = 1. Substituting these values gives the mass loss rate equation given in the main text.

\section{Mass loss rate comparison} \label{sec:jmlr_comparison_appendix}

\begin{figure*}
    \centering
    \subfigure{\includegraphics[width=0.49\textwidth]{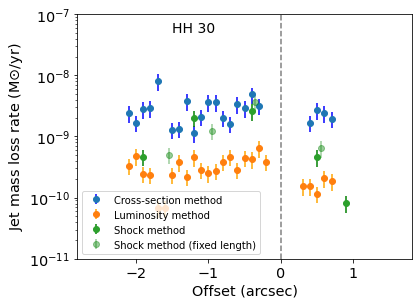}}
    \subfigure{\includegraphics[width=0.49\textwidth]{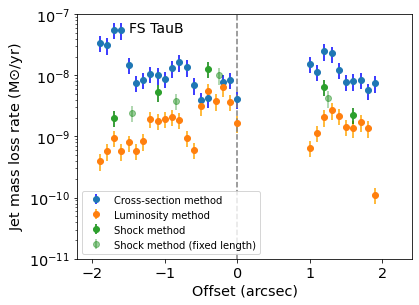}}\\ 
    \subfigure{\includegraphics[width=0.49\textwidth]{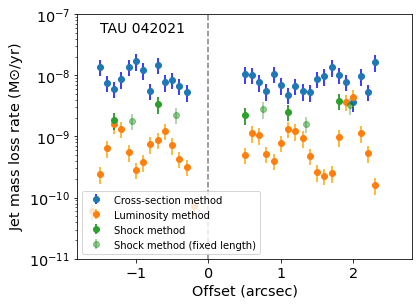}}
    \subfigure{\includegraphics[width=0.49\textwidth]{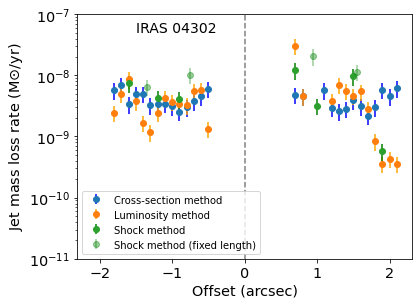}}
    \caption{Each sub-plot shows the jet mass loss rate estimates as a function of distance from the source using three different methods. In blue is method 1 (Section \ref{sec:jmlr_method1}), in orange is method 2 (Section \ref{sec:jmlr_method2}), and in green is method 3 (Section \ref{sec:jmlr_method3}). The error bars represent 30\% uncertainty for HH~30 and FS~TauB and 50\% for Tau~042021 and IRAS~04302, as described in the text.}
    \label{fig:jmlr}
\end{figure*}

The jet mass loss rate profiles along the jet for all the sources are shown in Figure \ref{fig:jmlr}. The error bars represent an uncertainty of 30\% for HH~30 and FS~TauB, and 50\% for Tau~042021 and IRAS~04302 and act as minimum uncertainty propagated through the jet velocities.

\section{Shock Identification} \label{sec:shock_id}

\begin{figure*}
    \centering
    \subfigure{\includegraphics[width=0.49\textwidth]{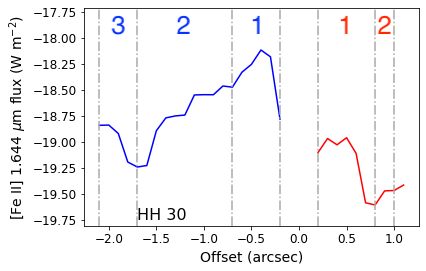}}
    \subfigure{\includegraphics[width=0.49\textwidth]{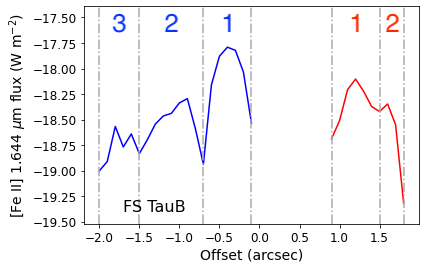}}\\ 
    \subfigure{\includegraphics[width=0.49\textwidth]{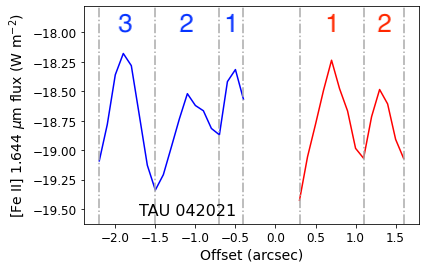}}
    \subfigure{\includegraphics[width=0.49\textwidth]{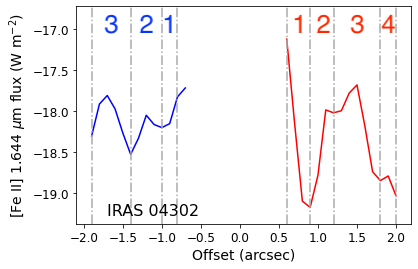}}
    \caption{Each sub-plot shows the shock regions in the blue- and red-shifted jet lobes, identified in the extinction-corrected \feii{} intensity (log) profile plotted as a function of distance from the star. Gray dash-dot lines highlight the local minima used to locate the shock regions.}
    \label{fig:shock_id}
\end{figure*}

As mentioned in Section \ref{sec:jmlr_method3}, we identify a shock region as that between two adjacent local minima. In Figure \ref{fig:shock_id}, we mark these regions, with shock~1 being closest to the source in either direction. E.g., in Tau~042021, the emission between 0.4\arc{} and 1.1\arc{} is one complete shock (shock~1), and that between 1.1\arc{} and 1.6\arc{} is the next (shock~2). To avoid false classification of small flux variations as a shock, we constrain the minimum distance between two consecutive shocks to be $>$28~au (2 pixels). The total luminosity from a shock is calculated using the total flux from a marked region.

\end{document}